\documentclass{article}

\usepackage{ssArxiv}

\usepackage[utf8]{inputenc}

\usepackage[title]{appendix}

\usepackage{amsmath,amsbsy,amsgen,amscd,amssymb,amsthm,amsfonts}
\usepackage{mathtools}
\usepackage{booktabs}
\usepackage{bm}
\usepackage{enumitem}
\usepackage{subcaption, geometry, graphicx, lipsum}
\usepackage{microtype}
\usepackage{multirow}
\usepackage[
  pagebackref=true,
  colorlinks=true,
  citecolor=MidnightBlue,
  linkcolor=Fuchsia,
]{hyperref}
\usepackage{pifont}
\usepackage[nameinlink]{cleveref}
\usepackage{url} 
\usepackage{graphicx}
\usepackage{algorithm,algorithmic}
\usepackage{natbib}
\usepackage{xspace}
\usepackage{caption}
\usepackage{listings}
\definecolor{codegreen}{rgb}{0,0.6,0}
\lstdefinestyle{mystyle}{  
    commentstyle=\color{codegreen},
    keywordstyle=\color{blue},
    basicstyle=\ttfamily\footnotesize,
    breakatwhitespace=false,         
    breaklines=true,                 
    captionpos=b,                    
    keepspaces=true,                 
    showspaces=false,                
    showstringspaces=false,
    showtabs=false,                  
    tabsize=2
}
\usepackage[dvipsnames]{xcolor}
\usepackage[most]{tcolorbox}

\definecolor{lightgreen}{HTML}{C4FFCF}

\crefname{assumption}{assumption}{assumptions}

\newcommand{\red}[1]{\textbf{\textcolor{red}{#1}}}
\newcommand{\green}[1]{\textbf{\textcolor{ForestGreen}{#1}}}

\newcommand{\cmark}{\ding{51}}
\newcommand{\xmark}{\ding{55}}
\newcommand{\benchmark}{\textsc{CRUXEval}\xspace}
\newcommand{\benchmarki}{\textsc{CRUXEval-I}\xspace}
\newcommand{\benchmarko}{\textsc{CRUXEval-O}\xspace}

\newcommand{\codellamamed}{\textsc{Code Llama 13B}\xspace}
\newcommand{\codellamalarge}{\textsc{Code Llama 34B}\xspace}

\makeatletter
\newcommand\nnfootnote[1]{%
  \begin{NoHyper}
  \renewcommand\thefootnote{}\footnote{#1}%
  \addtocounter{footnote}{-1}%
  \end{NoHyper}
}

\makeatother
  
\renewcommand*\backref[1]{\ifx#1\relax \else (Cited on pg. #1) \fi}
\title{CRUXEval: A Benchmark for Code Reasoning, Understanding and Execution \\ \
}

\author{\name Alex Gu$^{\star}$
\email{gua@mit.edu} \\
\addr{MIT CSAIL} \\
\name Baptiste Rozière
\email{broz@meta.com} \\
\addr{Meta AI} \\
\name Hugh Leather
\email{hleather@meta.com} \\
\addr{Meta AI} \\
\name Armando Solar-Lezama
\email{asolar@csail.mit.edu} \\
\addr{MIT CSAIL} \\
\name Gabriel Synnaeve
\email{gab@meta.com} \\
\addr{Meta AI} \\
\name Sida I. Wang
\email{sida@meta.com} \\
\addr{Meta AI} \\
}

\begin{document}

\maketitle
\raggedbottom

\nnfootnote{$^{\star}$ Work primarily done during an internship at Meta AI}

\begin{abstract}
We present \benchmark (\textbf{C}ode \textbf{R}easoning, \textbf{U}nderstanding, and e\textbf{X}ecution \textbf{Eval}uation), a benchmark consisting of 800 Python functions (3-13 lines). Each function comes with an input-output pair, leading to two natural tasks: input prediction and output prediction. First, we propose a generic recipe for generating our execution benchmark which can be used to create future variation of the benchmark. Second, we evaluate twenty code models on our benchmark and discover that many recent high-scoring models on HumanEval do not show the same improvements on our benchmark. Third, we show that simple CoT and fine-tuning schemes can improve performance on our benchmark but remain far from solving it. The best setup, GPT-4 with chain of thought (CoT), achieves a pass@1 of 75\% and 81\% on input and output prediction, respectively. In contrast, Code Llama 34B achieves a pass@1 of 50\% and 46\% on input and output prediction, highlighting the gap between open and closed source models. As no model is close to acing \benchmark, we provide examples of consistent GPT-4 failures on simple programs as a lens into its code reasoning capabilities and areas for improvement. 
\end{abstract}

\section{Introduction} \label{sec:introduction}
In recent months, software engineering and programming have become increasingly mainstream domains for language models (LMs) as they attempt to conquer a potpourri of tasks including code completion, program repair, debugging, test case generation, and code optimization (see \citet{zan2023large} and \citet{fan2023large} for surveys). Recent models including Code Llama \citep{roziere2023code}, GPT-3.5 \citep{brown2020language, ouyang2022training}, and GPT-4 \citep{openai2023gpt} have shown promise in code-related tasks and are being used to develop tools to help programmers write code more efficiently.

The primary way that the community has been evaluating code LMs is via benchmarks such as HumanEval \citep{chen2021evaluating} and MBPP \citep{austin2021program}, which test the ability to generate short code snippets from natural language specifications. While HumanEval and MBPP capture code generation abilities on simple and fundamental tasks, there is an absence of benchmarks capturing other fundamental dimensions of code LMs such as code understanding and execution. 

Motivated by this, we contribute a new benchmark, \benchmark (\textbf{C}ode \textbf{R}easoning, \textbf{U}nderstanding, and e\textbf{X}ecution \textbf{Eval}uation) with two tasks: 1) output prediction, \benchmarko to measure code execution following and 2) input prediction, \benchmarki to measure code reasoning and understanding. An example of a sample in \benchmark is shown in Listings \ref{lst:benchmark-example1} and \ref{lst:benchmark-example} (modified for readability). \benchmark examines the abilities of code LMs to reason about the execution behaviour of \textit{simple} Python programs. While LMs shouldn't be expected to replace an interpreter on arbitrarily complex problems, we ensure the samples in our benchmark are simple (maximum 13 lines, no complex arithmetic) and solvable by a university-level CS graduate without needing more memory (in our opinion). \benchmark provides a useful and important probe for better understanding the capabilities of code LMs, as following a few simple steps of code execution should be a basic requirement for these models. The ability to reason about the execution behavior of code also paves the way to tackling more difficult tasks such as code repair with execution feedback and code summarization.

\begin{minipage}{.48\textwidth}
\begin{lstlisting}[caption={Sample problem},label={lst:benchmark-example1}, captionpos=t, breaklines=true, language=Python]
def f(string):
    string_x = string.rstrip("a")
    string = string_x.rstrip("e")
    return string

# output prediction, CRUXEval-O
assert f("xxxxaaee") == ??
## GPT4: "xxxx", incorrect

# input prediction, CRUXEval-I
assert f(??) == "xxxxaa"
## GPT4: "xxxxaae", correct
\end{lstlisting}
\end{minipage}\hfill
\begin{minipage}{.48\textwidth}
\begin{lstlisting}[caption={Sample problem},label={lst:benchmark-example}, captionpos=t, breaklines=true, language=python]
def f(nums):
    count = len(nums)
    for i in range(-count+1, 0):
        nums.append(nums[i])
    return nums
# output prediction, CRUXEval-O
assert f([2, 6, 1, 3, 1]) == ??
# GPT4: [2, 6, 1, 3, 1, 6, 1, 3, 1], incorrect

# input prediction, CRUXEval-I
assert f(??) == [2, 6, 1, 3, 1, 6, 3, 6, 6]
# GPT4: [2, 6, 1], incorrect
\end{lstlisting}
\end{minipage}

At a high level, our benchmark is constructed as follows. First, we use \codellamalarge to generate a large set of functions and inputs. The outputs are generated by executing the functions on the inputs. Second, we filter the set so that our benchmark only consists of short problems with low computation and memory requirements, problems which a good human programmer should be able to do without extra memory in a minute or so. Third, we randomly select 800 samples passing the filter, ensuring the benchmark is both small enough to easily run but large enough to reliably see performance differences among various models. We use this approach because while it is difficult to manually come up with example where the strongest models like GPT-4 fail completely, we observe that they fail quite often on random yet reasonable programs. We also highlight that as models improve, this generate-and-filter approach can be used to create future benchmarks that are more difficult and test different aspects of program execution.

The best model, GPT-4, achieves a pass@1 of 67\% on \benchmarki and 63\% on \benchmarko. In contrast, the best open-source models only achieve 47\% on \benchmarki and 44\% on \benchmarko, failing over half the time at simple execution prediction and code reasoning despite being trained on ~100G of Python code and ~1T of code data. We also observe that for base models, stronger HumanEval performance correlates with stronger \benchmark performance. However, this trend breaks down for models distilled on GPT-4 like data such as WizardCoder, Phind, and Phi, which have impressively high HumanEval scores but no better than CodeLlama on \benchmark.

We also observe that CoT and fine-tuning on input-output assertions are effective techniques for improving performance on \benchmark, but are far from enough to ace it. Overall, our benchmark reveals that the gap between GPT-4 and open source models reflects GPT-4's stronger ability to reason about the behavior of code. As existing benchmarks like HumanEval and MBPP are insufficient for measuring code understanding and execution ability, capturing it through our benchmark is critical to make progress towards closing the gap between open models and GPT-4. Finally, we discover that despite its impressive abilities, GPT-4 consistently fails to understand the execution behavior of some surprisingly simple Python programs.

\section{Related Work} \label{sec:related-work}

\noindent \textbf{LMs for Code Generation}: There have been many efforts training LMs to generate code. Base models include Codex \citep{chen2021evaluating}, CodeGeeX \citep{zheng2023codegeex}, SantaCoder \citep{allal2023santacoder}, PolyCoder \citep{xu2022systematic}, InCoder \citep{fried2022incoder}, CodeGen \citep{nijkamp2022codegen}, StarCoder \citep{li2023starcoder}, DeepSeek-Coder \citep{deepseek-coder}, and Code Llama \citep{roziere2023code}. Later, some of these models were fine-tuned on instruction-like data distilled from GPT-3.5 and GPT-4, resulting in models like Phind \citep{Phind}, WizardCoder \citep{luo2023wizardcoder}, and Phi-1/Phi-1.5 \citep{li2023textbooks, gunasekar2023textbooks}. We evaluate the performance of a selection of these models on our \benchmark.

\noindent \textbf{Benchmarks for Evaluating Code LMs}: There are various benchmarks serving to evaluate different aspects of these code LMs. We survey a handful here and refer readers to the survey \citep{zhang2023survey} for more. HumanEval \citep{chen2021evaluating} and MBPP \citep{austin2021program} evaluate Python code generation on relatively simple functions. HumanEval+ \citep{liu2023your} augments HumanEval with better test cases after discovering many passing solutions are incorrect. ReCode \citep{wang2022recode} is a variant of HumanEval with perturbed function names and docstrings. HumanEval-X  \citep{zheng2023codegeex}, MultiPLe \citep{cassano2022multipl}, and MBXP \citep{athiwaratkun2022multi} are extensions of HumanEval and MBPP with a focus on including programming languages outside of Python. APPS \citep{hendrycks2021measuring}, CodeContests \citep{li2022competition}, and LeetCode-Hard \citep{shinn2023reflexion} evaluate code generation on more difficult, interview or competition style problems.

There are also benchmarks to evaluate code generation in data science applications, such as DS-1000 \citep{lai2023ds}, ARCADE \citep{yin2022natural}, NumpyEval \citep{zhang2023toolcoder}, and PandasEval \citep{jain2022jigsaw}. Going one step further, some benchmarks also measure ability to use API's or perform more general software engineering tasks, such as JuICe \citep{agashe2019juice}, APIBench \citep{patil2023gorilla}, RepoBench \citep{liu2023repobench}, ODEX \citep{wang2022execution}, SWE-Bench \citep{jimenez2023swe}, GoogleCodeRepo \citep{shrivastava2023repository}, RepoEval \citep{zhang2023repocoder}, and Cocomic-Data \citep{ding2022cocomic}.

Finally, there are a variety of benchmarks for other tasks, such as code translation \citep{roziere2020unsupervised, zhu2022xlcost, ahmad-etal-2021-avatar}, test case generation \citep{tufano2022methods2test, watson2020learning}, code search \citep{husain2019codesearchnet}, type prediction \citep{mir2022type4py, wei2023typet5, malik2019nl2type}, commit message generation \citep{liu2020atom}, code summarization \citep{leclair2019neural, iyer2016summarizing, barone2017parallel, hasan2021codesc, alon2018code2seq}, code security \citep{liguori2022can, pearce2022asleep, tony2023llmseceval}, program repair \citep{jiang2023impact, xia2022practical, tufano2019empirical, haque2022fixeval, jin2023inferfix, gupta2017deepfix, berabi2021tfix}, performance optimization \citep{garg2022deepperf, madaan2023learning}, and so on.

To our knowledge, our \benchmark is the first publicly available benchmark to measure the execution ability of code LMs. While some prior work has measured the output prediction ability of code LMs, we leverage our \benchmarko to perform a more thorough investigation of these capabilities. Our \benchmarki is the first to measure the ability of code LMs to perform input prediction.

\textbf{Leveraging Test Cases and Code Execution}: Another line of work uses test cases and code execution information to improve code generation. Some examples include Speculyzer \citep{key2022speak}, CodeT \citep{chen2022codet}, CodeGen-Test \citep{zhong2022codegen}, Coder-Reviewer reranking \citep{zhang2023coder}, MBR-EXEC \citep{shi2022natural}
TCoT \citep{tian2023test},
Algo \citep{zhang2023algo}, Pangu-Coder2 \citep{shen2023pangu}, LEVER \cite{ni2023lever}, and Self-Play \citep{haluptzok2022language}.
The idea of these works is to generate many programs and many test cases and select which programs and test cases seem correct based on the execution results. 
using execution info. Other works use RL-style execution feedback to improve code generation, including CodeRL \citep{le2022coderl}, Reflexion \citep{shinn2023reflexion}, and PG-TD \citep{zhang2023planning}. \citep{chen2023teaching, olausson2023demystifying, madaan2023self, peng2023check, zhang2023self} investigate self-repair, using error messages as feedback for models to improve.

Most relevant to our work, a handful of works examine and improve the execution ability of code LMs. \cite{austin2021program}, 
Scratchpad \citep{nye2021show}, and CodeExecutor \citep{liu2023code} train code LMs on execution information. Inspired by these works, we briefly touch on two primitive ways to improve performance on our benchmark, chain-of-thought and fine-tuning. Moving forward, we believe our \benchmark could serve as a useful reference point as more techniques are designed to improve code execution abilities.

\textbf{Failure modes of LM Reasoning}: Another dream of the community is to better understand the failure modes of LMs on reasoning tasks. \cite{bubeck2023sparks, liu2023evaluating, arkoudas2023gpt, zhang2022paradox, dziri2023faith, olausson2023linc, lee2023teaching, zhang2023can} all investigate and point out various failure modes of LMs on a wide variety of reasoning tasks. Other examples of reasoning failures include 1) understanding negation \citep{hosseini2021understanding}, 2) ignoring irrelevant context \citep{shi2023large}, 3) operating under counterfactual situations such as 1-indexed Python or base-9 addition \citep{wu2023reasoning}, and 4) generating Python code after identifier swaps like \texttt{print, len = len, print} \citep{miceli2023larger}. Taking a more theoretical perspective, \cite{dziri2023faith, zhou2023algorithms, merrill2023expresssive, giannou2023looped} characterize the types of reasoning tasks transformers can and cannot be expected to carry out.
\cite{merrill2021provable} argues that it is not possible to learn meaning from ungrounded form with context dependence and assuming that syntax is independent of semantics. 
In this work, we use \benchmark to empirically examine failures in code execution / reasoning. 

% CodeScore \citep{dong2023codescore} trains a model to learn. 

\section{Benchmark Construction} \label{sec:benchmark-construction}

\benchmark consists of 800 distinct functions, each with an input-output pair such that executing the function on the input deterministically produces the output. Using these functions and input-output pairs, we derive two benchmark tasks. In the \textit{output prediction} task, the goal is to predict the output of executing the function on its associated input. In the \textit{input prediction} task, the goal is to find any input such that executing the function on that input produces the output. For both tasks, we use an execution-based correctness metric. For input prediction, a generated input passes if \texttt{assert f(generated\_input) == output} passes, and for output prediction, a generated output passes if \texttt{assert f(input) == generated\_output} passes. A few statistics about the samples of \benchmark can be found in Appendix \ref{appendix:benchmark-dataset-statistics}.

\subsection{Generating Candidates} \label{subsec:generating-candidates}
We use \codellamalarge to generate all the candidate functions and inputs of \benchmark. To do so, we prompt it with the name of a function in the Python standard library such as \texttt{str.zfill} and ask it to generate a Python function that makes use of the library function in addition to 5 test inputs. We provide two varying few-shot examples in our prompt for improved diversity of generations (see Appendix \ref{appendix:benchmark-generation-fewshot} for more details). A sample prompt is shown in Listing \ref{lst:benchmark-generation-prompt}.

We use a total of 69 different functions from the standard library: 47 from the \texttt{str}, 11 from \texttt{dict}, and 11 from \texttt{list} (see Appendix \ref{appendix:benchmark-generation-functions} for the full list of functions). Overall, we generate a total of 102000 functions (46\% \texttt{str}, 27\% \texttt{dict}, 27\% \texttt{list}) and 489306 input-output pairs.

% String method count:  47 47000
% Dict method coun  11 27500
% List method count:  11 27500
% 102000

\subsection{Filtering Candidates} \label{subsec:filtering-candidates}
Next, we filter the generated candidates to ensure that the samples in the dataset are reasonable and of high quality. In order to avoid forcing the model to perform tasks such as arithmetic calculation, we design filters so that the benchmark only consists of samples that are solveable by a human without extra memory.

Concretely, we filter based on the following criteria. 
\begin{itemize}
    \item Compile time: all arguments of the function must be used in the function, length of code is between 75 and 300 characters, no syntax errors, proper assertion \texttt{assert f(input) == output}.

    \item Runtime: no float point operations, true division, exp, other integer operations must have at least one argument $\leq 3$, string and list operations must have at least one argument with length $\leq 3$, finish running in 2 seconds, no uncaught exceptions.

    \item Best effort to remove other undesirable code: function cannot have any imports (such as os, random), must be deterministic (random, set ordering), and cannot have side effects such as input, \texttt{\_\_builtins\_\_}.
\end{itemize}

\subsection{Data size and measuring noise} \label{subsec:data-size-noise}
The success of HumanEval (164 examples) shows that evaluation benchmarks can be small where faster and cheaper evaluation is an overlooked advantage. Since additional examples are easy to generate, we first overgenerate and then measure if the noise is sufficiently small on a smaller dataset. 

% When determining the final size of our benchmark, we balanced two factors: the benchmark should be small enough to easily run and iterate on but large enough so that signal overcomes noise. 

\begin{figure}[H]
    \centering
    \includegraphics[width=0.49\textwidth] {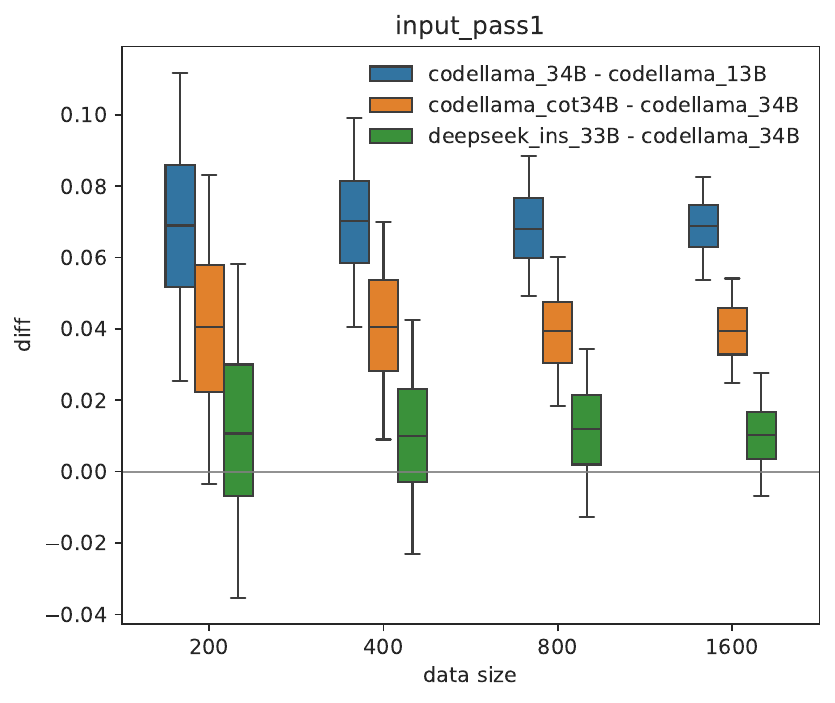}
    \includegraphics[width=0.49\textwidth]{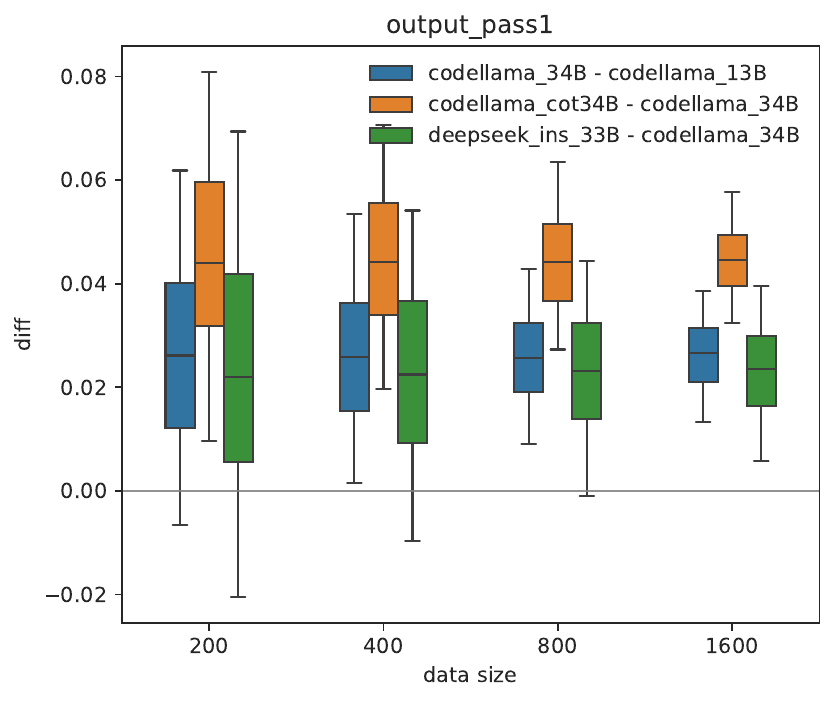}
    \caption{Difference between model pairs on bootstrap samples of various sizes. The whiskers show (2.5, 97.5) and boxes show (25, 75) percentiles. }
    \label{fig:benchmark-filtering-size-selection}
\end{figure}

Out of all the samples, \codellamalarge outperforms \codellamamed as expected and we would like to retain this property with high confidence in a smaller dataset. To do this, we took bootstrap samples of size $N$ out of $\sim$1700 samples to measure the probability that the performance would be reversed, shown in Fig. \ref{fig:benchmark-filtering-size-selection}. 800 examples are enough to test that \codellamalarge $>$ \codellamamed, \textsc{Code Llama cot} $>$ \textsc{Code Llama} and  as well as between \textsc{Deepseek 33B} $>$ \codellamalarge (output).

We measure two sources of noise: 1) sampling which data points to include in the benchmark, and 2) sampling candidates from models for each data point (temperature $>0$). Of these, 1) dominates 2). For 1) since model $A$ does not always outperform model $B$ on all data points even if $A>B$ in aggregate, the measured performance depends on which data points are included. We can measure both noise on each model individually, and also measure type 1) noise on pairs of models using bootstrap. Fortunately, we do not see major differences between models and the most important factor is just the size of dataset. Type 1) noise is generally around 1.5\% for each model whereas type 2) is around 0.2\% at $N=800$. Type 1) noise usually becomes smaller on pairs of models due to correlation, yielding statistically significant results at the $\alpha=0.05$ level for many model pairs.

\section{Evaluation} \label{sec:evaluation}

We evaluate a selection of models on \benchmark: StarCoder (Base 7B, 15.5B) \citep{li2023starcoder}, Mistral (7B) \citep{jiang2023mistral}, WizardCoder (13B, 34B) \citep{luo2023wizardcoder}, Phi-1 \cite{gunasekar2023textbooks} and Phi-1.5 \citep{li2023textbooks} (1.3B), Phind v2 \citep{Phind} (34B), Code Llama \citep{roziere2023code} (Base and Python 7B, 13B, 34B), DeepSeek Coder (Base and Instruct 6.7B, 33B), GPT-3.5 \citep{brown2020language, ouyang2022training}, and GPT-4 \citep{openai2023gpt}. To facilitate reproducibility, the HuggingFace checkpoints of non-GPT models are in Appendix \ref{sec:appendix-models} and all prompts are in Appendix \ref{sec:appendix-direct-prompts}. 

We use $N=100$ samples for all non-GPT models and $N=10$ samples for GPT models. We report both pass@1 scores $(T=0.2)$ and pass@5 scores $(T=0.8)$. The results are shown in Fig.~\ref{fig:main-results}, and raw scores are provided in the Appendix in Table \ref{tab:benchmark-results}. In Fig. \ref{fig:main-results}, we show the intervals generated by 10000 bootstrap samples from the dataset, where non-overlapping whiskers would be significant at the 2.5\% level. To get more statistical power, we compare pairs of models on each bootstrapped sample. We show how each model compares to \codellamalarge in Fig.~\ref{fig:main-results-all-appendix}.
The intervals generally decreases due to correlations. On all models vs. \codellamalarge, if the median bar clears the whisker in Fig.~\ref{fig:main-results}, then the difference actually holds with $>$97.5\% probability under paired bootstrap. For example, \codellamalarge is better than \textsc{wizard\_34B} on input and \codellamalarge is worse than \textsc{deepseek\_33B} on output prediction with $>$97.5\% probability.

\begin{figure}[H]
     \centering
     \begin{subfigure}[b]{0.49\textwidth}
         \centering
         \includegraphics[width=\textwidth]{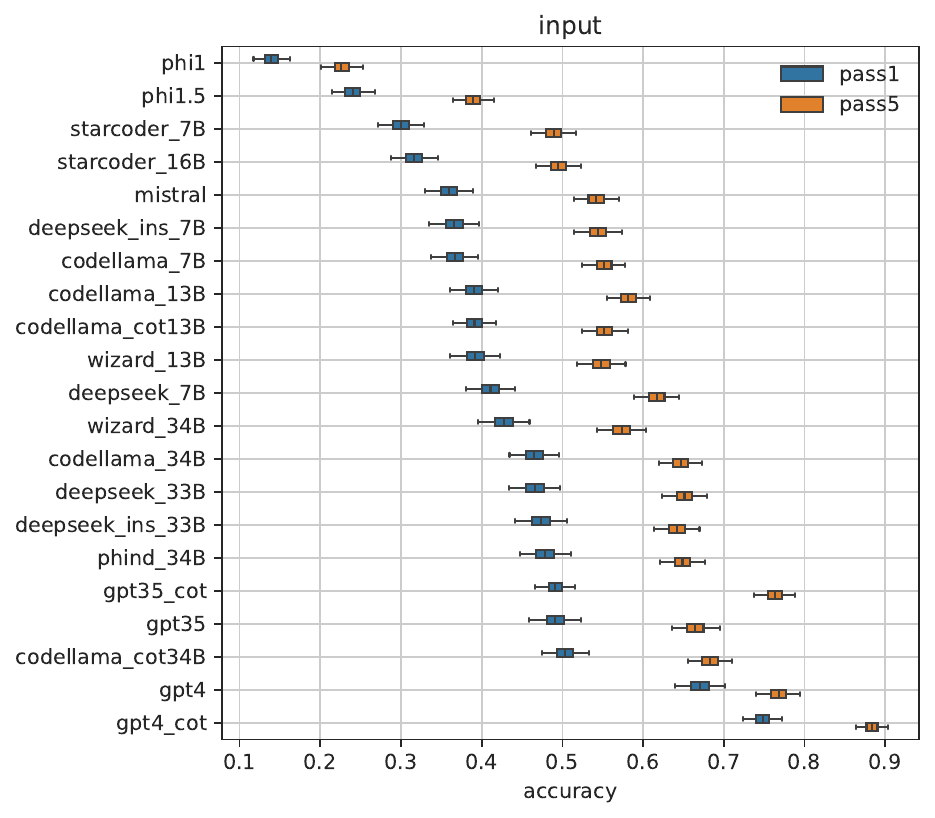}
         \caption{\benchmarki Performance}
         \label{fig:main-results-pass1-input}
     \end{subfigure}
     \hfill
     \begin{subfigure}[b]{0.49\textwidth}
         \centering
         \includegraphics[width=\textwidth]{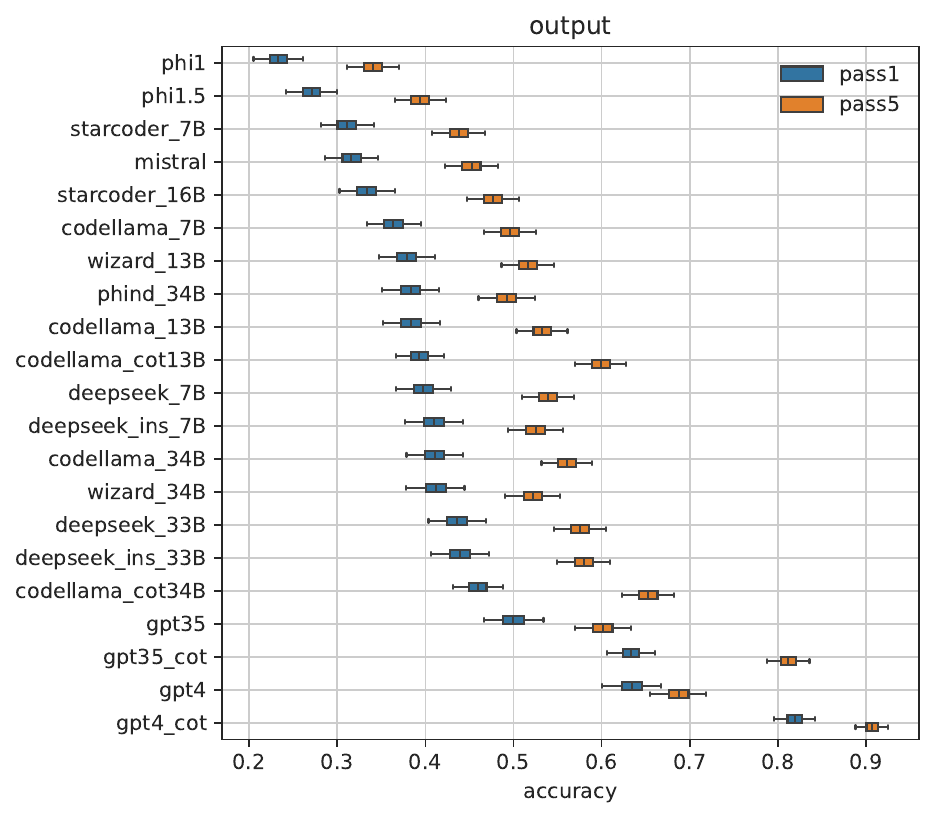}
         \caption{\benchmarko Performance}
         \label{fig:main-results-pass1-output}
     \end{subfigure}
     \caption{Main Results. Boxes show (25, 75) percentiles, whiskers show (2.5, 97.5), and the middle bar shows the median ($\approx$ mean). }
     \label{fig:main-results}
\end{figure}

\section{Quantitative Analysis} \label{sec:quantitative-analysis}

\textbf{Correlation between scores on HumanEval and \benchmark}: After the release of Code Llama's model and GPT-3.5 and GPT-4's APIs, there have been many creative efforts to take data distilled from GPT models and use them to train more powerful code models such as WizardCoder \citep{luo2023wizardcoder}, Phi-1 \citep{gunasekar2023textbooks}, Phi-1.5 \citep{gunasekar2023textbooks}, and Phind \citep{Phind}. For example, WizardCoder 34B started with the Code Llama 34B base model and improved the HumanEval pass@1 score from 53.7\% to 73.2\%, a significant and impressive achievement. There remains curiosity about whether these models show more general improvements in other aspects of programming or code understanding \citep{gudibande2023false}. We measure this through \benchmark.

\begin{figure}[H]
     \centering
     \begin{subfigure}[b]{0.49\textwidth}
         \centering
         \includegraphics[scale=0.49]{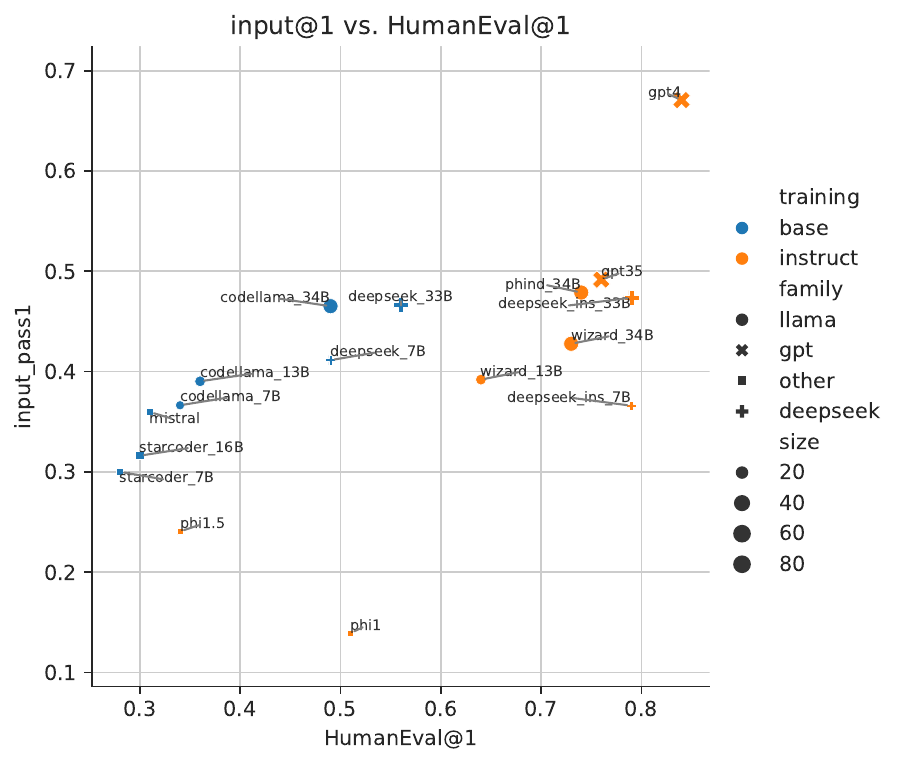}
         \caption{\benchmarki vs. HumanEval}
         \label{fig:quant-analysis-humaneval-correlation-input}
     \end{subfigure}
     \hfill
     \begin{subfigure}[b]{0.49\textwidth}
         \centering
         \includegraphics[scale=0.49]{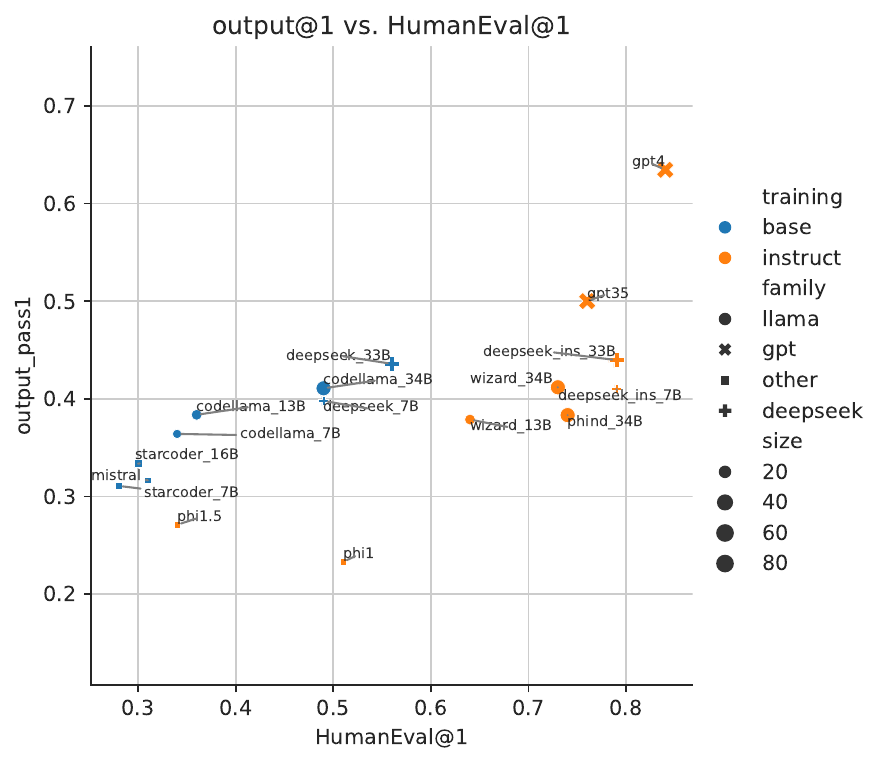}
         \caption{\benchmarko vs. HumanEval}
         \label{fig:quant-analysis-humaneval-correlation-output}
     \end{subfigure}
     \caption{Correlation between HumanEval pass@1 scores and \benchmarko pass@1 scores}
     \label{fig:quant-analysis-humaneval-correlation-1}
\end{figure}

In Fig. \ref{fig:quant-analysis-humaneval-correlation-1}, we plot reported HumanEval scores (we did not reproduce them ourselves) against scores on \benchmark. Indeed, we spot some interesting outliers: when comparing the distilled models WizardCoder 34B and Phind 34B to \codellamalarge, we see that the distilled models score over 20\% more than Code Llama on HumanEval but do not show this drastic improvement when evaluated on both input and output predictions. In addition, the Phi-1 model outperforms most of the bigger models on HumanEval, but performs among the worst of all our evaluated models on \benchmark. Overall, this suggests that models optimized for the HumanEval task by distilling data from GPT-3.5 and GPT-4 (WizardCoder, Phind, Phi) may not have learned other code reasoning capabilities along the way. On the other hand, for models such as StarCoder, Mistral, CodeLlama, and DeepSeek-Base, we still see a positive trend between HumanEval score and \benchmark score, suggesting that code generation and execution/understanding abilities are correlated.

\begin{tcolorbox}[colback=lightgreen, boxrule=0pt, arc=10pt, outer arc=10pt]
Base models show a weak correlation between HumanEval and \benchmark. For HumanEval, distilled models (WizardCoder, Phind, Phi) significantly beat their base models, but for \benchmark, no distilled model performs significantly better than \codellamalarge.
\end{tcolorbox}

\begin{figure}[H]
     \centering
     \begin{subfigure}[b]{0.49\textwidth}
         \centering
         \includegraphics[scale=0.49]{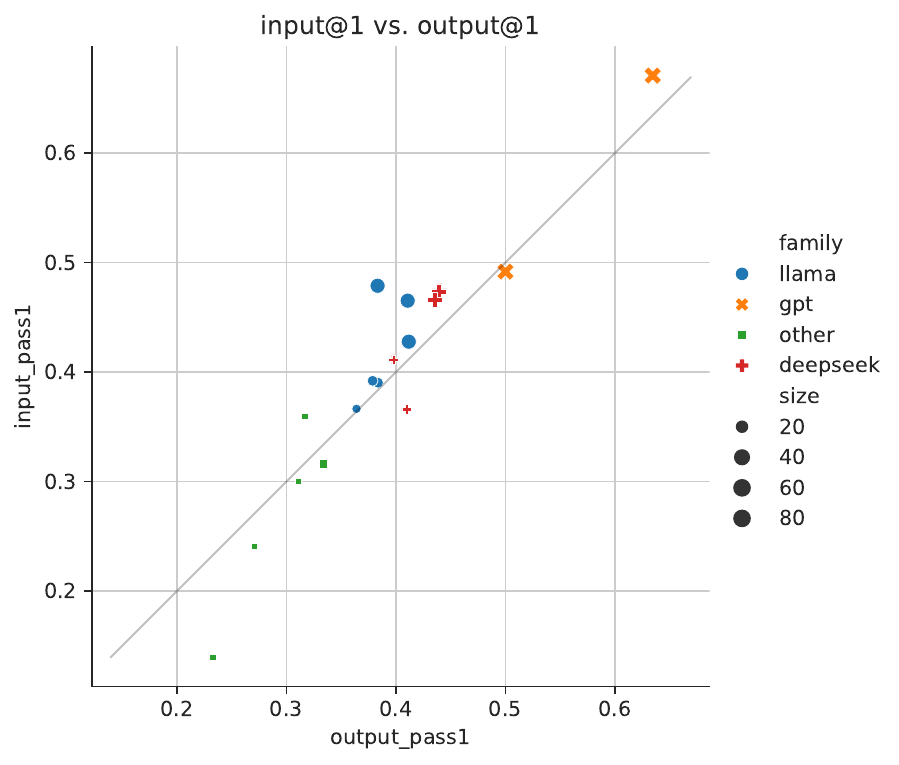}
         \caption{Input vs. Output Prediction, Direct}
         \label{fig:quant-analysis-input-output-direct}
     \end{subfigure}
     \hfill
     \begin{subfigure}[b]{0.49\textwidth}
         \centering
         \includegraphics[scale=0.5]{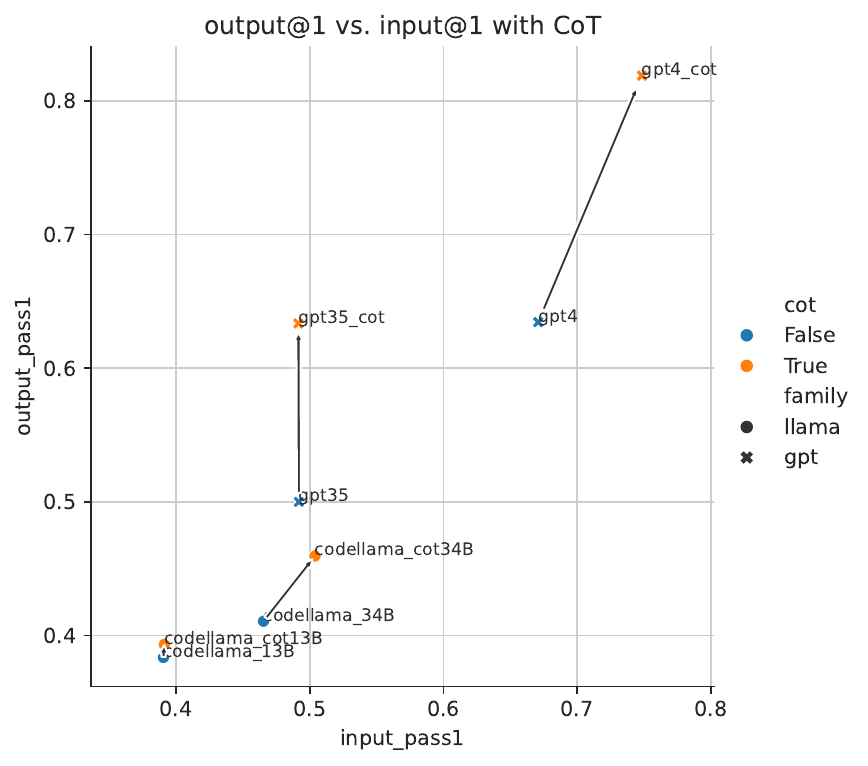}
         \caption{Output vs. Input Prediction, CoT}
         \label{fig:quant-analysis-input-output-cot}
     \end{subfigure}
     \caption{Correlation between Input and Output Prediction Scores, with and without CoT}
     \label{fig:quant-analysis-input-output}
\end{figure}

\textbf{Relationship between input prediction and output prediction:}
In Fig. \ref{fig:quant-analysis-input-output-direct}, we compare the input prediction and output prediction pass@1 scores with each other. Conceptually, the two tasks seem relatively different: output prediction is directly testing code execution ability, while input prediction requires a higher-level understanding of the code's functionality. However, we discover that there is a strong correlation between their performance. This suggests the hypothesis that performance on relatively distinct coding-related tasks may be closely correlated. In addition, we see a relatively clear impact of scaling the model size on our two tasks.

\begin{tcolorbox}[colback=lightgreen, boxrule=0pt, arc=10pt, outer arc=10pt]
With the potential exception of GPT models, performance on \benchmarki and \benchmarko seem to be very correlated. As the tasks seem relatively different, this suggests that the code reasoning capabilities of models may generalize from task to task.
\end{tcolorbox}

\begin{figure}[H]
     \centering
     \begin{subfigure}[b]{0.49\textwidth}
         \centering
         \includegraphics[width=1\textwidth]{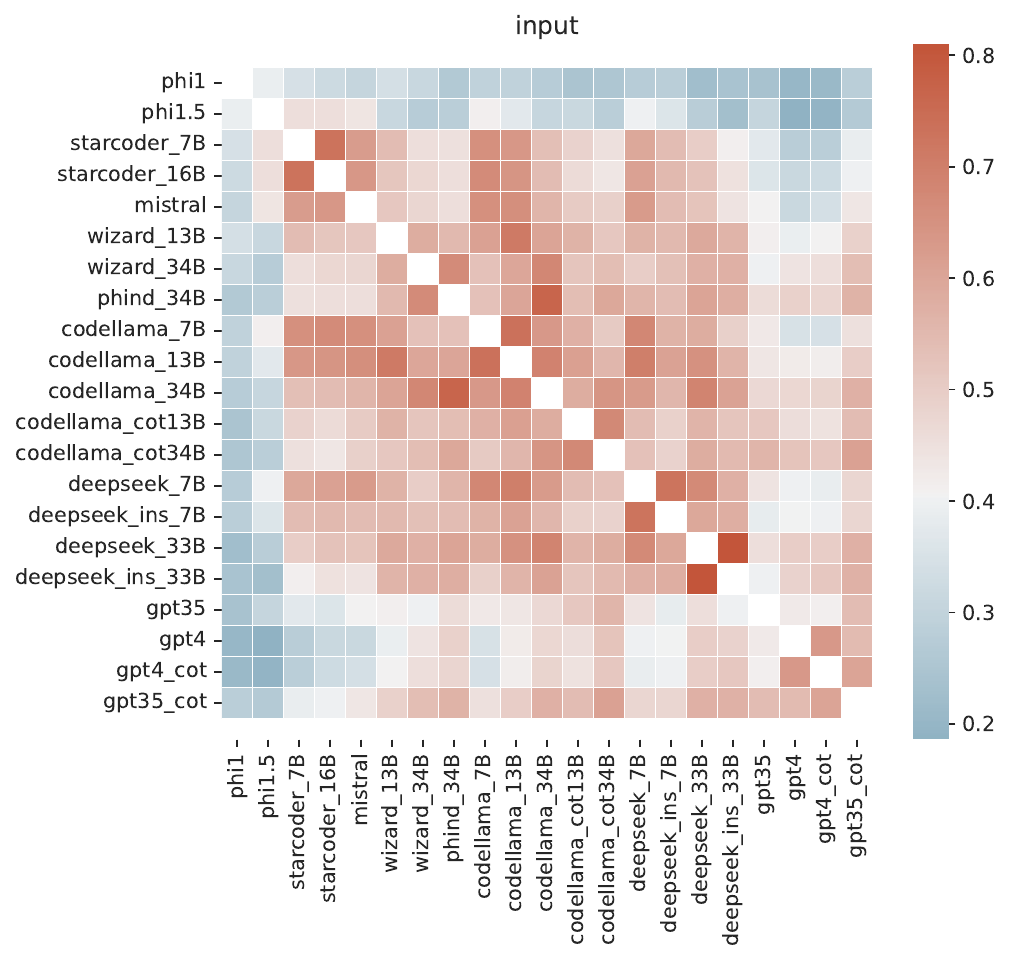}
     \end{subfigure}
     \hfill
     \begin{subfigure}[b]{0.49\textwidth}
         \centering
         \includegraphics[width=1\textwidth]{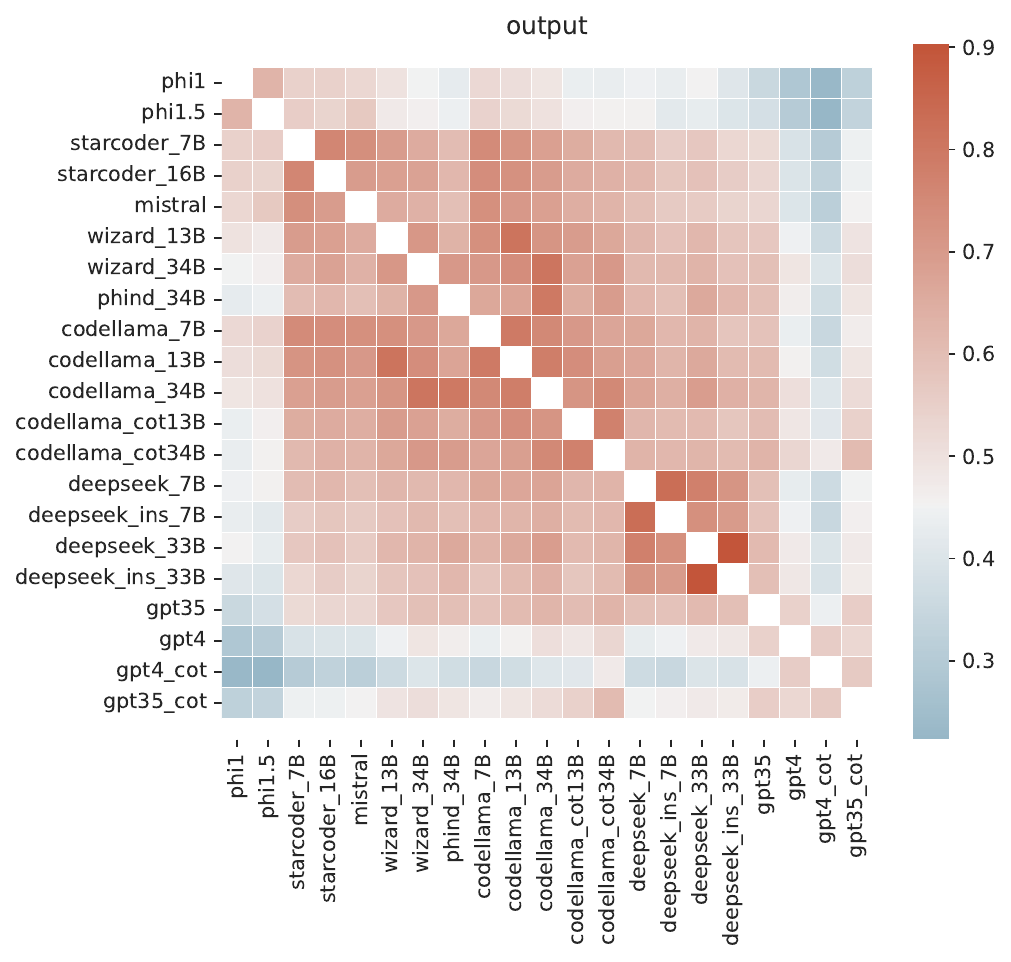}
     \end{subfigure}
     \caption{Correlation between predictions on input (left) and output (right)}
     \label{fig:error_corr-input-output}
\end{figure}

\textbf{Confusion matrix/error correlation for different models.}
Fig.~\ref{fig:error_corr-input-output} shows the pairwise correlation of pass@1 scores for each pair of models. The correlation is chosen based on its highest signal among cosine distance, Spearman and Kendall. The middle section of ``open''-ish models (StarCoder, Code Llama, DeepSeek, etc.) are strongly correlated with each other. Strong correlations are seen between sizes of the same model, between models of the same size, and between instruct and base (\textsc{Phind 34B, Wizard 34B} vs. \codellamalarge). CoT results also tend to have strong correlations with other CoT results, even GPT-4 vs Llama 13B. For the output task, \textsc{Deepseek} forms a small sub-cluster of especially strong associations.

\begin{tcolorbox}[colback=lightgreen, boxrule=0pt, arc=10pt, outer arc=10pt]
Looking at model predictions, strong correlations are seen between sizes of the same model, between models of the same size, and between instruct and base models. Although what is hard for a better model tend to be hard for worse models on average, worse models succeeded on some examples where the better models fail completely.
\end{tcolorbox}

\subsection{Chain of Thought Prompting} \label{subsec:cot}
Next, we evaluate how the popular chain-of-thought (CoT) prompting method \citep{wei2022chain} affects the performance of Code Llama, GPT-3.5, and GPT-4 models on \benchmark. The full prompts can be found in Appendix \ref{sec:appendix-cot-prompts}. All results are reported using $N=10$ samples other than CodeLlama 13B and 34B without CoT, which are reported with $N=100$ samples. As before, pass@1 is reported with $T=0.2$ and pass@5 with $T=0.8$. Additional results can be found in Appendix \ref{subsec:appendix-cot}.

\textbf{Impact of CoT}: We begin by focusing our attention on the pass@1 scores of models with and without CoT. In Fig. \ref{fig:quant-analysis-input-output-cot}, we plot the input and output prediction scores of each model with and without CoT. First, GPT-4 benefits significantly more than other models. Second, output prediction boosts are generally larger than input prediction. In fact, CoT does not seem to improve Code Llama 13B and GPT-3.5 performance on input prediction. This is intuitive, as input prediction involves a more difficult reasoning task, while output prediction only requires executing the program step by step. We defer raw numbers to the Appendix in Table \ref{tab:benchmark-results-cot}.

\begin{tcolorbox}[colback=lightgreen, boxrule=0pt, arc=10pt, outer arc=10pt]
CoT helps Code Llama 34B and GPT-4 on both input and output prediction, GPT-3.5 on only output prediction, and Code Llama 13B on neither task. CoT also leads to larger boosts on output prediction than input prediction. GPT-4 benefits significantly more from CoT than other models, achieving the highest pass@1 of 74.8\% on input prediction and 81.9\% on output prediction but still far from acing the benchmark.
\end{tcolorbox}

\begin{figure}[H]
     \centering
     \begin{subfigure}[b]{0.49\textwidth}
         \centering
         \includegraphics[scale=0.45]{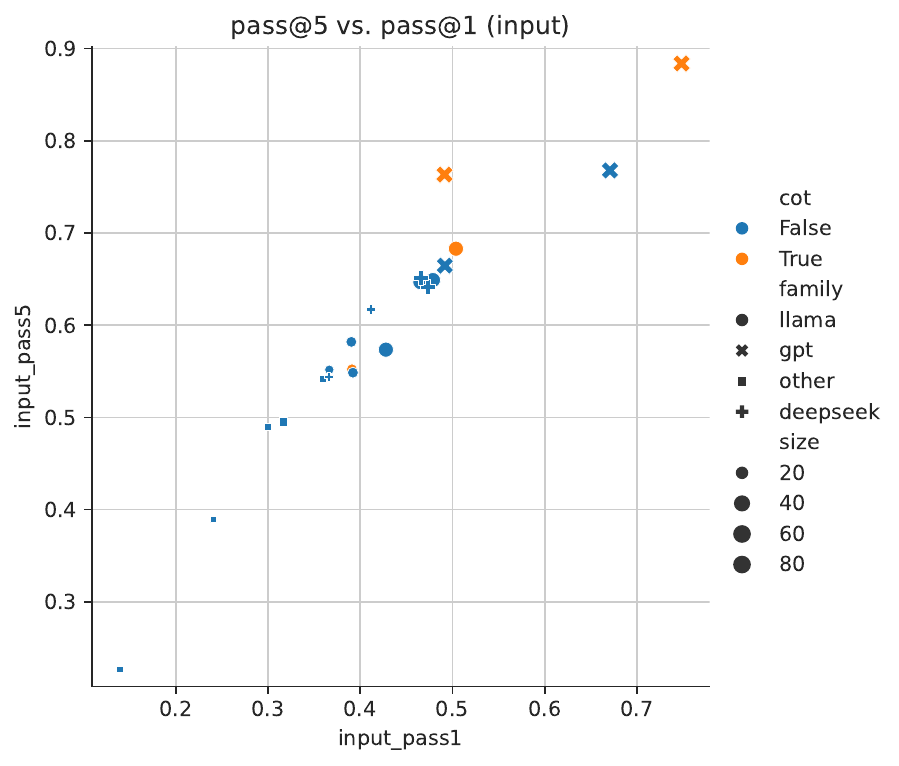}
         \caption{Input prediction}
         \label{fig:pass5-vs-pass1-cot-input}
     \end{subfigure}
     \hfill
     \begin{subfigure}[b]{0.49\textwidth}
         \centering
         \includegraphics[scale=0.45]{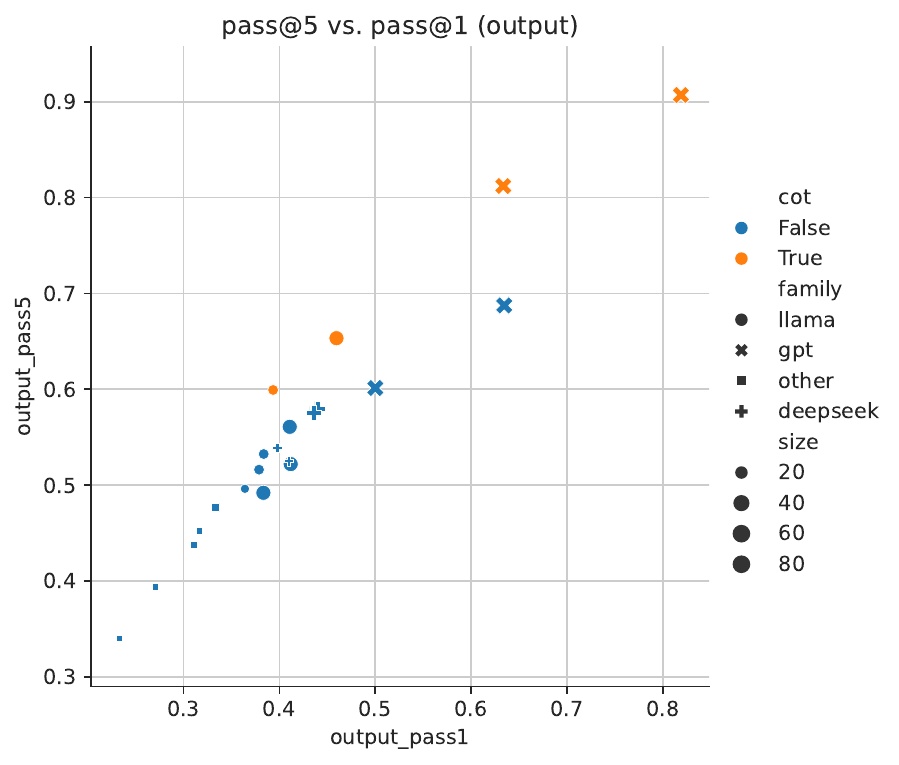}
         \caption{Output prediction}
         \label{fig:pass5-vs-pass1-cot-output}
     \end{subfigure}
     \caption{pass@5 score vs. pass@1 score with and without CoT}
     \label{fig:pass5-vs-pass1-cot}
\end{figure}

\textbf{CoT widens the gap between pass@5 and pass@1 scores}: In Fig. \ref{fig:pass5-vs-pass1-cot}, we plot the pass@5 scores against the pass@1 scores for all models. For models without CoT (shown in blue), there is a positive correlation between pass@1 and pass@5 scores. For models with CoT (shown in orange), we see an increase in the gap between pass@5 and pass@1 scores. We believe this phenomenon may be due to the additional diversity induced by CoT, which we analyze in detail in Appendix \ref{sec:appendix-diversity}.

\begin{tcolorbox}[colback=lightgreen, boxrule=0pt, arc=10pt, outer arc=10pt]
Because CoT increases the diversity of generated inputs and outputs, models with CoT see a larger gap between pass@1 and pass@5 score compared to models without.
\end{tcolorbox}

\begin{figure}[H]
     \centering
     \begin{subfigure}[t]{0.49\textwidth}
         \centering
         \includegraphics[width=\textwidth]{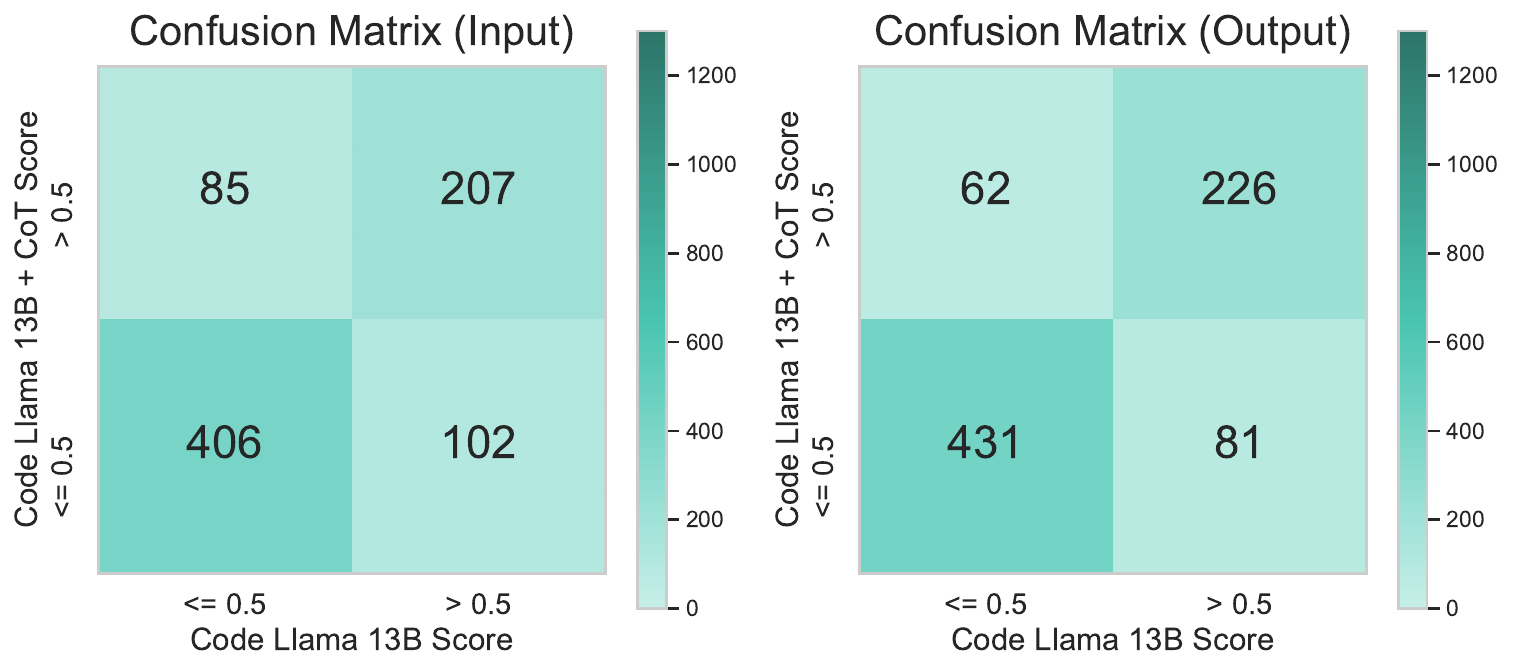}
         \caption{Code Llama 13B}
         \label{fig:confusion-cot-codellama-13b}
     \end{subfigure}%
     \hfill
     \begin{subfigure}[t]{0.49\textwidth}
         \centering
         \includegraphics[width=\textwidth]{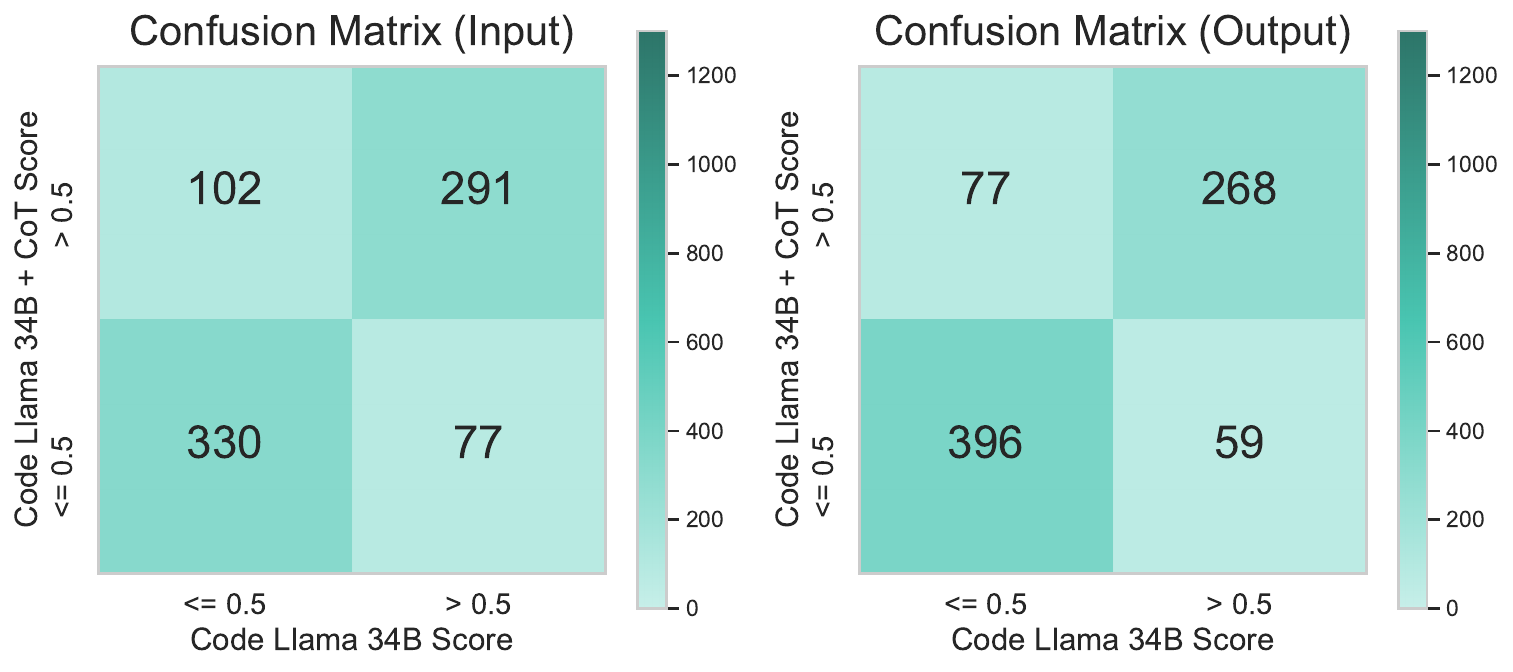}
         \caption{Code Llama 34B}
         \label{fig:confusion-cot-codellama-34b}
     \end{subfigure}
     \newline
     \newline
     \newline
     \begin{subfigure}[t]{0.49\textwidth}
         \centering
         \includegraphics[width=\textwidth]{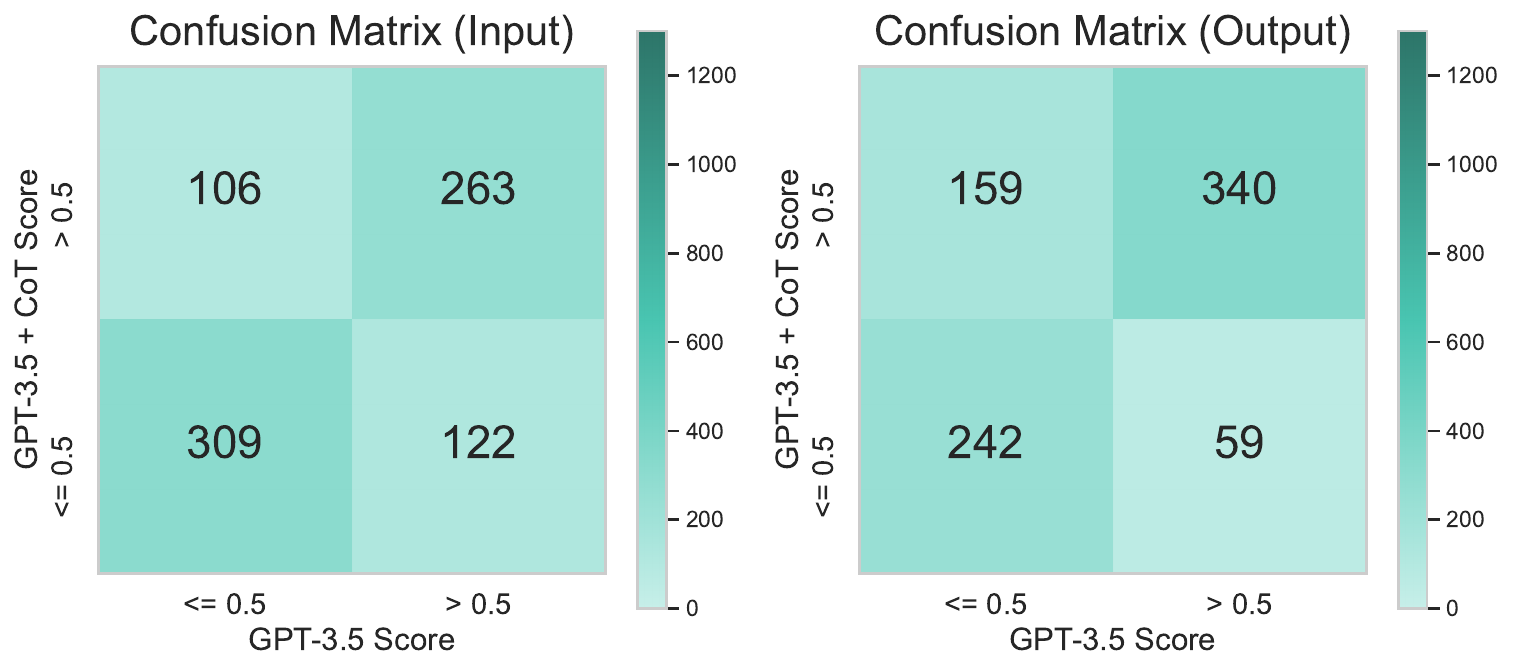}
         \caption{GPT-3.5}
         \label{fig:confusion-cot-gpt35}
     \end{subfigure}%
     \hfill
     \begin{subfigure}[t]{0.49\textwidth}
         \centering
         \includegraphics[width=\textwidth]{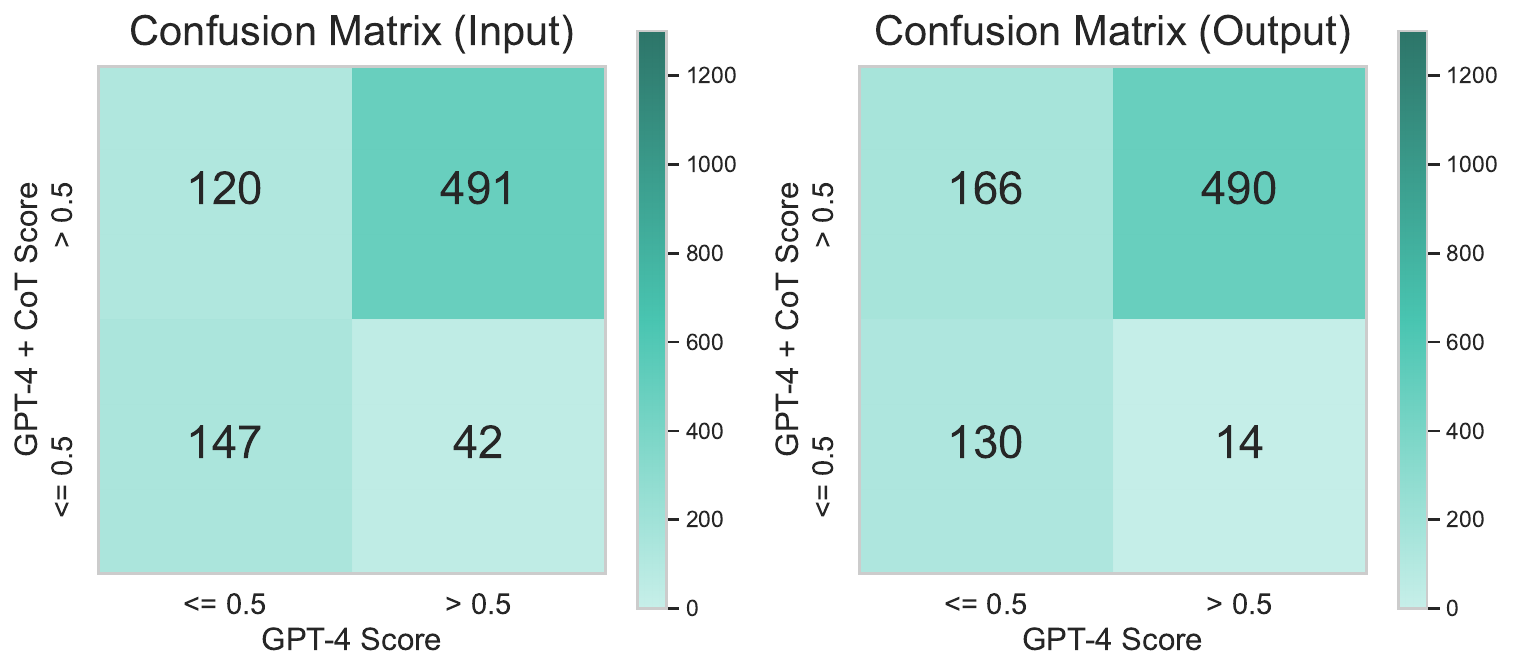}
         \caption{GPT-4}
         \label{fig:confusion-cot-gpt4}
     \end{subfigure}
     \caption{Confusion Matrix of Direct Prediction vs. CoT Prediction $(T=0.2)$}
     \label{fig:confusion-cot-all}
\end{figure}

\textbf{Predictions of CoT vs. Base Model}: In Fig. \ref{fig:confusion-cot-all}, we show a confusion matrix over samples to better understand the correlations between direct output predictions and CoT predictions. For CodeLlama 13B, 34B, and GPT-3.5, we observe a large number of samples where direct prediction succeeds but CoT fails. However, with GPT-4, we observe that there are relatively few samples where this is the case.

\begin{tcolorbox}[colback=lightgreen, boxrule=0pt, arc=10pt, outer arc=10pt]
While CoT generally improves performance overall, there are many individual samples where CoT actually hurts the prediction accuracy for Code Llama 13B/34B and GPT-3.5 on both input and output prediction. For GPT-4, CoT generally improves individual sample accuracy, more so for output prediction than for input prediction.
\end{tcolorbox}

\subsection{Fine-tuning Experiments} \label{subsec:finetuning}
Next, we do a preliminary analysis to understand the effect of simple fine-tuning schemes on \benchmark performance. We fine-tuned \codellamalarge on nearly 140K samples of Python functions distilled with the procedure outlined in Sec. \ref{sec:benchmark-construction}, without filtering. We perform weak decontamination, only removing samples where both the function and input-output pairs match samples in the benchmark. 

\begin{figure}[H]
     \centering
     \begin{subfigure}[b]{0.48\textwidth}
         \centering
         \includegraphics[scale=0.4]{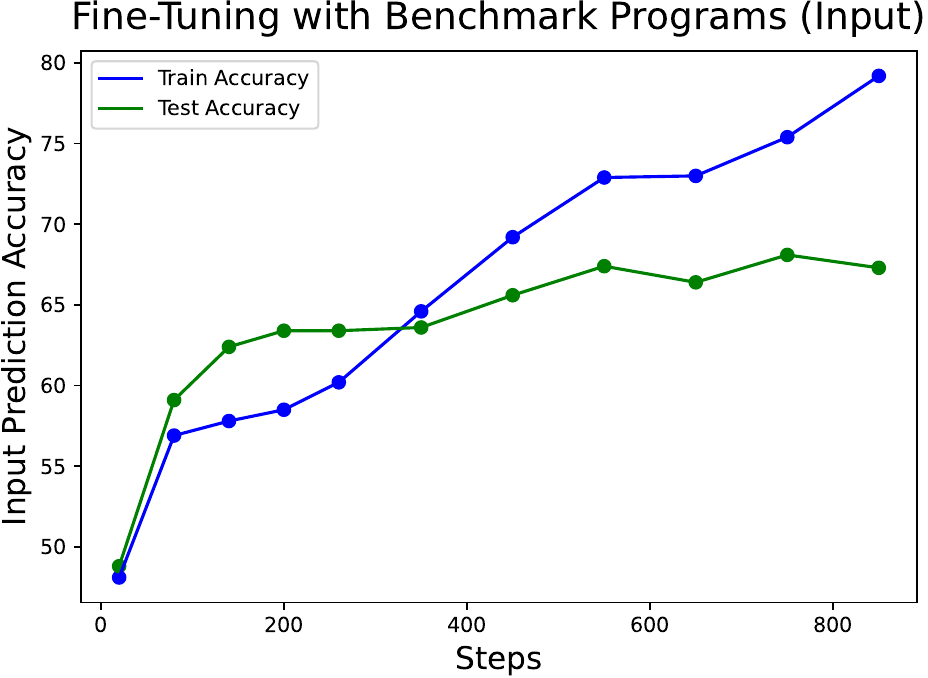}
         \caption{Input prediction}
         \label{fig:finetuning-accuracy-samples-plot-input-main}
     \end{subfigure}
     \hfill
     \begin{subfigure}[b]{0.48\textwidth}
         \centering
         \includegraphics[scale=0.4]{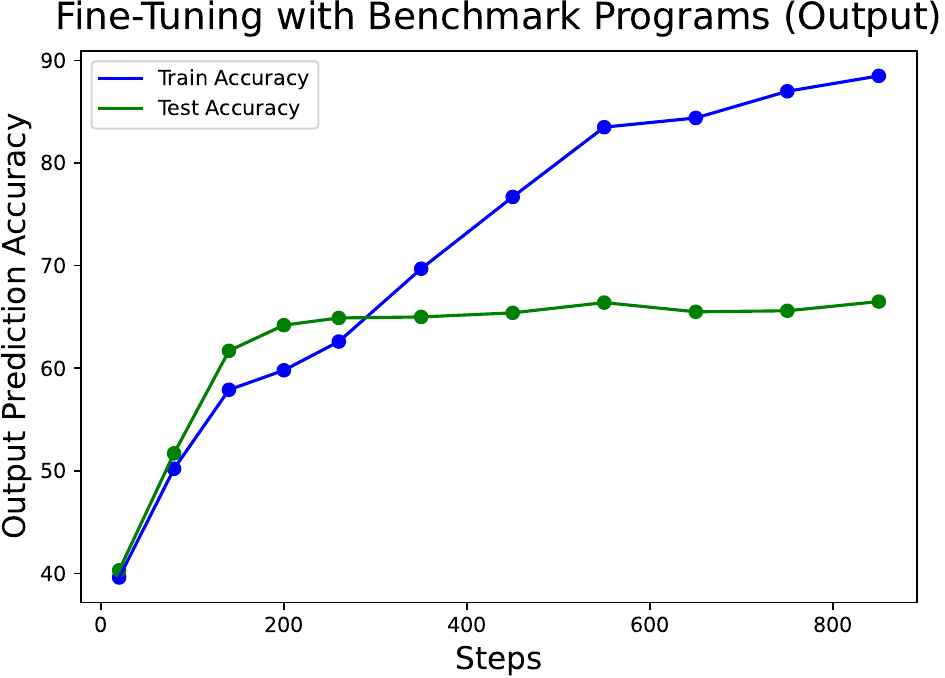}
         \caption{Output prediction}
         \label{fig:finetuning-accuracy-samples-plot-output-main}
     \end{subfigure}
     \caption{Improvements and Limits of \benchmark Performance after Fine-Tuning}
     \label{fig:finetuning-accuracy-samples-plot-main}
\end{figure}

In particular, we finetune on a mixture of 50\% samples where the function is not in the benchmark and 50\% samples where the function is in the benchmark but input-output pairs are not, a very liberal setup. The training and testing accuracy over time is shown in Fig. \ref{fig:finetuning-accuracy-samples-plot-main}. Despite finetuning on programs very similar to the benchmark, we still observe a plateauing effect in the test accuracy, suggesting that our execution tasks may be too difficult to learn from this simple fine-tuning scheme. We defer a few other insights from fine-tuning to Appendix \ref{subsec:appendix-finetuning} and suggest a few fine-tuning ideas for improving our benchmark in Sec. \ref{sec:limitations-future-work}.

\begin{tcolorbox}[colback=lightgreen, boxrule=0pt, arc=10pt, outer arc=10pt]
After fine-tuning on samples very similar to those in our benchmark, Code Llama 34B can match the performance of GPT-4 on both input and output prediction. However, accuracy plateaus at under 70\% for both tasks, so simple finetuning is far from solving the benchmark.
\end{tcolorbox}

\section{Qualitative Analysis} \label{sec:qualitative-analysis}
All models except GPT4 has over 50\% failure rate, showing they cannot do simple executions. 
In this section, we focus on GPT4 with CoT and verify that the remaining 20\% failures are due to the model, are consistent and are indeed on \emph{simple} programs. We refer the reader to Appendix \ref{sec:appendix-qualitative} for further examples of the failures highlighted in this section and impressive successes.

\paragraph{Failures of GPT-4 CoT.}
GPT-4 Cot scored 0/10 on 54 output prediction tasks and 65 input prediction tasks. On 22 problem, it scored 0/10 on both input and output prediction tasks. 
We manually check the 22 problems if they pass our criteria of being simple problems. Most are indeed simple (Listings~\ref{lst:gpt-4-simple1}, ~\ref{lst:gpt-4-simple2}). There are 2 problems that require counting to around 30 (Listing~\ref{lst:gpt-4-hard1}) and 2 problems (Listing~\ref{lst:gpt-4-hard2}) that require simulating a few actual steps, which might be difficult for direct generation but within scope for CoT.

\begin{minipage}{.48\textwidth} % sample_218 steps: 37
\begin{lstlisting}[caption=GPT-4 has the right idea but cannot do the string concatenation correctly,label={lst:gpt-4-simple1}, breaklines=true, language=python]
def f(string, sep):
    cnt = string.count(sep)
    return((string+sep) * cnt)[::-1]
assert f('caabcfcabfc', 'ab') == 'bacfbacfcbaacbacfbacfcbaac'
# GPT4+CoT preds:
# f('baa', 'c')
# f('cba', 'f')
# 'bacfcabcfcaabcabfcabcfcaac'
# 'bacfbacfcbaabcabfbacfcbaabc'
\end{lstlisting}
\end{minipage}\hfill
\begin{minipage}{.48\textwidth} % sample_209	steps: 26	
\begin{lstlisting}[caption=GPT-4 might have been misled by the variable name prefix,label={lst:gpt-4-simple2}, breaklines=true, language=python]
def f(prefix, s):
    return str.removeprefix(prefix, s)
assert f('hymi', 'hymifulhxhzpnyihyf') == 'hymi'
# GPT4+CoT preds:
# input
# f('p', 'phymi')
# f('', 'hymi')
# output
# 'fulhxhzpnyihyf'
# 'fulhxhzpnyihyf'
\end{lstlisting}
\end{minipage}

\begin{minipage}{.48\textwidth} % sample_129	steps: 132	
\begin{lstlisting}[caption={GPT-4 CoT failures where solutions requires counting to 30}, label={lst:gpt-4-hard1}, breaklines=true, language=python]
def f(text, search_string):
    indexes = []
    while search_string in text:
        indexes.append(text.rindex(search_string))
        text = text[:text.rindex(search_string)]
    return indexes
assert f('ONBPICJOHRHDJOSNCPNJ9ONTHBQCJ', 'J') == [28, 19, 12, 6]

# GPT4+CoT preds:
# input
# f("0000123000012300001230000123", "123")
# f('bbbbbbabbbbbbabbbbbbbabbbbbbab', 'a')
# f("abcxxxxxxabcxxxxxxabcxxxxxxabc","abc")
# output
# [23, 13, 8, 5]
# [25, 18, 15, 11, 6]
# [7, 10, 14, 18]
\end{lstlisting}
\end{minipage}\hfill %sample_185	steps: 275	
\begin{minipage}{.48\textwidth}
\begin{lstlisting}[caption={GPT-4 CoT failure, cannot easily tell the answer without running the loops}, label={lst:gpt-4-hard2}, breaklines=true, language=python]
def f(L):
    N = len(L)
    for k in range(1, N//2 + 1):
        i = k - 1
        j = N - k
        while i < j:
            # swap elements:
            L[i], L[j] = L[j], L[i]
            # update i, j:
            i += 1
            j -= 1
    return L
assert f([16, 14, 12, 7, 9, 11]) == [11, 14, 7, 12, 9, 16]
# GPT4+CoT preds:
# f([16, 9, 12, 7, 14, 11])
# f([16, 9, 7, 12, 14, 11])
# [11, 9, 7, 12, 14, 16]
# [11, 9, 7, 12, 14, 16]
\end{lstlisting}
\end{minipage}

Listings \ref{lst:gpt-4-output-fail1} and \ref{lst:gpt-4-output-fail2} show GPT-4 CoT failures for output prediction only. In Listing \ref{lst:gpt-4-output-fail2}, the model fails because it concludes that 6173 is not less than 1000. Small perturbations like changing to \texttt{0 < num and num < 1000} or changing the strings also failed. Both problems only have 2 possible answers and other models sometimes get them correctly whereas GPT4 CoT is consistently wrong. We manually tested scratchpad \cite{nye2021show} style prompts, which failed in the same way as regular CoT (Appendix~\ref{sub:output_cot}). 

\begin{minipage}{.48\textwidth}
\begin{lstlisting}[caption=GPT-4 CoT output,label={lst:gpt-4-output-fail1}, breaklines=true, language=python]
def f(text, suffix):
    if suffix == '':
        suffix = None
    return text.endswith(suffix)
assert f('uMeGndkGh', 'kG') == ??
# GPT-4 CoT: True
# should be False
\end{lstlisting}
\end{minipage}\hfill
\begin{minipage}{.48\textwidth}
\begin{lstlisting}[caption=GPT-4 CoT output,label={lst:gpt-4-output-fail2}, breaklines=true, language=python]
def f(num):
    if 0 < num < 1000 and num != 6174:
        return 'Half Life'
    return 'Not found'
assert f(6173) == ??
# GPT-4 CoT: 'Half Life'
# should be 'Not found'
\end{lstlisting}
\end{minipage}

\textbf{Failures of GPT-4, Input Prediction}: Here are two simple failures on input prediction. Listings \ref{lst:gpt-4-input-fail1} and \ref{lst:gpt-4-input-fail2} show input prediction failures for concise and simple Python programs with and without CoT, respectively.

\begin{minipage}{.48\textwidth}
\begin{lstlisting}[caption=GPT-4 CoT input,label={lst:gpt-4-input-fail1}, breaklines=true, language=python]
def f(text, repl):
    trans = str.maketrans(text.lower(), repl.lower())
    return text.translate(trans)
assert f(??) == 'lwwer case'

# GPT4 CoT: 'lower case', 'ow'
# could be 'lower case', 'lwwer case'
\end{lstlisting}
\end{minipage}\hfill
\begin{minipage}{.48\textwidth}
\begin{lstlisting}[caption=GPT-4 CoT input,label={lst:gpt-4-input-fail2}, breaklines=true, language=python]
def f(text):
    string = ''
    for char in text:
        string += char + char.lower()
    return string
assert f(??) == 'llaallaakk'
# GPT-4 CoT: 'LAK'
# should be 'lalak'
\end{lstlisting}
\end{minipage}

\textbf{Other GPT-4 Failures}: Finally, we conclude with a set of six relatively simple string manipulation tasks that we discovered that GPT-4 fails on. We suspect the errors could be partially due to tokenization.
The full GPT-4 outputs of these tasks can be found in Appendix \ref{appendix:simple-isolated-failures}.

\begin{lstlisting}
- What is a string containing 'a' three times, 'b' three times, 'c' twice, 'd' three times, and 'z' twice?
- In Python, what is " BaB ".rfind(" B ")?
- In Python, if I have a string s = 'iabnm~~~~~~~~~~', what is s[1::2]?
- In Python, what is "+".join(['*', '+', 'n', 'z', 'o', 'h'])?
- In Python, if text = "!123Leap and the net will appear" and res = 123, what is text[len(str(res)):]?
- What is "pomodoro".replace("or", "pomodoro")?
\end{lstlisting}

\section{Limitations and Future Work} \label{sec:limitations-future-work}

\textbf{Correlations between various code tasks}: While our benchmark serves as an interesting lens to analyze code LMs, one might object that output prediction can simply be done with a Python interpreter and that input prediction abilities can be greatly enhanced by equipping a LM with an interpreter, like in GPT-4 Code Interpreter mode. While this is true, we believe that a good code LM still ought to have good code understanding and execution capabilities, similar to that of a strong programmer. We see that base models have a reasonably strong correlation between HumanEval, input prediction, and output prediction score. An interesting future direction is to more deeply investigate the correlations between performance on various code-related tasks such as code completion, execution, bug-finding, and code summarization.

\textbf{Distilling future execution benchmarks}: Our benchmark only measures the input and output prediction accuracy of relatively simple and self-contained Python functions distilled from a single model (Code Llama 34B). It would also be interesting to measure these capabilities on longer and more difficult code snippets, open-domain code samples, or code in other programming languages. As our distillation technique is relatively general, we welcome others to create their own benchmarks measuring the execution of code snippets from other distributions.

\textbf{Variation due to prompt and temperature}: The accuracy of a model on our benchmark may be very sensitive to the prompt and task format \citep{mizrahi2023state}. We try our best to address this by using prompts that are similar as possible across models (see Appendix \ref{sec:appendix-direct-prompts} and \ref{sec:appendix-cot-prompts}) but understand that some prompts may improve the performance of certain models while decrease the performance on others. There are also countless prompting techniques (see \citep{liu2023pre} for a comprehensive survey) that can be tried to improve the performance. We also run all our experiments with $T=0.2$ and $T=0.8$ due to budget constraints, but different temperatures will lead to different performance for all models. One must always be cautious and critical when using benchmarks to compare models. For example, for input prediction, while Phind v2's 47.9\% pass@1 may seem to beat CodeLlama's 46.5\%, the standard deviations of both models with respect to the $800$ samples selected turns out to be around 1.5\%, so this conclusion cannot be made.

\textbf{Information loss due to pass@1}: While the average pass@k metric is common in the code generation literature, it compresses a large amount of information into one number. While we suggest reporting pass@1 and pass@5 for our benchmark, we comment that pass@k is only one perspective of measuring execution ability. We try to shed more light on behaviour by including a bit more analysis throughout this work, but encourage the development of different evaluation and analysis techniques.

\textbf{Fine-tuning}: In our first fine-tuning experiment, we only check for exact string match when decontaminating the fine-tuning set, so there may still be semantic duplication or similar programs with small modifications, which may lead to a higher performance than if those examples were removed. In this work, we only consider the most direct and straightforward fine-tuning scheme. We believe there is room for improvement via more sophisticated techniques, such as using process supervision \citep{uesato2022solving}, fine-tuning on correct CoT generations, or fine-tuning on snippets of code while including the program state after each step. Seeing that models like Phi, WizardCoder, and Phind outperformed Code Llama on HumanEval but not \benchmark inspires the need for a deeper investigation of the utility of finetuning on distilled data from a more powerful model. Lastly, it remains a curiosity whether fine-tuning on execution information can help code generation abilities.

\textbf{Jointly improving code generation and code execution}: As we discovered, distilled models like Phi, Phind, and WizardCoder that are fine-tuned on code generation do not improve significantly on \benchmark compared to their base models. It is unknown whether the opposite is true: does improved fine-tuning on code execution lead to better code generation abilities? It would also be interesting to explore techniques that can lead to improved performance on both code generation and code execution simultaneously.

\textbf{Understanding reasoning from the lens of code}: As future work, we believe that our benchmark serves as a good starting point towards understanding the code reasoning abilities of LM. Many further execution evaluations may be possible, such as testing execution of recursive functions, execution from a natural language description and an input, or execution of a composition of two functions. We find that output prediction serves as a good testbed for understanding CoT failures, because each step clearly corresponds to an operation with a ground truth, so reasoning failures can be pinpointed. We observed many examples of CoT failures due to simple mistakes that the model seems to have knowledge about (see Appendix \ref{sec:appendix-gpt4-cot-output} for examples), and it should be possible to analyze and characterize this behaviour more systematically. 

\textbf{Self-repair}: Lately, self-repair has been used to improve the reasoning and programming abilities of LLMs \citep{chen2023teaching, olausson2023demystifying, madaan2023self, peng2023check, zhang2023self, tyen2023llms}. From our qualitative analysis, we find that when using CoT, many output prediction failures are recitation errors of information the model may already understand. Therefore, we believe that these mistakes may be easier to repair than when the correct reasoning path is not found in the first place, and that \benchmark may be a simpler task to better understand model repair capabilities.

\section{Conclusion}

We propose \benchmark, a new benchmark consisting of simple Python functions to evaluate the input and output prediction abilities of code LMs. First, we propose a three-part recipe to distill our benchmark consisting of large-scale distillation, filtering, and data size selection via a statistical noise analysis (Sec. \ref{sec:benchmark-construction}). Second, we conduct a qualitative analysis by evaluating 20 models on our benchmark (Sec. \ref{sec:evaluation}). Our analysis leads to insights regarding the correlation between HumanEval and our benchmark, the correlation between input and output prediction, differences between various code LMs, and the diversity of different models. Third, we explore the potential of CoT (Sec. \ref{subsec:cot}) and fine-tuning (Sec. \ref{subsec:finetuning}) for improving performance. Fourth, we provide a qualitative analysis showcasing successes and failures of GPT-4 on our benchmark (Sec. \ref{sec:qualitative-analysis} and Appendix \ref{sec:appendix-qualitative}). Overall, we believe that \benchmark provides a complimentary perspective to classical code LM evaluation such as HumanEval and MBPP and encourage creators of future code LMs to try out our benchmark!

\section{Acknowledgements}
In alphabetical order by first name, we thank Chris Cummings, Dylan Zhang, Jonas Gehring, Kunhao Zheng, Luyu Gao, Naman Jain, Nicolas Usunier, Ofir Press, Robin Jia, Scott Yih, Theo Olausson, and Wen-Ding Li for very helpful suggestions and support that influenced the trajectory of this work.

A. Gu is supported by the National Science Foundation (NSF) Graduate Research Fellowship under Grant No. 2141064. A. Solar-Lezama is supported by the National Science Foundation (NSF) and Intel Corporation through NSF Grant CCF:2217064.

\bibliography{bibliography}

\begin{thebibliography}{109}
\providecommand{\natexlab}[1]{#1}
\providecommand{\url}[1]{\texttt{#1}}
\expandafter\ifx\csname urlstyle\endcsname\relax
  \providecommand{\doi}[1]{doi: #1}\else
  \providecommand{\doi}{doi: \begingroup \urlstyle{rm}\Url}\fi

\bibitem[Agashe et~al.(2019)Agashe, Iyer, and Zettlemoyer]{agashe2019juice}
Agashe, R., Iyer, S., and Zettlemoyer, L.
\newblock Juice: A large scale distantly supervised dataset for open domain context-based code generation.
\newblock \emph{arXiv preprint arXiv:1910.02216}, 2019.

\bibitem[Ahmad et~al.(2021)Ahmad, Tushar, Chakraborty, and Chang]{ahmad-etal-2021-avatar}
Ahmad, W.~U., Tushar, M. G.~R., Chakraborty, S., and Chang, K.-W.
\newblock Avatar: A parallel corpus for java-python program translation.
\newblock \emph{arXiv preprint arXiv:2108.11590}, 2021.

\bibitem[AI(2023)]{deepseek-coder}
AI, D.
\newblock Deepseek coder: Let the code write itself.
\newblock \url{https://github.com/deepseek-ai/DeepSeek-Coder}, 2023.

\bibitem[Allal et~al.(2023)Allal, Li, Kocetkov, Mou, Akiki, Ferrandis, Muennighoff, Mishra, Gu, Dey, et~al.]{allal2023santacoder}
Allal, L.~B., Li, R., Kocetkov, D., Mou, C., Akiki, C., Ferrandis, C.~M., Muennighoff, N., Mishra, M., Gu, A., Dey, M., et~al.
\newblock Santacoder: don't reach for the stars!
\newblock \emph{arXiv preprint arXiv:2301.03988}, 2023.

\bibitem[Alon et~al.(2018)Alon, Brody, Levy, and Yahav]{alon2018code2seq}
Alon, U., Brody, S., Levy, O., and Yahav, E.
\newblock code2seq: Generating sequences from structured representations of code.
\newblock \emph{arXiv preprint arXiv:1808.01400}, 2018.

\bibitem[Arkoudas(2023)]{arkoudas2023gpt}
Arkoudas, K.
\newblock Gpt-4 can't reason.
\newblock \emph{arXiv preprint arXiv:2308.03762}, 2023.

\bibitem[Athiwaratkun et~al.(2022)Athiwaratkun, Gouda, Wang, Li, Tian, Tan, Ahmad, Wang, Sun, Shang, et~al.]{athiwaratkun2022multi}
Athiwaratkun, B., Gouda, S.~K., Wang, Z., Li, X., Tian, Y., Tan, M., Ahmad, W.~U., Wang, S., Sun, Q., Shang, M., et~al.
\newblock Multi-lingual evaluation of code generation models.
\newblock \emph{arXiv preprint arXiv:2210.14868}, 2022.

\bibitem[Austin et~al.(2021)Austin, Odena, Nye, Bosma, Michalewski, Dohan, Jiang, Cai, Terry, Le, et~al.]{austin2021program}
Austin, J., Odena, A., Nye, M., Bosma, M., Michalewski, H., Dohan, D., Jiang, E., Cai, C., Terry, M., Le, Q., et~al.
\newblock Program synthesis with large language models.
\newblock \emph{arXiv preprint arXiv:2108.07732}, 2021.

\bibitem[Barone \& Sennrich(2017)Barone and Sennrich]{barone2017parallel}
Barone, A. V.~M. and Sennrich, R.
\newblock A parallel corpus of python functions and documentation strings for automated code documentation and code generation.
\newblock \emph{arXiv preprint arXiv:1707.02275}, 2017.

\bibitem[Berabi et~al.(2021)Berabi, He, Raychev, and Vechev]{berabi2021tfix}
Berabi, B., He, J., Raychev, V., and Vechev, M.
\newblock Tfix: Learning to fix coding errors with a text-to-text transformer.
\newblock In \emph{International Conference on Machine Learning}, pp.\  780--791. PMLR, 2021.

\bibitem[Berglund et~al.(2023)Berglund, Tong, Kaufmann, Balesni, Stickland, Korbak, and Evans]{berglund2023reversal}
Berglund, L., Tong, M., Kaufmann, M., Balesni, M., Stickland, A.~C., Korbak, T., and Evans, O.
\newblock The reversal curse: Llms trained on" a is b" fail to learn" b is a".
\newblock \emph{arXiv preprint arXiv:2309.12288}, 2023.

\bibitem[Brown et~al.(2020)Brown, Mann, Ryder, Subbiah, Kaplan, Dhariwal, Neelakantan, Shyam, Sastry, Askell, et~al.]{brown2020language}
Brown, T., Mann, B., Ryder, N., Subbiah, M., Kaplan, J.~D., Dhariwal, P., Neelakantan, A., Shyam, P., Sastry, G., Askell, A., et~al.
\newblock Language models are few-shot learners.
\newblock \emph{Advances in neural information processing systems}, 33:\penalty0 1877--1901, 2020.

\bibitem[Bubeck et~al.(2023)Bubeck, Chandrasekaran, Eldan, Gehrke, Horvitz, Kamar, Lee, Lee, Li, Lundberg, et~al.]{bubeck2023sparks}
Bubeck, S., Chandrasekaran, V., Eldan, R., Gehrke, J., Horvitz, E., Kamar, E., Lee, P., Lee, Y.~T., Li, Y., Lundberg, S., et~al.
\newblock Sparks of artificial general intelligence: Early experiments with gpt-4.
\newblock \emph{arXiv preprint arXiv:2303.12712}, 2023.

\bibitem[Cassano et~al.(2022)Cassano, Gouwar, Nguyen, Nguyen, Phipps-Costin, Pinckney, Yee, Zi, Anderson, Feldman, et~al.]{cassano2022multipl}
Cassano, F., Gouwar, J., Nguyen, D., Nguyen, S., Phipps-Costin, L., Pinckney, D., Yee, M.-H., Zi, Y., Anderson, C.~J., Feldman, M.~Q., et~al.
\newblock Multipl-e: A scalable and extensible approach to benchmarking neural code generation.
\newblock \emph{arXiv preprint arXiv:2208.08227}, 2022.

\bibitem[Chen et~al.(2022)Chen, Zhang, Nguyen, Zan, Lin, Lou, and Chen]{chen2022codet}
Chen, B., Zhang, F., Nguyen, A., Zan, D., Lin, Z., Lou, J.-G., and Chen, W.
\newblock Codet: Code generation with generated tests.
\newblock \emph{arXiv preprint arXiv:2207.10397}, 2022.

\bibitem[Chen et~al.(2021)Chen, Tworek, Jun, Yuan, Pinto, Kaplan, Edwards, Burda, Joseph, Brockman, et~al.]{chen2021evaluating}
Chen, M., Tworek, J., Jun, H., Yuan, Q., Pinto, H. P. d.~O., Kaplan, J., Edwards, H., Burda, Y., Joseph, N., Brockman, G., et~al.
\newblock Evaluating large language models trained on code.
\newblock \emph{arXiv preprint arXiv:2107.03374}, 2021.

\bibitem[Chen et~al.(2023)Chen, Lin, Sch{\"a}rli, and Zhou]{chen2023teaching}
Chen, X., Lin, M., Sch{\"a}rli, N., and Zhou, D.
\newblock Teaching large language models to self-debug.
\newblock \emph{arXiv preprint arXiv:2304.05128}, 2023.

\bibitem[Ding et~al.(2022)Ding, Wang, Ahmad, Ramanathan, Nallapati, Bhatia, Roth, and Xiang]{ding2022cocomic}
Ding, Y., Wang, Z., Ahmad, W.~U., Ramanathan, M.~K., Nallapati, R., Bhatia, P., Roth, D., and Xiang, B.
\newblock Cocomic: Code completion by jointly modeling in-file and cross-file context.
\newblock \emph{arXiv preprint arXiv:2212.10007}, 2022.

\bibitem[Dziri et~al.(2023)Dziri, Lu, Sclar, Li, Jian, Lin, West, Bhagavatula, Bras, Hwang, et~al.]{dziri2023faith}
Dziri, N., Lu, X., Sclar, M., Li, X.~L., Jian, L., Lin, B.~Y., West, P., Bhagavatula, C., Bras, R.~L., Hwang, J.~D., et~al.
\newblock Faith and fate: Limits of transformers on compositionality.
\newblock \emph{arXiv preprint arXiv:2305.18654}, 2023.

\bibitem[Fan et~al.(2023)Fan, Gokkaya, Harman, Lyubarskiy, Sengupta, Yoo, and Zhang]{fan2023large}
Fan, A., Gokkaya, B., Harman, M., Lyubarskiy, M., Sengupta, S., Yoo, S., and Zhang, J.~M.
\newblock Large language models for software engineering: Survey and open problems.
\newblock \emph{arXiv preprint arXiv:2310.03533}, 2023.

\bibitem[Fried et~al.(2022)Fried, Aghajanyan, Lin, Wang, Wallace, Shi, Zhong, tau Yih, Zettlemoyer, and Lewis]{fried2022incoder}
Fried, D., Aghajanyan, A., Lin, J., Wang, S., Wallace, E., Shi, F., Zhong, R., tau Yih, W., Zettlemoyer, L., and Lewis, M.
\newblock Incoder: A generative model for code infilling and synthesis.
\newblock \emph{preprint arXiv:2204.05999}, 2022.

\bibitem[Garg et~al.(2022)Garg, Moghaddam, Clement, Sundaresan, and Wu]{garg2022deepperf}
Garg, S., Moghaddam, R.~Z., Clement, C.~B., Sundaresan, N., and Wu, C.
\newblock Deepperf: A deep learning-based approach for improving software performance.
\newblock \emph{arXiv preprint arXiv:2206.13619}, 2022.

\bibitem[Giannou et~al.(2023)Giannou, Rajput, Sohn, Lee, Lee, and Papailiopoulos]{giannou2023looped}
Giannou, A., Rajput, S., Sohn, J.-y., Lee, K., Lee, J.~D., and Papailiopoulos, D.
\newblock Looped transformers as programmable computers.
\newblock \emph{arXiv preprint arXiv:2301.13196}, 2023.

\bibitem[Gudibande et~al.(2023)Gudibande, Wallace, Snell, Geng, Liu, Abbeel, Levine, and Song]{gudibande2023false}
Gudibande, A., Wallace, E., Snell, C., Geng, X., Liu, H., Abbeel, P., Levine, S., and Song, D.
\newblock The false promise of imitating proprietary llms.
\newblock \emph{arXiv preprint arXiv:2305.15717}, 2023.

\bibitem[Gunasekar et~al.(2023)Gunasekar, Zhang, Aneja, Mendes, Del~Giorno, Gopi, Javaheripi, Kauffmann, de~Rosa, Saarikivi, et~al.]{gunasekar2023textbooks}
Gunasekar, S., Zhang, Y., Aneja, J., Mendes, C. C.~T., Del~Giorno, A., Gopi, S., Javaheripi, M., Kauffmann, P., de~Rosa, G., Saarikivi, O., et~al.
\newblock Textbooks are all you need.
\newblock \emph{arXiv preprint arXiv:2306.11644}, 2023.

\bibitem[Gupta et~al.(2017)Gupta, Pal, Kanade, and Shevade]{gupta2017deepfix}
Gupta, R., Pal, S., Kanade, A., and Shevade, S.
\newblock Deepfix: Fixing common c language errors by deep learning.
\newblock In \emph{Proceedings of the aaai conference on artificial intelligence}, volume~31, 2017.

\bibitem[Haluptzok et~al.(2022)Haluptzok, Bowers, and Kalai]{haluptzok2022language}
Haluptzok, P., Bowers, M., and Kalai, A.~T.
\newblock Language models can teach themselves to program better.
\newblock \emph{arXiv preprint arXiv:2207.14502}, 2022.

\bibitem[Haque et~al.(2022)Haque, Ahmad, Lourentzou, and Brown]{haque2022fixeval}
Haque, M. M.~A., Ahmad, W.~U., Lourentzou, I., and Brown, C.
\newblock Fixeval: Execution-based evaluation of program fixes for competitive programming problems.
\newblock 2022.

\bibitem[Hasan et~al.(2021)Hasan, Muttaqueen, Ishtiaq, Mehrab, Haque, Hasan, Ahmad, Iqbal, and Shahriyar]{hasan2021codesc}
Hasan, M., Muttaqueen, T., Ishtiaq, A.~A., Mehrab, K.~S., Haque, M. M.~A., Hasan, T., Ahmad, W.~U., Iqbal, A., and Shahriyar, R.
\newblock Codesc: A large code-description parallel dataset.
\newblock \emph{arXiv preprint arXiv:2105.14220}, 2021.

\bibitem[Hendrycks et~al.(2021)Hendrycks, Basart, Kadavath, Mazeika, Arora, Guo, Burns, Puranik, He, Song, et~al.]{hendrycks2021measuring}
Hendrycks, D., Basart, S., Kadavath, S., Mazeika, M., Arora, A., Guo, E., Burns, C., Puranik, S., He, H., Song, D., et~al.
\newblock Measuring coding challenge competence with apps.
\newblock \emph{arXiv preprint arXiv:2105.09938}, 2021.

\bibitem[Hosseini et~al.(2021)Hosseini, Reddy, Bahdanau, Hjelm, Sordoni, and Courville]{hosseini2021understanding}
Hosseini, A., Reddy, S., Bahdanau, D., Hjelm, R.~D., Sordoni, A., and Courville, A.
\newblock Understanding by understanding not: Modeling negation in language models.
\newblock \emph{arXiv preprint arXiv:2105.03519}, 2021.

\bibitem[Husain et~al.(2019)Husain, Wu, Gazit, Allamanis, and Brockschmidt]{husain2019codesearchnet}
Husain, H., Wu, H.-H., Gazit, T., Allamanis, M., and Brockschmidt, M.
\newblock Codesearchnet challenge: Evaluating the state of semantic code search.
\newblock \emph{arXiv preprint arXiv:1909.09436}, 2019.

\bibitem[Iyer et~al.(2016)Iyer, Konstas, Cheung, and Zettlemoyer]{iyer2016summarizing}
Iyer, S., Konstas, I., Cheung, A., and Zettlemoyer, L.
\newblock Summarizing source code using a neural attention model.
\newblock In \emph{54th Annual Meeting of the Association for Computational Linguistics 2016}, pp.\  2073--2083. Association for Computational Linguistics, 2016.

\bibitem[Jain et~al.(2022)Jain, Vaidyanath, Iyer, Natarajan, Parthasarathy, Rajamani, and Sharma]{jain2022jigsaw}
Jain, N., Vaidyanath, S., Iyer, A., Natarajan, N., Parthasarathy, S., Rajamani, S., and Sharma, R.
\newblock Jigsaw: Large language models meet program synthesis.
\newblock In \emph{Proceedings of the 44th International Conference on Software Engineering}, pp.\  1219--1231, 2022.

\bibitem[Jiang et~al.(2023{\natexlab{a}})Jiang, Sablayrolles, Mensch, Bamford, Chaplot, Casas, Bressand, Lengyel, Lample, Saulnier, et~al.]{jiang2023mistral}
Jiang, A.~Q., Sablayrolles, A., Mensch, A., Bamford, C., Chaplot, D.~S., Casas, D. d.~l., Bressand, F., Lengyel, G., Lample, G., Saulnier, L., et~al.
\newblock Mistral 7b.
\newblock \emph{arXiv preprint arXiv:2310.06825}, 2023{\natexlab{a}}.

\bibitem[Jiang et~al.(2023{\natexlab{b}})Jiang, Liu, Lutellier, and Tan]{jiang2023impact}
Jiang, N., Liu, K., Lutellier, T., and Tan, L.
\newblock Impact of code language models on automated program repair.
\newblock \emph{arXiv preprint arXiv:2302.05020}, 2023{\natexlab{b}}.

\bibitem[Jimenez et~al.(2023)Jimenez, Yang, Wettig, Yao, Pei, Press, and Narasimhan]{jimenez2023swe}
Jimenez, C.~E., Yang, J., Wettig, A., Yao, S., Pei, K., Press, O., and Narasimhan, K.
\newblock Swe-bench: Can language models resolve real-world github issues?
\newblock \emph{arXiv preprint arXiv:2310.06770}, 2023.

\bibitem[Jin et~al.(2023)Jin, Shahriar, Tufano, Shi, Lu, Sundaresan, and Svyatkovskiy]{jin2023inferfix}
Jin, M., Shahriar, S., Tufano, M., Shi, X., Lu, S., Sundaresan, N., and Svyatkovskiy, A.
\newblock Inferfix: End-to-end program repair with llms.
\newblock \emph{arXiv preprint arXiv:2303.07263}, 2023.

\bibitem[Key et~al.(2022)Key, Li, and Ellis]{key2022speak}
Key, D., Li, W.-D., and Ellis, K.
\newblock I speak, you verify: Toward trustworthy neural program synthesis.
\newblock \emph{arXiv preprint arXiv:2210.00848}, 2022.

\bibitem[Lai et~al.(2023)Lai, Li, Wang, Zhang, Zhong, Zettlemoyer, Yih, Fried, Wang, and Yu]{lai2023ds}
Lai, Y., Li, C., Wang, Y., Zhang, T., Zhong, R., Zettlemoyer, L., Yih, W.-t., Fried, D., Wang, S., and Yu, T.
\newblock Ds-1000: A natural and reliable benchmark for data science code generation.
\newblock In \emph{International Conference on Machine Learning}, pp.\  18319--18345. PMLR, 2023.

\bibitem[Le et~al.(2022)Le, Wang, Gotmare, Savarese, and Hoi]{le2022coderl}
Le, H., Wang, Y., Gotmare, A.~D., Savarese, S., and Hoi, S. C.~H.
\newblock Coderl: Mastering code generation through pretrained models and deep reinforcement learning.
\newblock \emph{Advances in Neural Information Processing Systems}, 35:\penalty0 21314--21328, 2022.

\bibitem[LeClair et~al.(2019)LeClair, Jiang, and McMillan]{leclair2019neural}
LeClair, A., Jiang, S., and McMillan, C.
\newblock A neural model for generating natural language summaries of program subroutines.
\newblock In \emph{2019 IEEE/ACM 41st International Conference on Software Engineering (ICSE)}, pp.\  795--806. IEEE, 2019.

\bibitem[Lee et~al.(2023)Lee, Sreenivasan, Lee, Lee, and Papailiopoulos]{lee2023teaching}
Lee, N., Sreenivasan, K., Lee, J.~D., Lee, K., and Papailiopoulos, D.
\newblock Teaching arithmetic to small transformers.
\newblock \emph{arXiv preprint arXiv:2307.03381}, 2023.

\bibitem[Li et~al.(2023{\natexlab{a}})Li, Allal, Zi, Muennighoff, Kocetkov, Mou, Marone, Akiki, Li, Chim, et~al.]{li2023starcoder}
Li, R., Allal, L.~B., Zi, Y., Muennighoff, N., Kocetkov, D., Mou, C., Marone, M., Akiki, C., Li, J., Chim, J., et~al.
\newblock Starcoder: may the source be with you!
\newblock \emph{arXiv preprint arXiv:2305.06161}, 2023{\natexlab{a}}.

\bibitem[Li et~al.(2022)Li, Choi, Chung, Kushman, Schrittwieser, Leblond, Eccles, Keeling, Gimeno, Dal~Lago, et~al.]{li2022competition}
Li, Y., Choi, D., Chung, J., Kushman, N., Schrittwieser, J., Leblond, R., Eccles, T., Keeling, J., Gimeno, F., Dal~Lago, A., et~al.
\newblock Competition-level code generation with alphacode.
\newblock \emph{Science}, 378\penalty0 (6624):\penalty0 1092--1097, 2022.

\bibitem[Li et~al.(2023{\natexlab{b}})Li, Bubeck, Eldan, Del~Giorno, Gunasekar, and Lee]{li2023textbooks}
Li, Y., Bubeck, S., Eldan, R., Del~Giorno, A., Gunasekar, S., and Lee, Y.~T.
\newblock Textbooks are all you need ii: phi-1.5 technical report.
\newblock \emph{arXiv preprint arXiv:2309.05463}, 2023{\natexlab{b}}.

\bibitem[Liguori et~al.(2022)Liguori, Al-Hossami, Cotroneo, Natella, Cukic, and Shaikh]{liguori2022can}
Liguori, P., Al-Hossami, E., Cotroneo, D., Natella, R., Cukic, B., and Shaikh, S.
\newblock Can we generate shellcodes via natural language? an empirical study.
\newblock \emph{Automated Software Engineering}, 29\penalty0 (1):\penalty0 30, 2022.

\bibitem[Liu et~al.(2023{\natexlab{a}})Liu, Lu, Chen, Jiang, Svyatkovskiy, Fu, Sundaresan, and Duan]{liu2023code}
Liu, C., Lu, S., Chen, W., Jiang, D., Svyatkovskiy, A., Fu, S., Sundaresan, N., and Duan, N.
\newblock Code execution with pre-trained language models.
\newblock \emph{arXiv preprint arXiv:2305.05383}, 2023{\natexlab{a}}.

\bibitem[Liu et~al.(2023{\natexlab{b}})Liu, Ning, Teng, Liu, Zhou, and Zhang]{liu2023evaluating}
Liu, H., Ning, R., Teng, Z., Liu, J., Zhou, Q., and Zhang, Y.
\newblock Evaluating the logical reasoning ability of chatgpt and gpt-4.
\newblock \emph{arXiv preprint arXiv:2304.03439}, 2023{\natexlab{b}}.

\bibitem[Liu et~al.(2023{\natexlab{c}})Liu, Xia, Wang, and Zhang]{liu2023your}
Liu, J., Xia, C.~S., Wang, Y., and Zhang, L.
\newblock Is your code generated by chatgpt really correct? rigorous evaluation of large language models for code generation.
\newblock \emph{arXiv preprint arXiv:2305.01210}, 2023{\natexlab{c}}.

\bibitem[Liu et~al.(2023{\natexlab{d}})Liu, Yuan, Fu, Jiang, Hayashi, and Neubig]{liu2023pre}
Liu, P., Yuan, W., Fu, J., Jiang, Z., Hayashi, H., and Neubig, G.
\newblock Pre-train, prompt, and predict: A systematic survey of prompting methods in natural language processing.
\newblock \emph{ACM Computing Surveys}, 55\penalty0 (9):\penalty0 1--35, 2023{\natexlab{d}}.

\bibitem[Liu et~al.(2020)Liu, Gao, Chen, Nie, and Liu]{liu2020atom}
Liu, S., Gao, C., Chen, S., Nie, L.~Y., and Liu, Y.
\newblock Atom: Commit message generation based on abstract syntax tree and hybrid ranking.
\newblock \emph{IEEE Transactions on Software Engineering}, 48\penalty0 (5):\penalty0 1800--1817, 2020.

\bibitem[Liu et~al.(2023{\natexlab{e}})Liu, Xu, and McAuley]{liu2023repobench}
Liu, T., Xu, C., and McAuley, J.
\newblock Repobench: Benchmarking repository-level code auto-completion systems.
\newblock \emph{arXiv preprint arXiv:2306.03091}, 2023{\natexlab{e}}.

\bibitem[Luo et~al.(2023)Luo, Xu, Zhao, Sun, Geng, Hu, Tao, Ma, Lin, and Jiang]{luo2023wizardcoder}
Luo, Z., Xu, C., Zhao, P., Sun, Q., Geng, X., Hu, W., Tao, C., Ma, J., Lin, Q., and Jiang, D.
\newblock Wizardcoder: Empowering code large language models with evol-instruct.
\newblock \emph{arXiv preprint arXiv:2306.08568}, 2023.

\bibitem[Madaan et~al.(2023{\natexlab{a}})Madaan, Shypula, Alon, Hashemi, Ranganathan, Yang, Neubig, and Yazdanbakhsh]{madaan2023learning}
Madaan, A., Shypula, A., Alon, U., Hashemi, M., Ranganathan, P., Yang, Y., Neubig, G., and Yazdanbakhsh, A.
\newblock Learning performance-improving code edits.
\newblock \emph{arXiv preprint arXiv:2302.07867}, 2023{\natexlab{a}}.

\bibitem[Madaan et~al.(2023{\natexlab{b}})Madaan, Tandon, Gupta, Hallinan, Gao, Wiegreffe, Alon, Dziri, Prabhumoye, Yang, et~al.]{madaan2023self}
Madaan, A., Tandon, N., Gupta, P., Hallinan, S., Gao, L., Wiegreffe, S., Alon, U., Dziri, N., Prabhumoye, S., Yang, Y., et~al.
\newblock Self-refine: Iterative refinement with self-feedback.
\newblock \emph{arXiv preprint arXiv:2303.17651}, 2023{\natexlab{b}}.

\bibitem[Malik et~al.(2019)Malik, Patra, and Pradel]{malik2019nl2type}
Malik, R.~S., Patra, J., and Pradel, M.
\newblock Nl2type: inferring javascript function types from natural language information.
\newblock In \emph{2019 IEEE/ACM 41st International Conference on Software Engineering (ICSE)}, pp.\  304--315. IEEE, 2019.

\bibitem[Merrill \& Sabharwal(2023)Merrill and Sabharwal]{merrill2023expresssive}
Merrill, W. and Sabharwal, A.
\newblock The expresssive power of transformers with chain of thought.
\newblock \emph{arXiv preprint arXiv:2310.07923}, 2023.

\bibitem[Merrill et~al.(2021)Merrill, Goldberg, Schwartz, and Smith]{merrill2021provable}
Merrill, W.~C., Goldberg, Y., Schwartz, R., and Smith, N.~A.
\newblock Provable limitations of acquiring meaning from ungrounded form: What will future language models understand?
\newblock \emph{Transactions of the Association for Computational Linguistics}, 9:\penalty0 1047--1060, 2021.

\bibitem[Miceli-Barone et~al.(2023)Miceli-Barone, Barez, Konstas, and Cohen]{miceli2023larger}
Miceli-Barone, A.~V., Barez, F., Konstas, I., and Cohen, S.~B.
\newblock The larger they are, the harder they fail: Language models do not recognize identifier swaps in python.
\newblock \emph{arXiv preprint arXiv:2305.15507}, 2023.

\bibitem[Mir et~al.(2022)Mir, Lato{\v{s}}kinas, Proksch, and Gousios]{mir2022type4py}
Mir, A.~M., Lato{\v{s}}kinas, E., Proksch, S., and Gousios, G.
\newblock Type4py: Practical deep similarity learning-based type inference for python.
\newblock In \emph{Proceedings of the 44th International Conference on Software Engineering}, pp.\  2241--2252, 2022.

\bibitem[Mizrahi et~al.(2023)Mizrahi, Kaplan, Malkin, Dror, Shahaf, and Stanovsky]{mizrahi2023state}
Mizrahi, M., Kaplan, G., Malkin, D., Dror, R., Shahaf, D., and Stanovsky, G.
\newblock State of what art? a call for multi-prompt llm evaluation.
\newblock \emph{arXiv preprint arXiv:2401.00595}, 2023.

\bibitem[Ni et~al.(2023)Ni, Iyer, Radev, Stoyanov, Yih, Wang, and Lin]{ni2023lever}
Ni, A., Iyer, S., Radev, D., Stoyanov, V., Yih, W.-t., Wang, S., and Lin, X.~V.
\newblock Lever: Learning to verify language-to-code generation with execution.
\newblock In \emph{International Conference on Machine Learning}, pp.\  26106--26128. PMLR, 2023.

\bibitem[Nijkamp et~al.(2022)Nijkamp, Pang, Hayashi, Tu, Wang, Zhou, Savarese, and Xiong]{nijkamp2022codegen}
Nijkamp, E., Pang, B., Hayashi, H., Tu, L., Wang, H., Zhou, Y., Savarese, S., and Xiong, C.
\newblock Codegen: An open large language model for code with multi-turn program synthesis.
\newblock In \emph{The Eleventh International Conference on Learning Representations}, 2022.

\bibitem[Nye et~al.(2021)Nye, Andreassen, Gur-Ari, Michalewski, Austin, Bieber, Dohan, Lewkowycz, Bosma, Luan, et~al.]{nye2021show}
Nye, M., Andreassen, A.~J., Gur-Ari, G., Michalewski, H., Austin, J., Bieber, D., Dohan, D., Lewkowycz, A., Bosma, M., Luan, D., et~al.
\newblock Show your work: Scratchpads for intermediate computation with language models.
\newblock \emph{arXiv preprint arXiv:2112.00114}, 2021.

\bibitem[Olausson et~al.(2023{\natexlab{a}})Olausson, Gu, Lipkin, Zhang, Solar-Lezama, Tenenbaum, and Levy]{olausson2023linc}
Olausson, T.~X., Gu, A., Lipkin, B., Zhang, C.~E., Solar-Lezama, A., Tenenbaum, J.~B., and Levy, R.
\newblock Linc: A neurosymbolic approach for logical reasoning by combining language models with first-order logic provers.
\newblock \emph{arXiv preprint arXiv:2310.15164}, 2023{\natexlab{a}}.

\bibitem[Olausson et~al.(2023{\natexlab{b}})Olausson, Inala, Wang, Gao, and Solar-Lezama]{olausson2023demystifying}
Olausson, T.~X., Inala, J.~P., Wang, C., Gao, J., and Solar-Lezama, A.
\newblock Demystifying gpt self-repair for code generation.
\newblock \emph{arXiv preprint arXiv:2306.09896}, 2023{\natexlab{b}}.

\bibitem[OpenAI(2023)]{openai2023gpt}
OpenAI, R.
\newblock Gpt-4 technical report. arxiv 2303.08774.
\newblock \emph{View in Article}, 2023.

\bibitem[Ouyang et~al.(2022)Ouyang, Wu, Jiang, Almeida, Wainwright, Mishkin, Zhang, Agarwal, Slama, Ray, et~al.]{ouyang2022training}
Ouyang, L., Wu, J., Jiang, X., Almeida, D., Wainwright, C., Mishkin, P., Zhang, C., Agarwal, S., Slama, K., Ray, A., et~al.
\newblock Training language models to follow instructions with human feedback.
\newblock \emph{Advances in Neural Information Processing Systems}, 35:\penalty0 27730--27744, 2022.

\bibitem[Patil et~al.(2023)Patil, Zhang, Wang, and Gonzalez]{patil2023gorilla}
Patil, S.~G., Zhang, T., Wang, X., and Gonzalez, J.~E.
\newblock Gorilla: Large language model connected with massive apis.
\newblock \emph{arXiv preprint arXiv:2305.15334}, 2023.

\bibitem[Pearce et~al.(2022)Pearce, Ahmad, Tan, Dolan-Gavitt, and Karri]{pearce2022asleep}
Pearce, H., Ahmad, B., Tan, B., Dolan-Gavitt, B., and Karri, R.
\newblock Asleep at the keyboard? assessing the security of github copilot’s code contributions.
\newblock In \emph{2022 IEEE Symposium on Security and Privacy (SP)}, pp.\  754--768. IEEE, 2022.

\bibitem[Peng et~al.(2023)Peng, Galley, He, Cheng, Xie, Hu, Huang, Liden, Yu, Chen, and Gao]{peng2023check}
Peng, B., Galley, M., He, P., Cheng, H., Xie, Y., Hu, Y., Huang, Q., Liden, L., Yu, Z., Chen, W., and Gao, J.
\newblock Check your facts and try again: Improving large language models with external knowledge and automated feedback.
\newblock \emph{arXiv preprint arXiv:2302.12813}, 2023.

\bibitem[Royzen et~al.(2023)Royzen, Wei, and Coleman]{Phind}
Royzen, M., Wei, J., and Coleman, R.
\newblock Phind, 2023.
\newblock URL \url{https://www.phind.com}.

\bibitem[Roziere et~al.(2020)Roziere, Lachaux, Chanussot, and Lample]{roziere2020unsupervised}
Roziere, B., Lachaux, M.-A., Chanussot, L., and Lample, G.
\newblock Unsupervised translation of programming languages.
\newblock \emph{Advances in Neural Information Processing Systems}, 33:\penalty0 20601--20611, 2020.

\bibitem[Roziere et~al.(2023)Roziere, Gehring, Gloeckle, Sootla, Gat, Tan, Adi, Liu, Remez, Rapin, et~al.]{roziere2023code}
Roziere, B., Gehring, J., Gloeckle, F., Sootla, S., Gat, I., Tan, X.~E., Adi, Y., Liu, J., Remez, T., Rapin, J., et~al.
\newblock Code llama: Open foundation models for code.
\newblock \emph{arXiv preprint arXiv:2308.12950}, 2023.

\bibitem[Shen et~al.(2023)Shen, Zhang, Chen, Zan, Geng, Fu, Zeng, Yu, Ji, Zhao, et~al.]{shen2023pangu}
Shen, B., Zhang, J., Chen, T., Zan, D., Geng, B., Fu, A., Zeng, M., Yu, A., Ji, J., Zhao, J., et~al.
\newblock Pangu-coder2: Boosting large language models for code with ranking feedback.
\newblock \emph{arXiv preprint arXiv:2307.14936}, 2023.

\bibitem[Shi et~al.(2022)Shi, Fried, Ghazvininejad, Zettlemoyer, and Wang]{shi2022natural}
Shi, F., Fried, D., Ghazvininejad, M., Zettlemoyer, L., and Wang, S.~I.
\newblock Natural language to code translation with execution.
\newblock \emph{arXiv preprint arXiv:2204.11454}, 2022.

\bibitem[Shi et~al.(2023)Shi, Chen, Misra, Scales, Dohan, Chi, Sch{\"a}rli, and Zhou]{shi2023large}
Shi, F., Chen, X., Misra, K., Scales, N., Dohan, D., Chi, E.~H., Sch{\"a}rli, N., and Zhou, D.
\newblock Large language models can be easily distracted by irrelevant context.
\newblock In \emph{International Conference on Machine Learning}, pp.\  31210--31227. PMLR, 2023.

\bibitem[Shinn et~al.(2023)Shinn, Labash, and Gopinath]{shinn2023reflexion}
Shinn, N., Labash, B., and Gopinath, A.
\newblock Reflexion: an autonomous agent with dynamic memory and self-reflection.
\newblock \emph{arXiv preprint arXiv:2303.11366}, 2023.

\bibitem[Shrivastava et~al.(2023)Shrivastava, Larochelle, and Tarlow]{shrivastava2023repository}
Shrivastava, D., Larochelle, H., and Tarlow, D.
\newblock Repository-level prompt generation for large language models of code.
\newblock In \emph{International Conference on Machine Learning}, pp.\  31693--31715. PMLR, 2023.

\bibitem[Tian \& Chen(2023)Tian and Chen]{tian2023test}
Tian, Z. and Chen, J.
\newblock Test-case-driven programming understanding in large language models for better code generation.
\newblock \emph{arXiv preprint arXiv:2309.16120}, 2023.

\bibitem[Tony et~al.(2023)Tony, Mutas, Ferreyra, and Scandariato]{tony2023llmseceval}
Tony, C., Mutas, M., Ferreyra, N. E.~D., and Scandariato, R.
\newblock Llmseceval: A dataset of natural language prompts for security evaluations.
\newblock \emph{arXiv preprint arXiv:2303.09384}, 2023.

\bibitem[Tufano et~al.(2019)Tufano, Watson, Bavota, Penta, White, and Poshyvanyk]{tufano2019empirical}
Tufano, M., Watson, C., Bavota, G., Penta, M.~D., White, M., and Poshyvanyk, D.
\newblock An empirical study on learning bug-fixing patches in the wild via neural machine translation.
\newblock \emph{ACM Transactions on Software Engineering and Methodology (TOSEM)}, 28\penalty0 (4):\penalty0 1--29, 2019.

\bibitem[Tufano et~al.(2022)Tufano, Deng, Sundaresan, and Svyatkovskiy]{tufano2022methods2test}
Tufano, M., Deng, S.~K., Sundaresan, N., and Svyatkovskiy, A.
\newblock Methods2test: A dataset of focal methods mapped to test cases.
\newblock In \emph{Proceedings of the 19th International Conference on Mining Software Repositories}, pp.\  299--303, 2022.

\bibitem[Tyen et~al.(2023)Tyen, Mansoor, Chen, Mak, and C{\u{a}}rbune]{tyen2023llms}
Tyen, G., Mansoor, H., Chen, P., Mak, T., and C{\u{a}}rbune, V.
\newblock Llms cannot find reasoning errors, but can correct them!
\newblock \emph{arXiv preprint arXiv:2311.08516}, 2023.

\bibitem[Uesato et~al.(2022)Uesato, Kushman, Kumar, Song, Siegel, Wang, Creswell, Irving, and Higgins]{uesato2022solving}
Uesato, J., Kushman, N., Kumar, R., Song, F., Siegel, N., Wang, L., Creswell, A., Irving, G., and Higgins, I.
\newblock Solving math word problems with process-and outcome-based feedback.
\newblock \emph{arXiv preprint arXiv:2211.14275}, 2022.

\bibitem[Wang et~al.(2022{\natexlab{a}})Wang, Li, Qian, Yang, Wang, Shang, Kumar, Tan, Ray, Bhatia, et~al.]{wang2022recode}
Wang, S., Li, Z., Qian, H., Yang, C., Wang, Z., Shang, M., Kumar, V., Tan, S., Ray, B., Bhatia, P., et~al.
\newblock Recode: Robustness evaluation of code generation models.
\newblock \emph{arXiv preprint arXiv:2212.10264}, 2022{\natexlab{a}}.

\bibitem[Wang et~al.(2022{\natexlab{b}})Wang, Zhou, Fried, and Neubig]{wang2022execution}
Wang, Z., Zhou, S., Fried, D., and Neubig, G.
\newblock Execution-based evaluation for open-domain code generation.
\newblock \emph{arXiv preprint arXiv:2212.10481}, 2022{\natexlab{b}}.

\bibitem[Watson et~al.(2020)Watson, Tufano, Moran, Bavota, and Poshyvanyk]{watson2020learning}
Watson, C., Tufano, M., Moran, K., Bavota, G., and Poshyvanyk, D.
\newblock On learning meaningful assert statements for unit test cases.
\newblock In \emph{Proceedings of the ACM/IEEE 42nd International Conference on Software Engineering}, pp.\  1398--1409, 2020.

\bibitem[Wei et~al.(2022)Wei, Wang, Schuurmans, Bosma, Xia, Chi, Le, Zhou, et~al.]{wei2022chain}
Wei, J., Wang, X., Schuurmans, D., Bosma, M., Xia, F., Chi, E., Le, Q.~V., Zhou, D., et~al.
\newblock Chain-of-thought prompting elicits reasoning in large language models.
\newblock \emph{Advances in Neural Information Processing Systems}, 35:\penalty0 24824--24837, 2022.

\bibitem[Wei et~al.(2023)Wei, Durrett, and Dillig]{wei2023typet5}
Wei, J., Durrett, G., and Dillig, I.
\newblock Typet5: Seq2seq type inference using static analysis.
\newblock \emph{arXiv preprint arXiv:2303.09564}, 2023.

\bibitem[Wu et~al.(2023)Wu, Qiu, Ross, Aky{\"u}rek, Chen, Wang, Kim, Andreas, and Kim]{wu2023reasoning}
Wu, Z., Qiu, L., Ross, A., Aky{\"u}rek, E., Chen, B., Wang, B., Kim, N., Andreas, J., and Kim, Y.
\newblock Reasoning or reciting? exploring the capabilities and limitations of language models through counterfactual tasks.
\newblock \emph{arXiv preprint arXiv:2307.02477}, 2023.

\bibitem[Xia et~al.(2022)Xia, Wei, and Zhang]{xia2022practical}
Xia, C.~S., Wei, Y., and Zhang, L.
\newblock Practical program repair in the era of large pre-trained language models.
\newblock \emph{arXiv preprint arXiv:2210.14179}, 2022.

\bibitem[Xu et~al.(2022)Xu, Alon, Neubig, and Hellendoorn]{xu2022systematic}
Xu, F.~F., Alon, U., Neubig, G., and Hellendoorn, V.~J.
\newblock A systematic evaluation of large language models of code.
\newblock In \emph{Proceedings of the 6th ACM SIGPLAN International Symposium on Machine Programming}, pp.\  1--10, 2022.

\bibitem[Yin et~al.(2022)Yin, Li, Xiao, Rao, Wen, Shi, Howland, Bailey, Catasta, Michalewski, et~al.]{yin2022natural}
Yin, P., Li, W.-D., Xiao, K., Rao, A., Wen, Y., Shi, K., Howland, J., Bailey, P., Catasta, M., Michalewski, H., et~al.
\newblock Natural language to code generation in interactive data science notebooks.
\newblock \emph{arXiv preprint arXiv:2212.09248}, 2022.

\bibitem[Zan et~al.(2023)Zan, Chen, Zhang, Lu, Wu, Guan, Yongji, and Lou]{zan2023large}
Zan, D., Chen, B., Zhang, F., Lu, D., Wu, B., Guan, B., Yongji, W., and Lou, J.-G.
\newblock Large language models meet nl2code: A survey.
\newblock In \emph{Proceedings of the 61st Annual Meeting of the Association for Computational Linguistics (Volume 1: Long Papers)}, pp.\  7443--7464, 2023.

\bibitem[Zhang et~al.(2023{\natexlab{a}})Zhang, Chen, Zhang, Liu, Zan, Mao, Lou, and Chen]{zhang2023repocoder}
Zhang, F., Chen, B., Zhang, Y., Liu, J., Zan, D., Mao, Y., Lou, J.-G., and Chen, W.
\newblock Repocoder: Repository-level code completion through iterative retrieval and generation.
\newblock \emph{arXiv preprint arXiv:2303.12570}, 2023{\natexlab{a}}.

\bibitem[Zhang et~al.(2022)Zhang, Li, Meng, Chang, and Broeck]{zhang2022paradox}
Zhang, H., Li, L.~H., Meng, T., Chang, K.-W., and Broeck, G. V.~d.
\newblock On the paradox of learning to reason from data.
\newblock \emph{arXiv preprint arXiv:2205.11502}, 2022.

\bibitem[Zhang et~al.(2023{\natexlab{b}})Zhang, Li, Li, Li, and Jin]{zhang2023toolcoder}
Zhang, K., Li, G., Li, J., Li, Z., and Jin, Z.
\newblock Toolcoder: Teach code generation models to use apis with search tools.
\newblock \emph{arXiv preprint arXiv:2305.04032}, 2023{\natexlab{b}}.

\bibitem[Zhang et~al.(2023{\natexlab{c}})Zhang, Li, Li, Li, and Jin]{zhang2023self}
Zhang, K., Li, Z., Li, J., Li, G., and Jin, Z.
\newblock Self-edit: Fault-aware code editor for code generation.
\newblock In \emph{Proceedings of the 61st Annual Meeting of the Association for Computational Linguistics (Volume 1: Long Papers)}, pp.\  769--787, Toronto, Canada, July 2023{\natexlab{c}}. Association for Computational Linguistics.

\bibitem[Zhang et~al.(2023{\natexlab{d}})Zhang, Wang, Xia, Wang, and Li]{zhang2023algo}
Zhang, K., Wang, D., Xia, J., Wang, W.~Y., and Li, L.
\newblock Algo: Synthesizing algorithmic programs with generated oracle verifiers.
\newblock \emph{arXiv preprint arXiv:2305.14591}, 2023{\natexlab{d}}.

\bibitem[Zhang et~al.(2023{\natexlab{e}})Zhang, Chen, Shen, Ding, Tenenbaum, and Gan]{zhang2023planning}
Zhang, S., Chen, Z., Shen, Y., Ding, M., Tenenbaum, J.~B., and Gan, C.
\newblock Planning with large language models for code generation.
\newblock \emph{arXiv preprint arXiv:2303.05510}, 2023{\natexlab{e}}.

\bibitem[Zhang et~al.(2023{\natexlab{f}})Zhang, Tigges, Biderman, Raginsky, and Ringer]{zhang2023can}
Zhang, S.~D., Tigges, C., Biderman, S., Raginsky, M., and Ringer, T.
\newblock Can transformers learn to solve problems recursively?
\newblock \emph{arXiv preprint arXiv:2305.14699}, 2023{\natexlab{f}}.

\bibitem[Zhang et~al.(2023{\natexlab{g}})Zhang, Yu, Hashimoto, Lewis, Yih, Fried, and Wang]{zhang2023coder}
Zhang, T., Yu, T., Hashimoto, T., Lewis, M., Yih, W.-t., Fried, D., and Wang, S.
\newblock Coder reviewer reranking for code generation.
\newblock In \emph{International Conference on Machine Learning}, pp.\  41832--41846. PMLR, 2023{\natexlab{g}}.

\bibitem[Zhang et~al.(2023{\natexlab{h}})Zhang, Chen, Liu, Liao, Gong, Yu, Li, and Wang]{zhang2023survey}
Zhang, Z., Chen, C., Liu, B., Liao, C., Gong, Z., Yu, H., Li, J., and Wang, R.
\newblock A survey on language models for code.
\newblock 2023{\natexlab{h}}.

\bibitem[Zheng et~al.(2023)Zheng, Xia, Zou, Dong, Wang, Xue, Wang, Shen, Wang, Li, et~al.]{zheng2023codegeex}
Zheng, Q., Xia, X., Zou, X., Dong, Y., Wang, S., Xue, Y., Wang, Z., Shen, L., Wang, A., Li, Y., et~al.
\newblock Codegeex: A pre-trained model for code generation with multilingual evaluations on humaneval-x.
\newblock \emph{arXiv preprint arXiv:2303.17568}, 2023.

\bibitem[Zhong et~al.(2022)Zhong, Liu, Li, Kuang, Zeng, and Wang]{zhong2022codegen}
Zhong, M., Liu, G., Li, H., Kuang, J., Zeng, J., and Wang, M.
\newblock Codegen-test: An automatic code generation model integrating program test information.
\newblock \emph{arXiv preprint arXiv:2202.07612}, 2022.

\bibitem[Zhou et~al.(2023)Zhou, Bradley, Littwin, Razin, Saremi, Susskind, Bengio, and Nakkiran]{zhou2023algorithms}
Zhou, H., Bradley, A., Littwin, E., Razin, N., Saremi, O., Susskind, J., Bengio, S., and Nakkiran, P.
\newblock What algorithms can transformers learn? a study in length generalization.
\newblock \emph{arXiv preprint arXiv:2310.16028}, 2023.

\bibitem[Zhu et~al.(2022)Zhu, Jain, Suresh, Ravindran, Tipirneni, and Reddy]{zhu2022xlcost}
Zhu, M., Jain, A., Suresh, K., Ravindran, R., Tipirneni, S., and Reddy, C.~K.
\newblock Xlcost: A benchmark dataset for cross-lingual code intelligence.
\newblock \emph{arXiv preprint arXiv:2206.08474}, 2022.

\end{thebibliography}
\bibliographystyle{icml2021} 

\begin{appendices}
\section{Benchmark Construction and Statistics} \label{appendix:construction}
In this section, we provide more details about the process of constructing our benchmark. A sample prompt for generating functions and test cases is shown in Listing \ref{lst:benchmark-generation-prompt}. The prompt is constructed by including two few-shot examples, one containing a specified \texttt{str} function and one containing a specified \texttt{list} function. The full list of specified functions is given in \ref{appendix:benchmark-generation-functions}, and the full list of few-shot examples chosen from is given in \ref{appendix:benchmark-generation-fewshot}. We learned that having random-looking inputs instead of common words and phrases in the few-shot prompts significantly increased the difficulty of the benchmark.

\begin{lstlisting}[caption={Sample prompt for generating functions and test cases},label={lst:benchmark-generation-prompt}, captionpos=t, breaklines=true]
You will be given a function name between [TASK] and [/TASK] tags. Following the examples given, write a Python function that makes use of the given function and 5 test inputs for that function.

[TASK]
str.center
[/TASK]
[PYTHON]
def f(text):
    ls = list(text)
    for i in range(1, len(ls) - 1):
        ls.insert(i, '+')
    return ''.join(ls).center((len(ls) - 1) * 2)
[/PYTHON]
[TEST]
assert f('lynel') == ??
assert f('nzoh') == ??
assert f('u') == ??
assert f('anfsoixz') == ??
assert f('xzd') == ??
[/TEST]

[TASK]
list.append
[/TASK]
[PYTHON]
def f(nums):
    count = len(nums)
    for i in range(-count+1, 0):
        nums.append(nums[i])
    return nums
[/PYTHON]
[TEST]
assert f([2, 6, 1, 3, 1]) == ??
assert f([7, 1, 2, 6, 0, 2]) == ??
assert f([4, 3, 2, 1, 2, -1, 4, 2]) == ??
assert f([0, 6, 2, -1, -2]) == ??
assert f([-6, -2, 1, -3, 0, 1]) == ??
[/TEST]

[TASK]
str.zfill
[/TASK]
[PYTHON]
\end{lstlisting}

\subsection{Functions used in prompt} \label{appendix:benchmark-generation-functions}
For each of \texttt{str}, \texttt{list}, and \texttt{dict}, we use all the non-dunder methods under that class. The resulting list of methods is as follows:
\begin{itemize}
    \item \texttt{str}: capitalize, casefold, center, count, encode, endswith, expandtabs, find, format, format\_map, index, isalnum, isalpha, isascii, isdecimal, isdigit, isidentifier, islower, isnumeric, isprintable, isspace, istitle, isupper, join, ljust, lower, lstrip, maketrans, partition, removeprefix, removesuffix, replace, rfind, rindex, rjust, rpartition, rsplit, rstrip, split, splitlines, startswith, strip, swapcase, title, translate, upper, zfill
    \item \texttt{list}: append, clear, copy, count, extend, index, insert, pop, remove, reverse, sort
    \item \texttt{dict}: clear, copy, fromkeys, get, items, keys, pop, popitem, setdefault, update, values
\end{itemize}

Motivated by seeing a GPT-4 failure of treating the \texttt{\^} symbol as an exponential rather than an xor, we also attempted using all the non-dunder methods from \texttt{operator}. However, we found that the majority of the functions obtained were either too simple and uninteresting, or too computational, since many of the methods under \texttt{operator} are bit-manipulation or calculational operations. Therefore, we excluded it from our final benchmark.

\subsection{Few-shot Examples} \label{appendix:benchmark-generation-fewshot}
We use 10 handwritten few-shot examples, 5 using \texttt{str} functions and 5 using \texttt{list} functions. For each prompt, we include two few-shot examples, one string few-shot example and one list few-shot example, for a total of 25 different combinations of few-shot prompts. We generate programs and inputs using \codellamalarge with temperature $T=1$. 

One interesting observation is that for a fixed pair of few-shot examples, there seems to be a limit to the number of diverse functions that can be generated: after about $60000$ generations, only about $5000$ of them were unique. Using all 25 combinations of few-shot prompts helps to overcome this duplication bottleneck.

The full set of few-shot examples can be found in Listing \ref{lst:benchmark-generation-fewshot}.

\subsection{Dataset Statistics} \label{appendix:benchmark-dataset-statistics}

\label{appendix:benchmark-statistics}
\begin{figure}[H]
    \centering
    \includegraphics[width=0.9\textwidth]{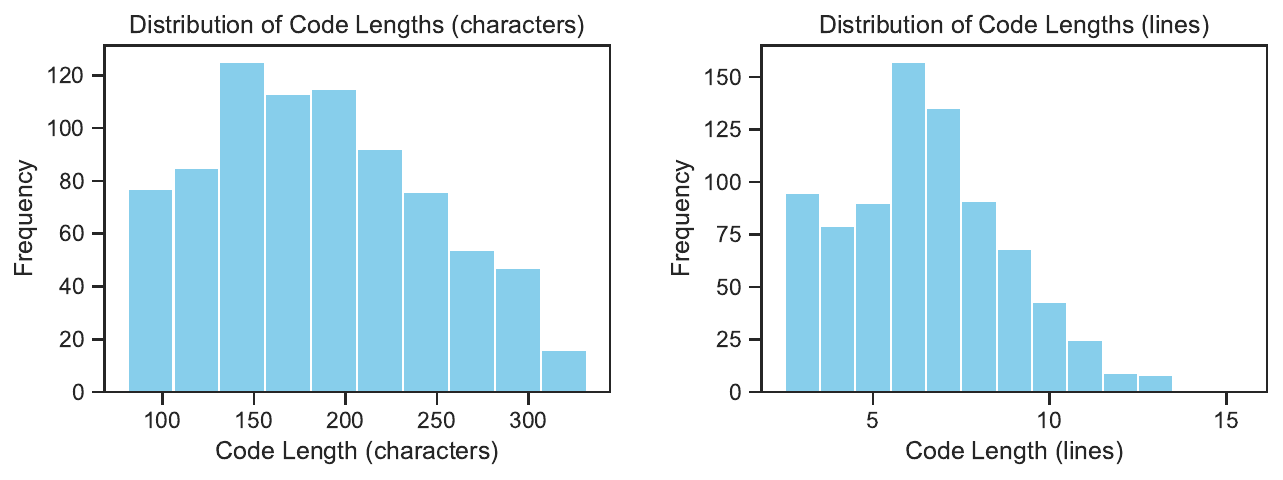}
    \caption{Dataset Distributions}
    \label{fig:dataset-distributions}
\end{figure}

In Fig. \ref{fig:dataset-distributions}, we show the distribution of character count and line count of the 800 samples in our benchmark.

\begin{figure}[H]
    \centering
    \includegraphics[width=0.5\textwidth]{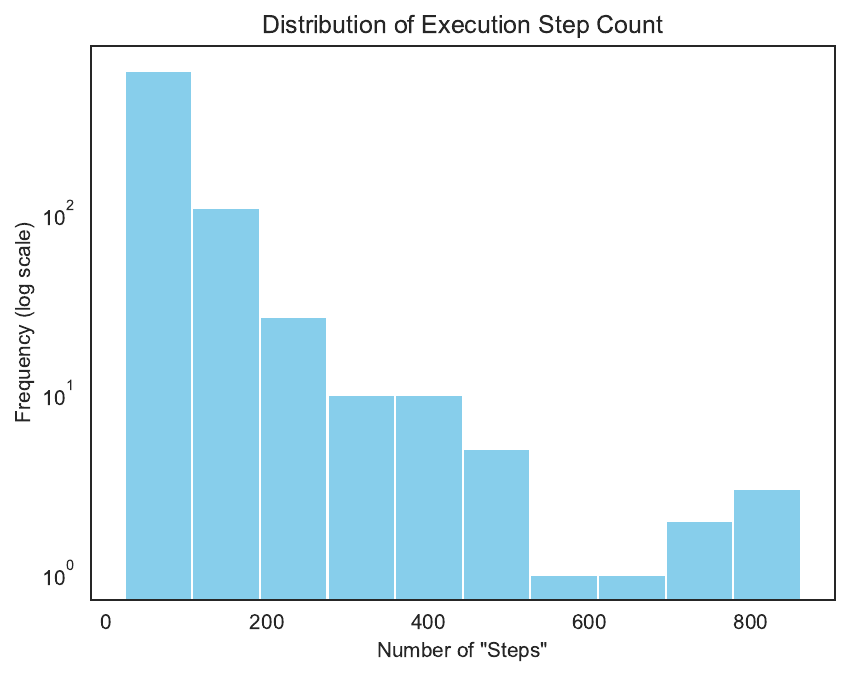}
    \caption{Number of steps}
    \label{fig:num-steps}
\end{figure}

In Fig. \ref{fig:num-steps}, we show the distribution of the ``step count'' of programs (with one outlier of 3175 steps excluded). Here, steps roughly correspond to Python bytecode operations during execution. The precise definition can be understood by checking the ``numsteps'' variable in our code \href{https://github.com/minimario/cruxeval/blob/main/data/filter/analyze_ops.py}{here}.

\begin{figure}[H]
    \centering
    \begin{subfigure}[b]{0.49\textwidth}
        \includegraphics[width=\textwidth]{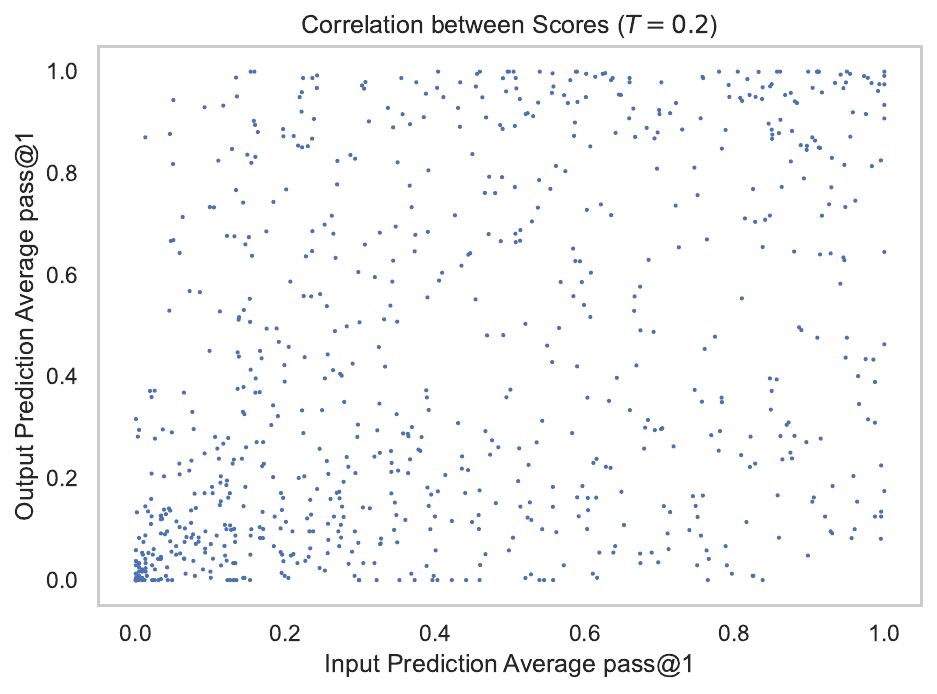}
        \caption{$T=0.2$}
    \end{subfigure}
    \begin{subfigure}[b]{0.49\textwidth}
        \includegraphics[width=\textwidth]{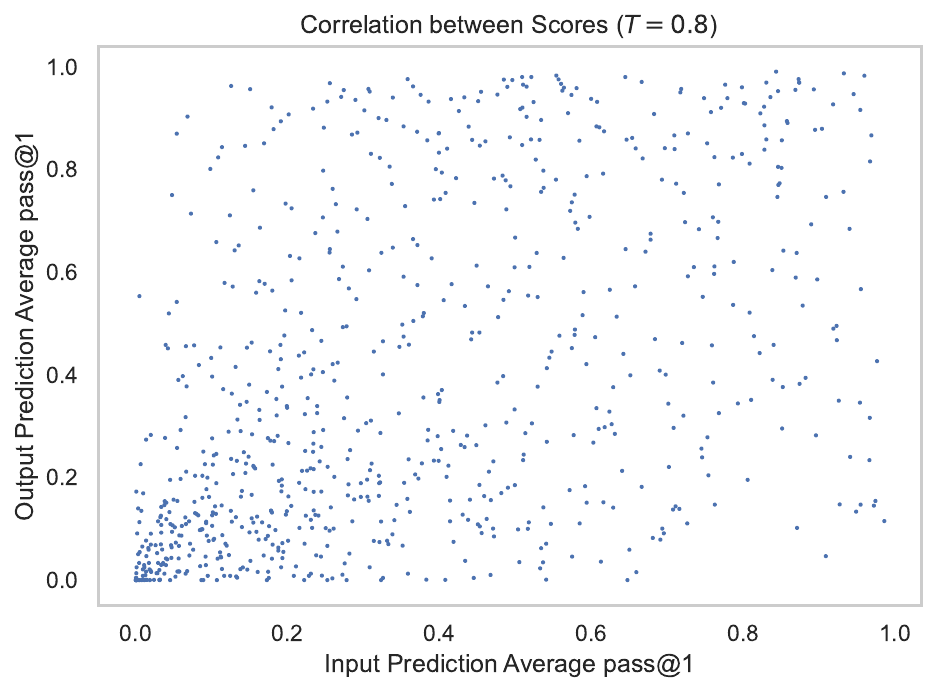}
        \caption{$T=0.8$}
    \end{subfigure}
    \caption{Sample-by-sample correlation between Input Prediction and Output Prediction}
    \label{fig:benchmark-sample-io-correlation}
\end{figure}

In Fig. \ref{fig:benchmark-sample-io-correlation}, we plot the output prediction pass@1 scores against the input prediction pass@1 scores for each sample, observing little to no correlation between the difficulty of the two.

\begin{figure}[H]
    \centering
    \includegraphics[width=\textwidth]{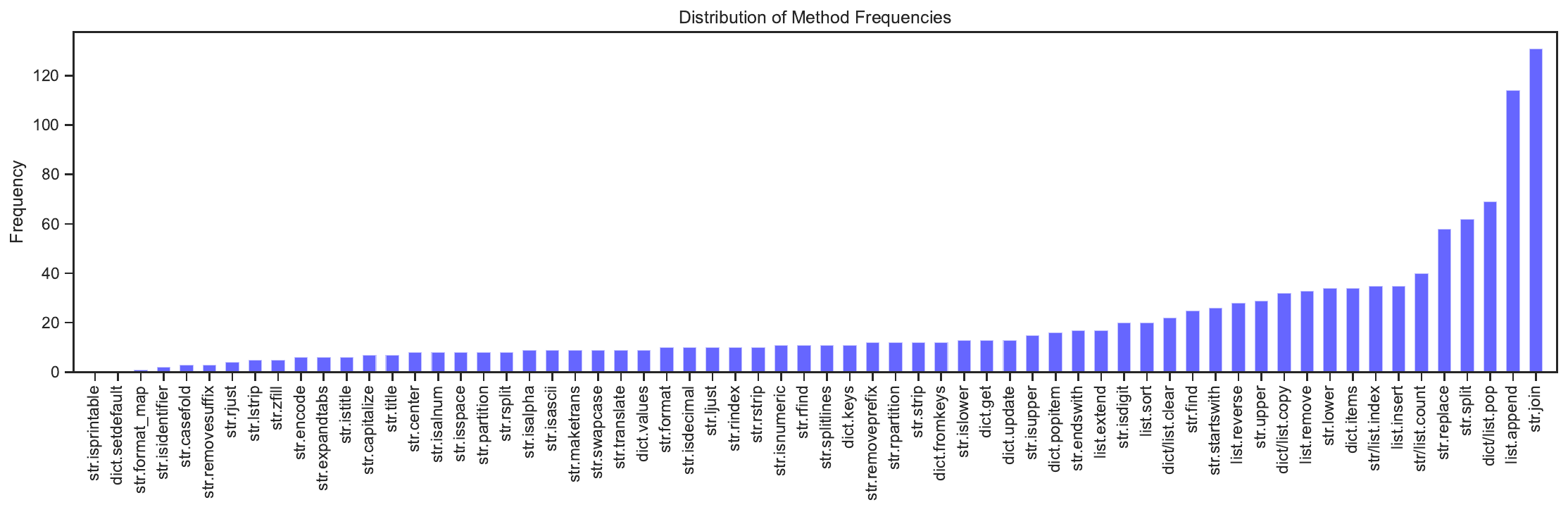}
    \caption{Frequency of various methods in \benchmark}
    \label{fig:benchmark-method-statistics}
\end{figure}

\textbf{Method-level statistics}: In Fig. \ref{fig:benchmark-method-statistics}, we show the number of samples containing each method in \texttt{str, list}, and \texttt{dict}. Even though we distilled the same number of samples using each \texttt{str} function and about twice as many for each \texttt{list} and \texttt{dict} functions, we observe that the resulting distribution is highly non-uniform. This is due to a few reasons. First, about 30\% of the time, \codellamalarge sometimes fails to follow the instruction of including the library method in the resulting function. Second, some functions naturally lead to more operations that are removed by the filter. Third, common functions such as \texttt{str/list.index} or \texttt{list.append} are used in methods they are not prompted in. 

\begin{figure}[H]
    \centering
    \begin{subfigure}[b]{0.49\textwidth}
        \includegraphics[width=\textwidth]{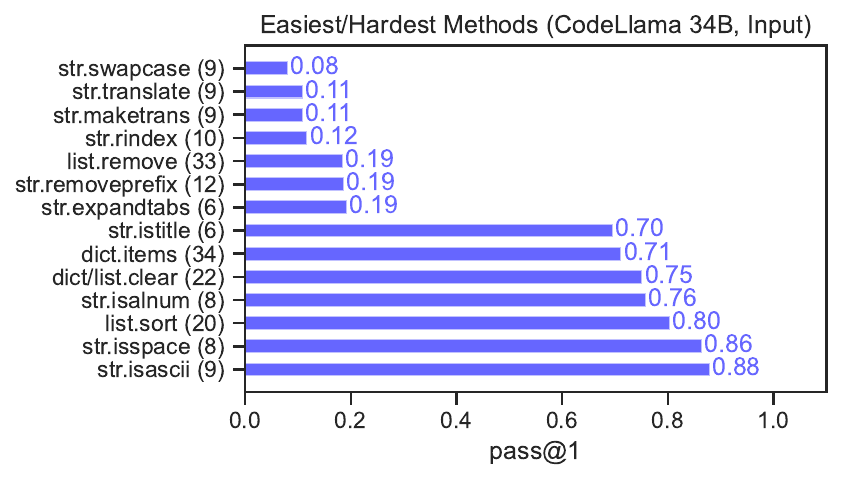}
        \caption{Input Prediction}
    \end{subfigure}
    \begin{subfigure}[b]{0.49\textwidth}
        \includegraphics[width=\textwidth]{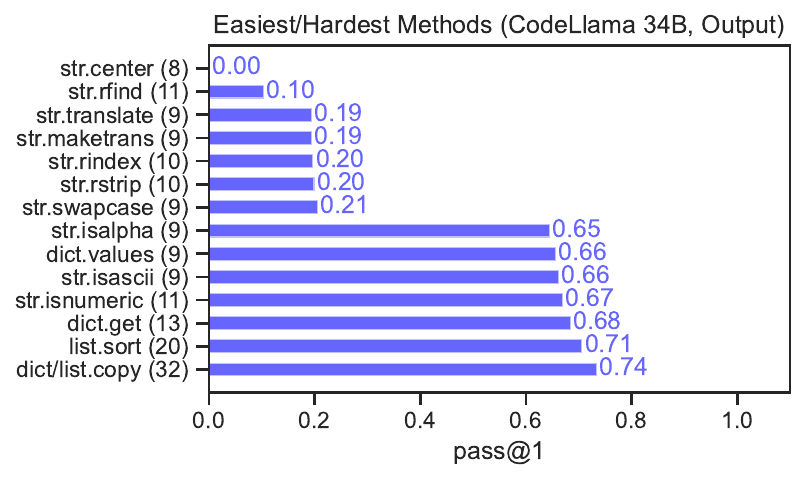}
        \caption{Output Prediction}
    \end{subfigure}
    \caption{Easiest and hardest methods for Code Llama 34B input and output prediction, by pass@1 score ($T=0.2$)}
    \label{fig:dataset-difficult-cl}
\end{figure}

\begin{figure}[H]
    \centering
    \begin{subfigure}[b]{0.49\textwidth}
        \includegraphics[width=\textwidth]{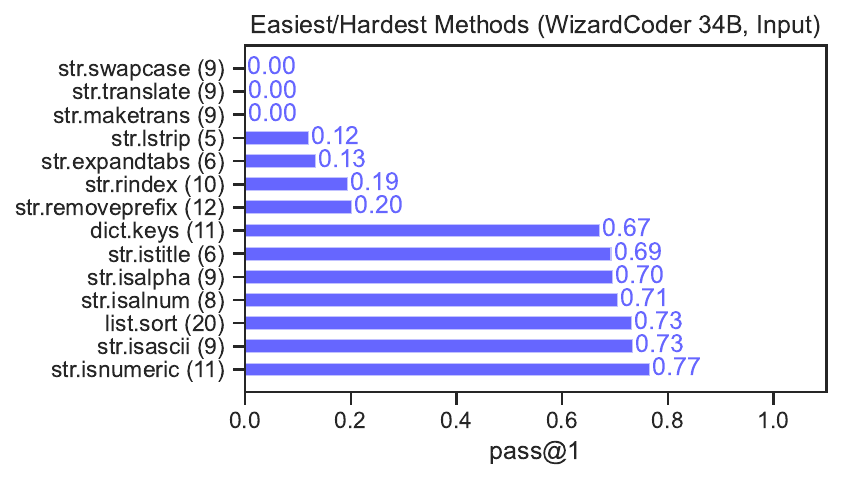}
        \caption{Input Prediction}
    \end{subfigure}
    \begin{subfigure}[b]{0.49\textwidth}
        \includegraphics[width=\textwidth]{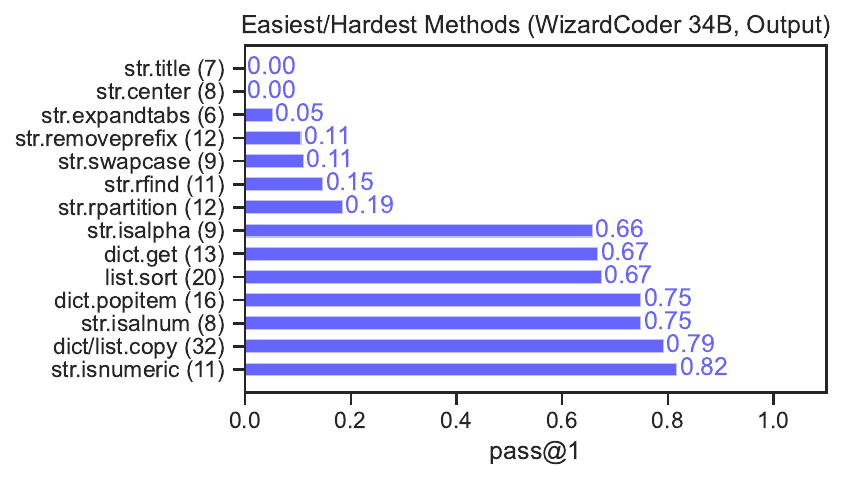}
        \caption{Output Prediction}
    \end{subfigure}
    \caption{Easiest and hardest methods for WizardCoder 34B input and output prediction, by pass@1 score ($T=0.2$)}
    \label{fig:dataset-difficult-wizardcoder}
\end{figure}

\begin{figure}[H]
    \centering
    \begin{subfigure}[b]{0.49\textwidth}
        \includegraphics[width=\textwidth]{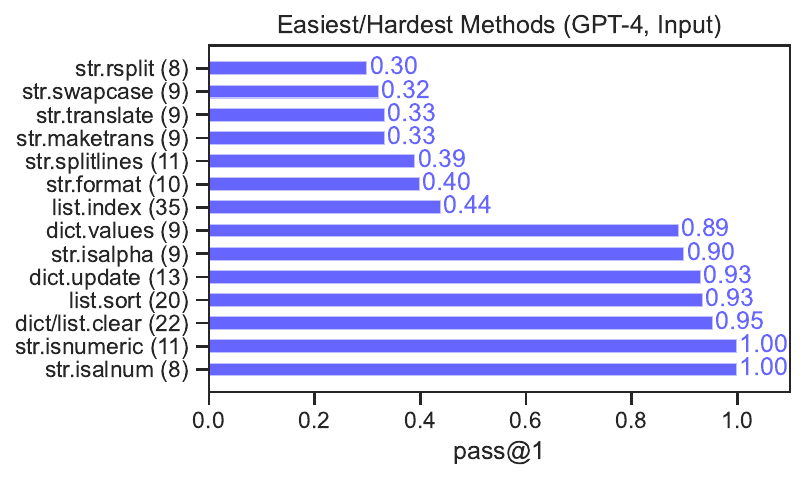}
        \caption{Input Prediction}
    \end{subfigure}
    \begin{subfigure}[b]{0.49\textwidth}
        \includegraphics[width=\textwidth]{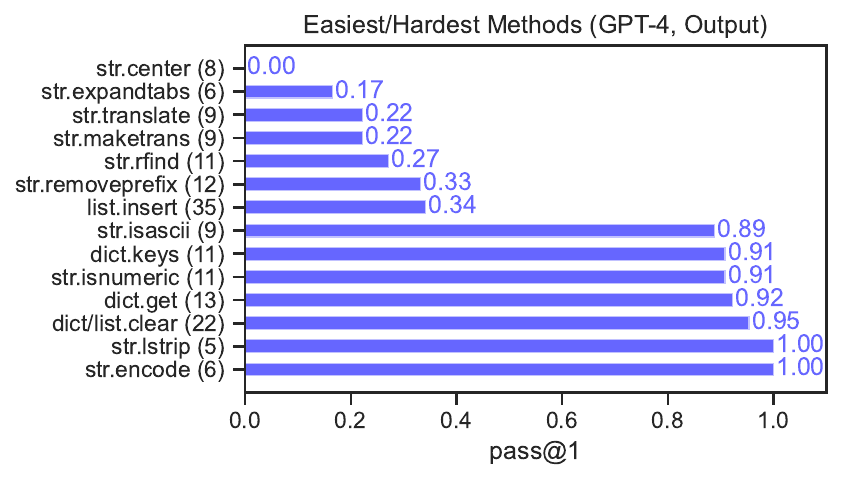}
        \caption{Output Prediction}
    \end{subfigure}
    \caption{Easiest and hardest methods for Code Llama 34B input and output prediction, by pass@1 score ($T=0.2$)}
    \label{fig:dataset-difficult-gpt4}
\end{figure}

Next, we investigate trends of which methods are easier/harder for code LMs. For each method in the \texttt{list}, \texttt{str}, and \texttt{dict} libraries listed in Appendix \ref{appendix:benchmark-generation-functions} with at least 5 samples, we calculate the average input prediction and output prediction score of benchmark samples containing that method. We show the 7 easiest and hardest methods for Code Llama 34B (Fig. \ref{fig:dataset-difficult-cl}), WizardCoder 34B (Fig. \ref{fig:dataset-difficult-wizardcoder}), and GPT-4 (Fig. \ref{fig:dataset-difficult-gpt4}). Some of the hardest methods, such as \texttt{str.rsplit, str.maketrans, str.rfind}, seem to be more obscure. We hypothesize that they may be underrepresented in pretraining corpora, explaining models' worse performance. While the distilled datasets of models like Phi, Phind, and WizardCoder are not yet available to the public, we speculate that they may include of fewer instances of these underrepresented functions and that distilling more obscure methods may help the model better learn their syntax and semantics.

\section{Model URLs}\label{sec:appendix-models}
For evaluation, we used the \texttt{gpt-3.5-turbo} and \texttt{gpt-4} models on October 26, 2023. Note that this is before the OpenAI developer day release of GPT-4-Turbo. The corresponding HuggingFace model URLs for the rest of the evaluated models are listed in Table \ref{tab:model-url-list}.

\begin{table}[h]
    \centering
    \small
    \caption{Models and HuggingFace URLs}
    \begin{tabular}{|ll|}
        \hline
        \textbf{Model Name} & \textbf{HuggingFace URL} \\
        \hline
        Mistral (7B) & \url{https://huggingface.co/mistralai/Mistral-7B-v0.1} \\
        Phi-1 (1.3B) & \url{https://huggingface.co/microsoft/phi-1} \\
        Phi-1.5 (1.3B) & \url{https://huggingface.co/microsoft/phi-1_5} \\
        DeepSeek Instruct (6.7B) & \url{https://huggingface.co/deepseek-ai/deepseek-coder-6.7b-instruct} \\
        DeepSeek Instruct (33B) & \url{https://huggingface.co/deepseek-ai/deepseek-coder-33b-instruct} \\
        DeepSeek Base (6.7B) & \url{https://huggingface.co/deepseek-ai/deepseek-coder-6.7b-base} \\
        DeepSeek Base (33B) & \url{https://huggingface.co/deepseek-ai/deepseek-coder-33b-base} \\
        StarCoderBase (15.5B) & \url{https://huggingface.co/bigcode/starcoderbase} \\
        StarCoderBase (7B) & \url{https://huggingface.co/bigcode/starcoderbase-7b} \\
        WizardCoder (13B) & \url{https://huggingface.co/WizardLM/WizardCoder-Python-13B-V1.0} \\
        WizardCoder (34B) & \url{https://huggingface.co/WizardLM/WizardCoder-Python-34B-V1.0} \\
        Phind (34B) & \url{https://huggingface.co/Phind/Phind-CodeLlama-34B-v2} \\
        CodeLlama 7B & \url{https://huggingface.co/codellama/CodeLlama-7b-hf} \\
        CodeLlama (13B) & \url{https://huggingface.co/codellama/CodeLlama-13b-hf} \\
        CodeLlama (34B) & \url{https://huggingface.co/codellama/CodeLlama-34b-hf} \\
        CodeLlama Python (7B) & \url{https://huggingface.co/codellama/CodeLlama-7b-Python-hf} \\
        CodeLlama Python (13B) & \url{https://huggingface.co/codellama/CodeLlama-13b-Python-hf} \\
        CodeLlama Python (34B) & \url{https://huggingface.co/codellama/CodeLlama-34b-Python-hf} \\
        \hline
    \end{tabular}
    \label{tab:model-url-list}
\end{table}

\section{Evaluation Results} \label{sec:appendix-quantitative}

\subsection{Main Results}
\begin{figure}[H]
     \centering
     \begin{subfigure}[b]{0.49\textwidth}
         \centering
         \includegraphics[scale=0.4]{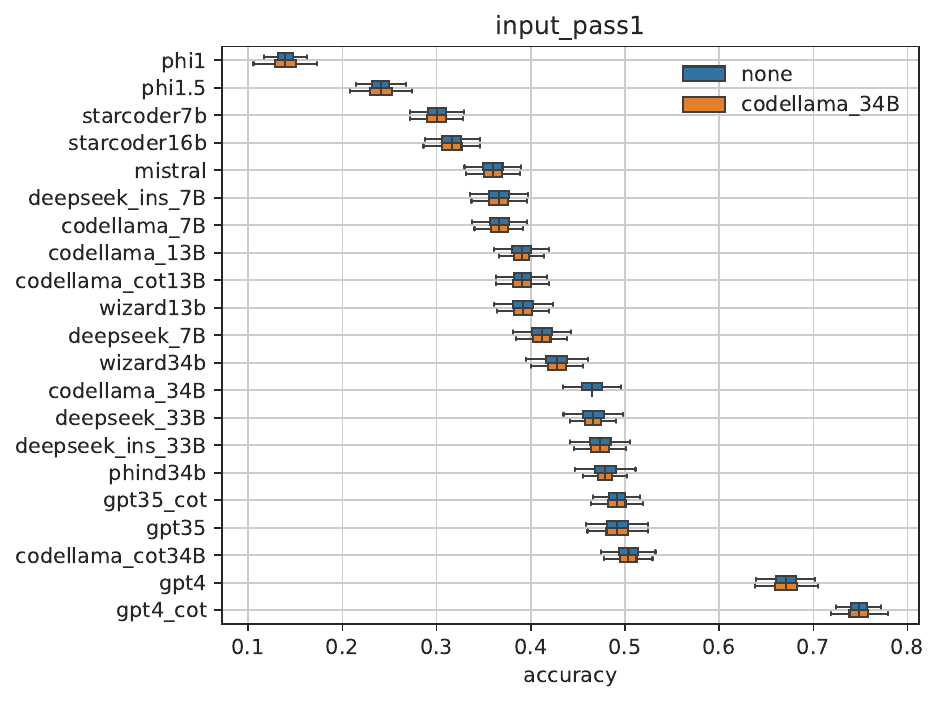}
         \caption{Main Results, pass@1 (Input)}
         \label{fig:main-results-pass1-input-appendix}
     \end{subfigure}
     \hfill
     \begin{subfigure}[b]{0.49\textwidth}
         \centering
         \includegraphics[scale=0.4]{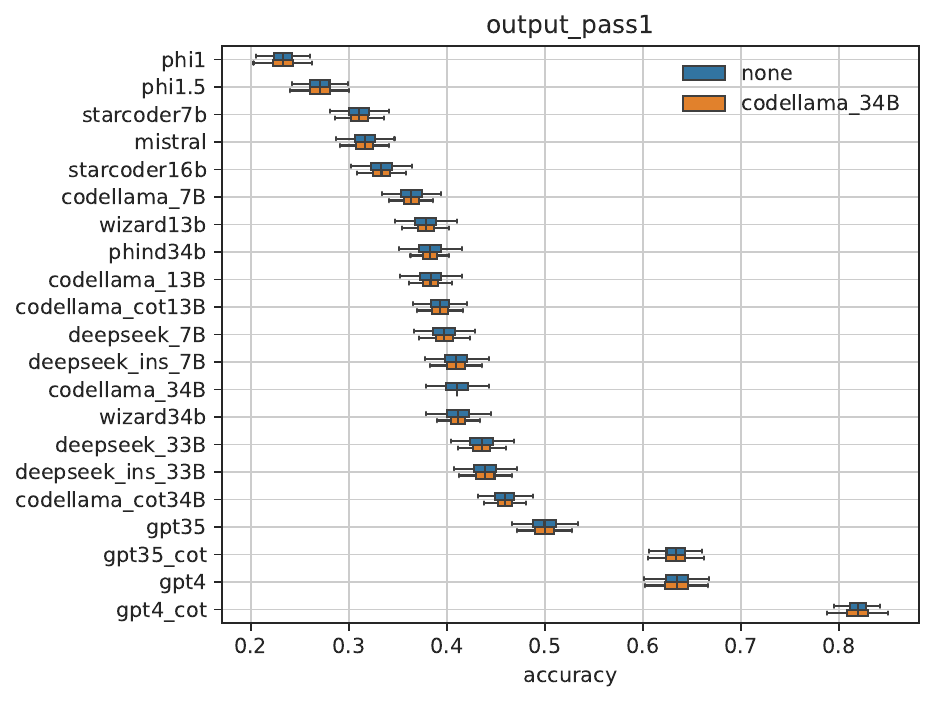}
         \caption{Main Results, pass@1 (Output)}
         \label{fig:main-results-pass1-output-appendix}
     \end{subfigure}
     \newline
     \newline
     \begin{subfigure}[b]{0.49\textwidth}
         \centering
         \includegraphics[scale=0.4]{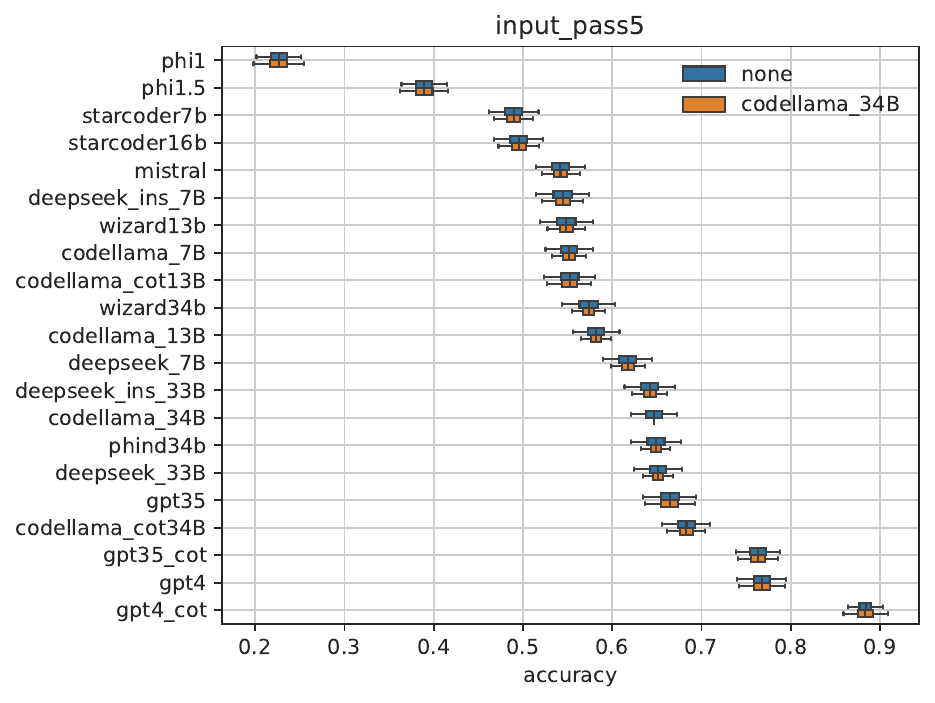}
         \caption{Main Results, pass@5 (Input)}
         \label{fig:main-results-pass5-input-appendix}
     \end{subfigure}
     \hfill
     \begin{subfigure}[b]{0.49\textwidth}
         \centering
         \includegraphics[scale=0.4]{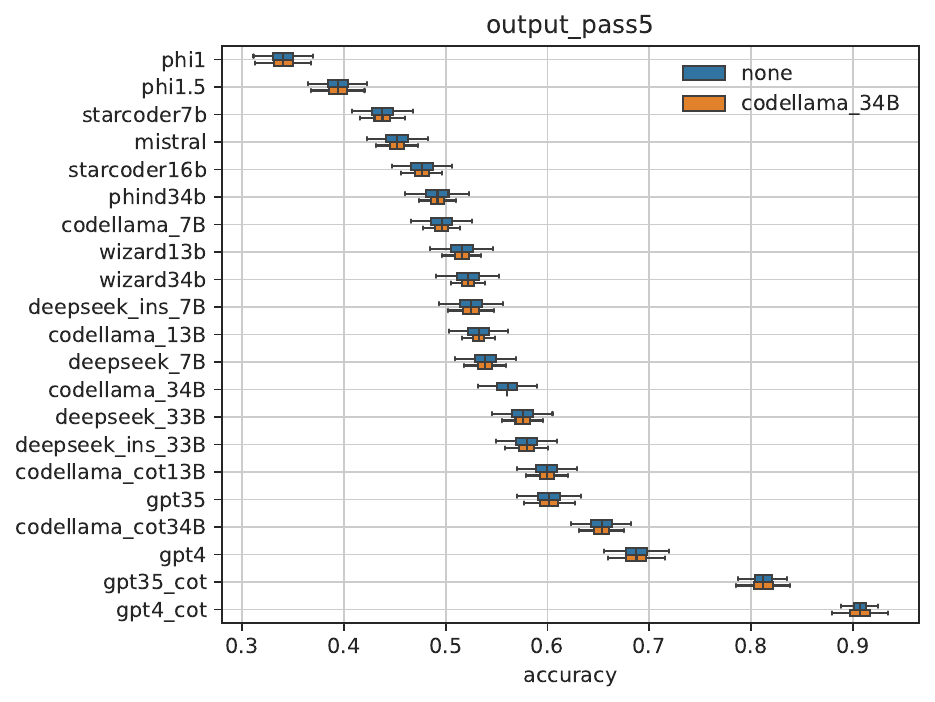}
         \caption{Main Results, pass@5 (Output)}
         \label{fig:main-results-pass5-output-appendix}
     \end{subfigure}
     \caption{Main Results with confidence intervals compared to codellama\_34B.}
     \label{fig:main-results-all-appendix}
\end{figure}

Table \ref{tab:benchmark-results} shows the pass@1 and pass@5 results of all evaluated models on \benchmark, and Fig. \ref{fig:main-results-all-appendix} shows them in box-plot form.

\begin{table}[htbp]
    \centering
    \caption{Results of all models on \benchmark}
    \begin{tabular}{cccccc}
        \toprule
        \multirow{2}{*}{\textbf{Model}} & \multirow{2}{*}{\textbf{Size}} & \multicolumn{2}{c}{\textbf{Input Prediction}} & \multicolumn{2}{c}{\textbf{Output Prediction}} \\
        \cmidrule(lr){3-4} \cmidrule(lr){5-6}
        & & \textbf{Pass@1} & \textbf{Pass@5} & \textbf{Pass@1} & \textbf{Pass@5} \\
	\midrule
	\multirow{3}{*}{CodeLlama}
	& 7B & 36.6\% & 55.2\% & 36.4\% & 49.6\% \\
	& 13B & 39.0\% & 58.2\% & 38.4\% & 53.2\% \\
	& 34B & 46.5\% & 64.7\% & 41.1\% & 56.1\% \\
	\midrule
	\multirow{3}{*}{CodeLlama Python}
	& 7B & 36.3\% & 56.0\% & 36.4\% & 49.7\% \\
	& 13B & 40.5\% & 58.0\% & 37.8\% & 50.8\% \\
	& 34B & 41.5\% & 59.2\% & 40.7\% & 53.7\% \\
	\midrule
	\multirow{2}{*}{StarCoder-Base}
	& 7B & 30.0\% & 48.9\% & 31.1\% & 43.8\% \\
	& 15.5B & 31.6\% & 49.5\% & 33.3\% & 47.7\% \\
	\midrule
	\multirow{2}{*}{WizardCoder}
	& 13B & 39.2\% & 54.8\% & 37.9\% & 51.6\% \\
	& 34B & 42.8\% & 57.3\% & 41.2\% & 52.2\% \\
	\midrule
	Phi-1
	& 1.3B & 13.9\% & 22.6\% & 23.3\% & 34.0\% \\
	\midrule
	Phi-1.5
	& 1.3B & 24.1\% & 38.9\% & 27.1\% & 39.4\% \\
	\midrule
	Phind v2
	& 34B & 47.9\% & 64.9\% & 38.3\% & 49.2\% \\
	\midrule
	\multirow{2}{*}{Deepseek Coder-Base}
	& 6.7B & 41.1\% & 61.7\% & 39.8\% & 53.9\% \\
	& 33B & 46.6\% & 65.1\% & 43.6\% & 57.6\% \\
	\midrule
	\multirow{2}{*}{Deepseek Coder-Instruct}
	& 6.7B & 36.6\% & 54.4\% & 41.0\% & 52.5\% \\
	& 33B & 47.4\% & 64.2\% & 44.0\% & 58.0\% \\
	\midrule
	Mistral
	& 7B & 36.0\% & 54.2\% & 31.7\% & 45.2\% \\
	\midrule
	GPT-3.5
	& - & 49.2\% & 66.5\% & 50.0\% & 60.1\% \\
	\midrule
	GPT-4
	& - & 67.1\% & 76.8\% & 63.4\% & 68.7\% \\
    \bottomrule
    \end{tabular}
    \label{tab:benchmark-results}
\end{table}

% \begin{figure}[H]
%      \centering
%      \begin{subfigure}[b]{0.49\textwidth}
%          \centering
%          \includegraphics[width=\textwidth]{figs/main_results/main_box_input_pass1.pdf}
%          \caption{Performance, pass@1 (Input)}
%          \label{fig:main-results-pass1-input}
%      \end{subfigure}
%      \hfill
%      \begin{subfigure}[b]{0.49\textwidth}
%          \centering
%          \includegraphics[width=\textwidth]{figs/main_results/main_box_output_pass1.pdf}
%          \caption{Performance, pass@1 (Output)}
%          \label{fig:main-results-pass1-output}
%      \end{subfigure}
%      \caption{Boxes show (25, 75) percentiles and whiskers show (2.5, 97.5) independently and when all models are compared against \codellamalarge as the baseline where the difference between more pairs are significant. }
%      \label{fig:baseline_results}
% \end{figure}

\subsection{Additional Results on Impact of CoT} \label{subsec:appendix-cot}
Table \ref{tab:benchmark-results-cot} shows the results of including CoT on Code Llama 13B, 34B, GPT-3.5, and GPT-4.

\begin{table}[h]
    \centering
    \caption{Impact of CoT on \benchmark}
    \begin{tabular}{cccccc}
        \toprule
        \multirow{2}{*}{\textbf{Model}} & \multirow{2}{*}{\textbf{CoT}} & \multicolumn{2}{c}{\textbf{Input Prediction}} & \multicolumn{2}{c}{\textbf{Output Prediction}} \\
        \cmidrule(lr){3-4} \cmidrule(lr){5-6}
        & & \textbf{Pass@1} & \textbf{Pass@5} & \textbf{Pass@1} & \textbf{Pass@5} \\
	\midrule
	\multirow{3}{*}{Code Llama 13B}
	& \xmark & 39.0\% & 58.2\% & 38.4\% & 53.2\% \\
	& \cmark & 39.1\% & 55.2\% & 39.3\% & 59.9\% \\
	& - & \green{+0.1\%} & \red{-3.0\%} & \green{+0.9\%} & \green{+6.7\%} \\
	\midrule
	\multirow{3}{*}{Code Llama 34B}
	& \xmark & 46.5\% & 64.7\% & 41.1\% & 56.1\% \\
	& \cmark & 50.4\% & 68.3\% & 46.0\% & 65.3\% \\
	& - & \green{+3.9\%} & \green{+3.6\%} & \green{+4.9\%} & \green{+9.2\%} \\
	\midrule
	\multirow{3}{*}{GPT-3.5}
	& \xmark & 49.2\% & 66.5\% & 50.0\% & 60.1\% \\
	& \cmark & 49.1\% & 76.3\% & 63.3\% & 81.2\% \\
	& - & \red{-0.1\%} & \green{+9.8\%} & \green{+13.3\%} & \green{+21.1\%} \\
	\midrule
	\multirow{3}{*}{GPT-4}
	& \xmark & 67.1\% & 76.8\% & 63.4\% & 68.7\% \\
	& \cmark & 74.8\% & 88.4\% & 81.9\% & 90.7\% \\
	& - & \green{+7.7\%} & \green{+11.6\%} & \green{+18.5\%} & \green{+22.0\%} \\
        \bottomrule
    \end{tabular}
    \label{tab:benchmark-results-cot}
\end{table}

\begin{figure}[H]
     \centering
     \begin{subfigure}[t]{0.49\textwidth}
         \centering
         \includegraphics[width=\textwidth]{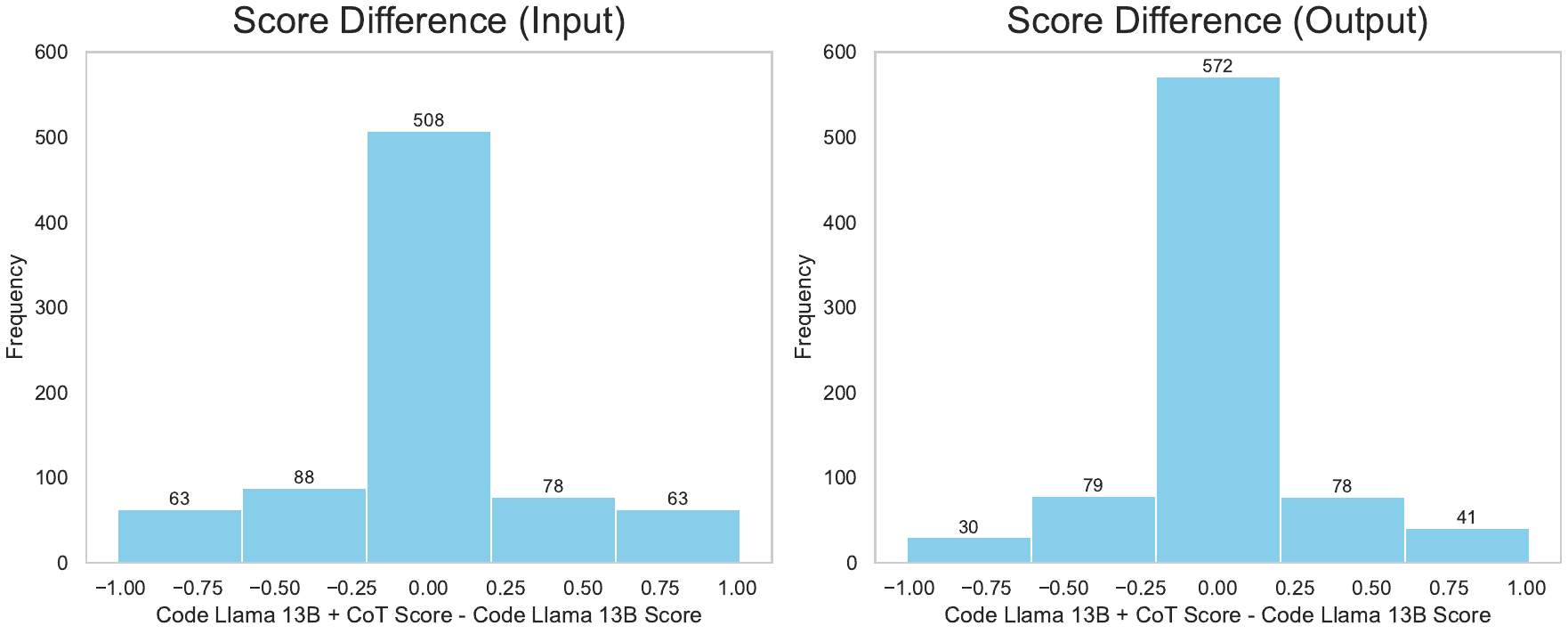}
         \caption{Code Llama 13B}
     \end{subfigure}%
     \hfill
     \begin{subfigure}[t]{0.49\textwidth}
         \centering
         \includegraphics[width=\textwidth]{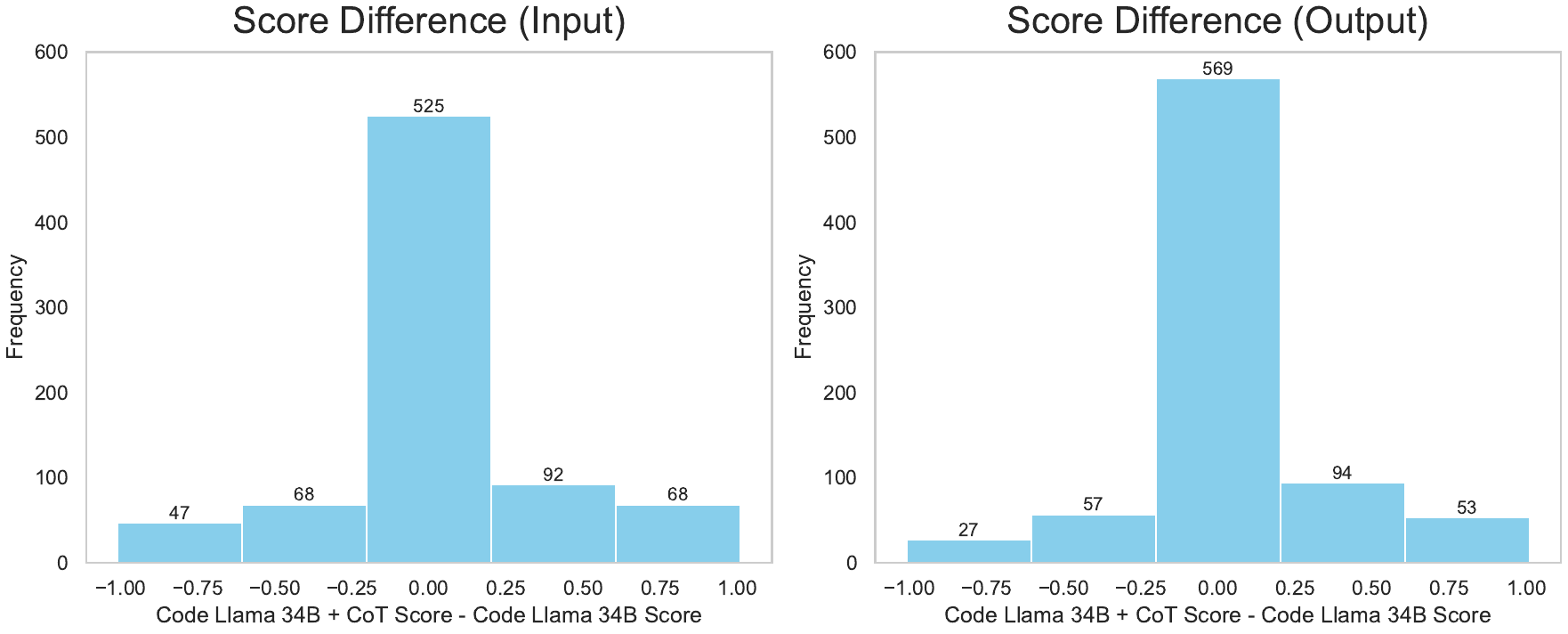}
         \caption{Code Llama 34B}
     \end{subfigure}
     \newline
     \newline
     \newline
     \begin{subfigure}[t]{0.49\textwidth}
         \centering
         \includegraphics[width=\textwidth]{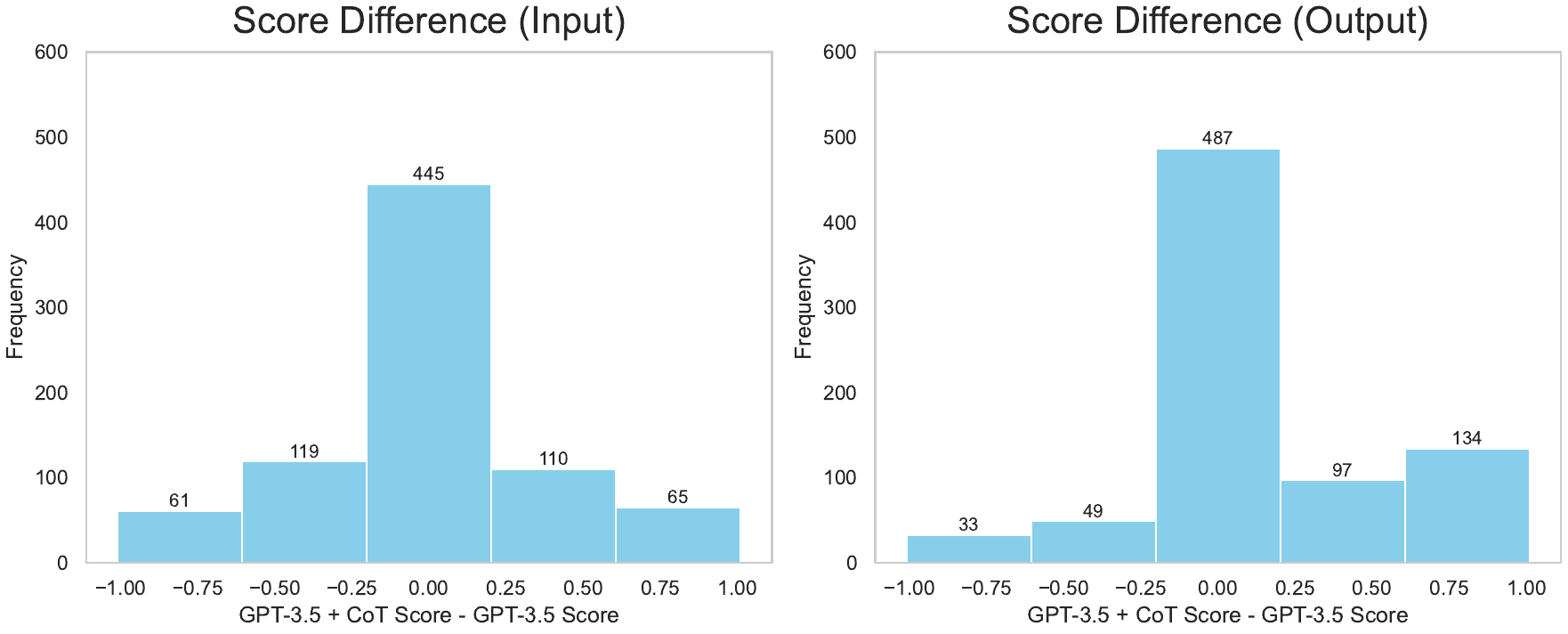}
         \caption{GPT-3.5}
     \end{subfigure}%
     \hfill
     \begin{subfigure}[t]{0.49\textwidth}
         \centering
         \includegraphics[width=\textwidth]{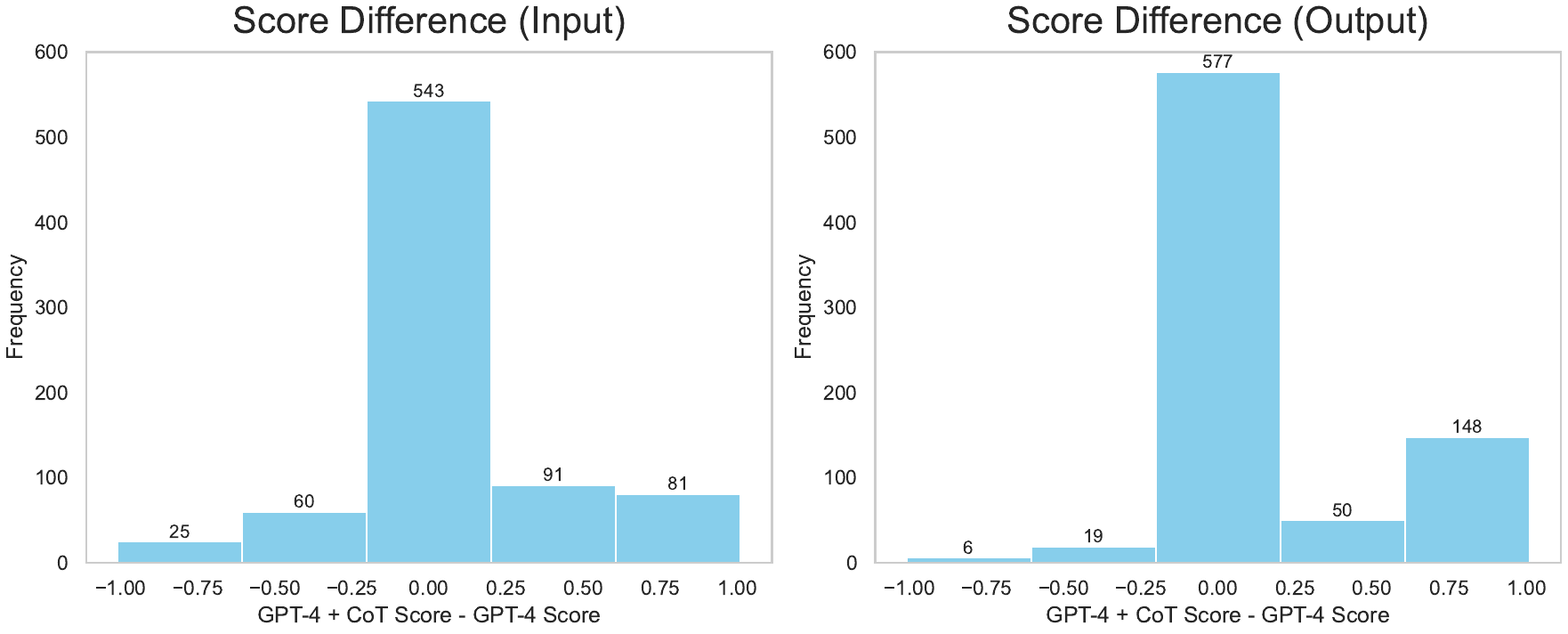}
         \caption{GPT-4}
         \label{fig:difference-hist-all-gpt4}
     \end{subfigure}
     \caption{Histogram of Score Differences between CoT and Original $(T=0.2)$}
     \label{fig:difference-hist-all}
\end{figure}

\textbf{Sample-wide improvements from CoT}: In Fig. \ref{fig:difference-hist-all}, we show a histogram of how much CoT improves the pass@1 score of each sample (negative values means that CoT decreased the accuracy). We observe that CoT leads to little improvement for the majority of samples, this effect is partly due to samples already having high pass@1 scores. As evidenced by Fig. \ref{fig:difference-hist-all-gpt4}, we see that CoT is much more effective for GPT-4 output prediction compared to both GPT-4 input prediction and other models. For the other models, however, we observe a large proportion of samples for which CoT actually decreases the pass@1. 

\begin{figure}[H]
     \centering
     \begin{subfigure}[t]{0.49\textwidth}
         \centering
         \includegraphics[width=\textwidth]{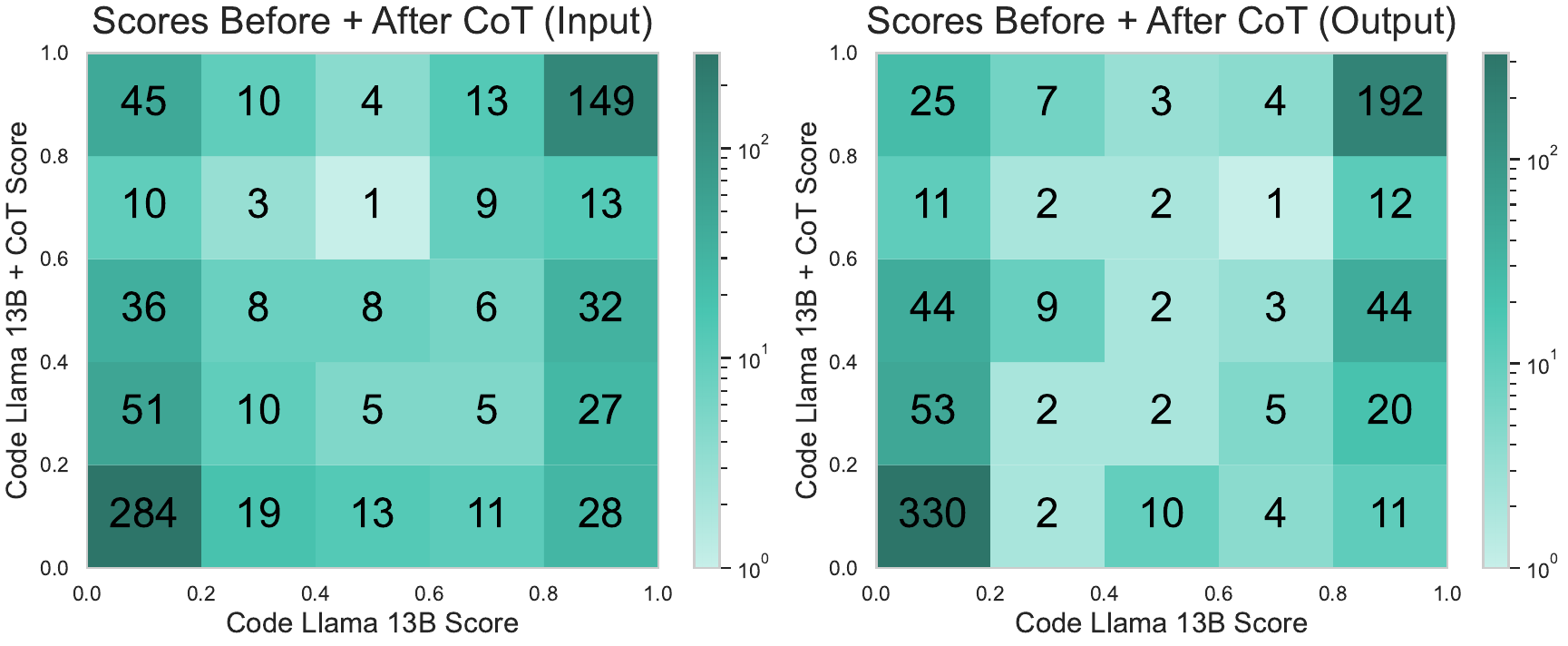}
         \caption{Code Llama 13B}
     \end{subfigure}%
     \hfill
     \begin{subfigure}[t]{0.49\textwidth}
         \centering
         \includegraphics[width=\textwidth]{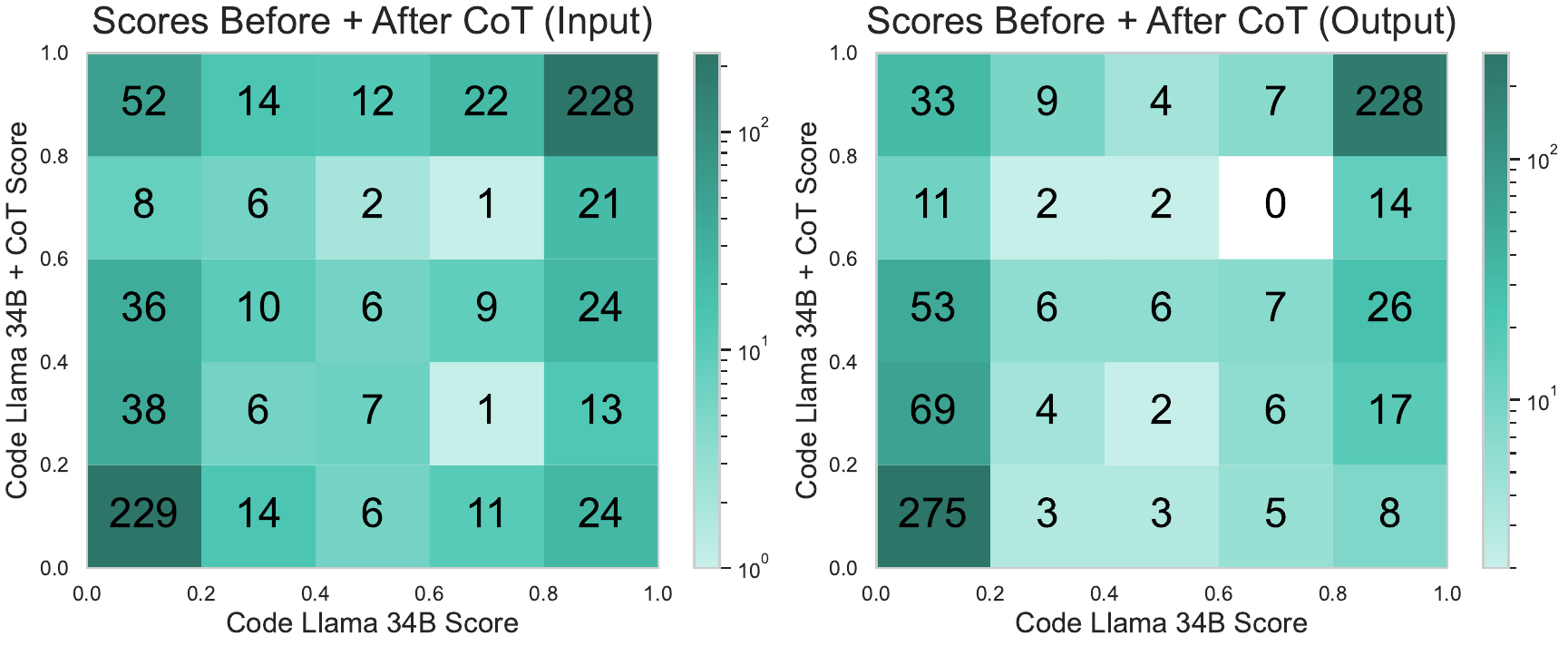}
         \caption{Code Llama 34B}
     \end{subfigure}
     \newline
     \newline
     \newline
     \begin{subfigure}[t]{0.49\textwidth}
         \centering
         \includegraphics[width=\textwidth]{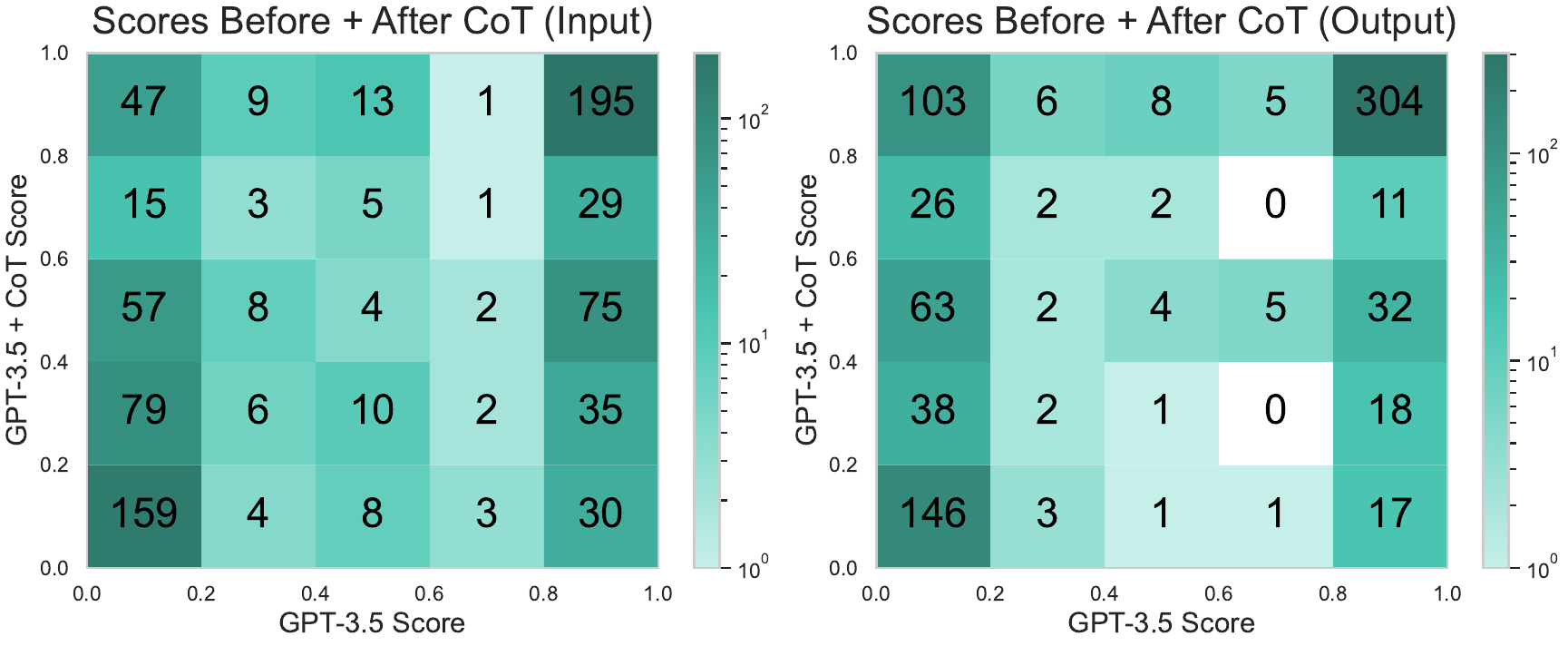}
         \caption{GPT-3.5}
     \end{subfigure}%
     \hfill
     \begin{subfigure}[t]{0.49\textwidth}
         \centering
         \includegraphics[width=\textwidth]{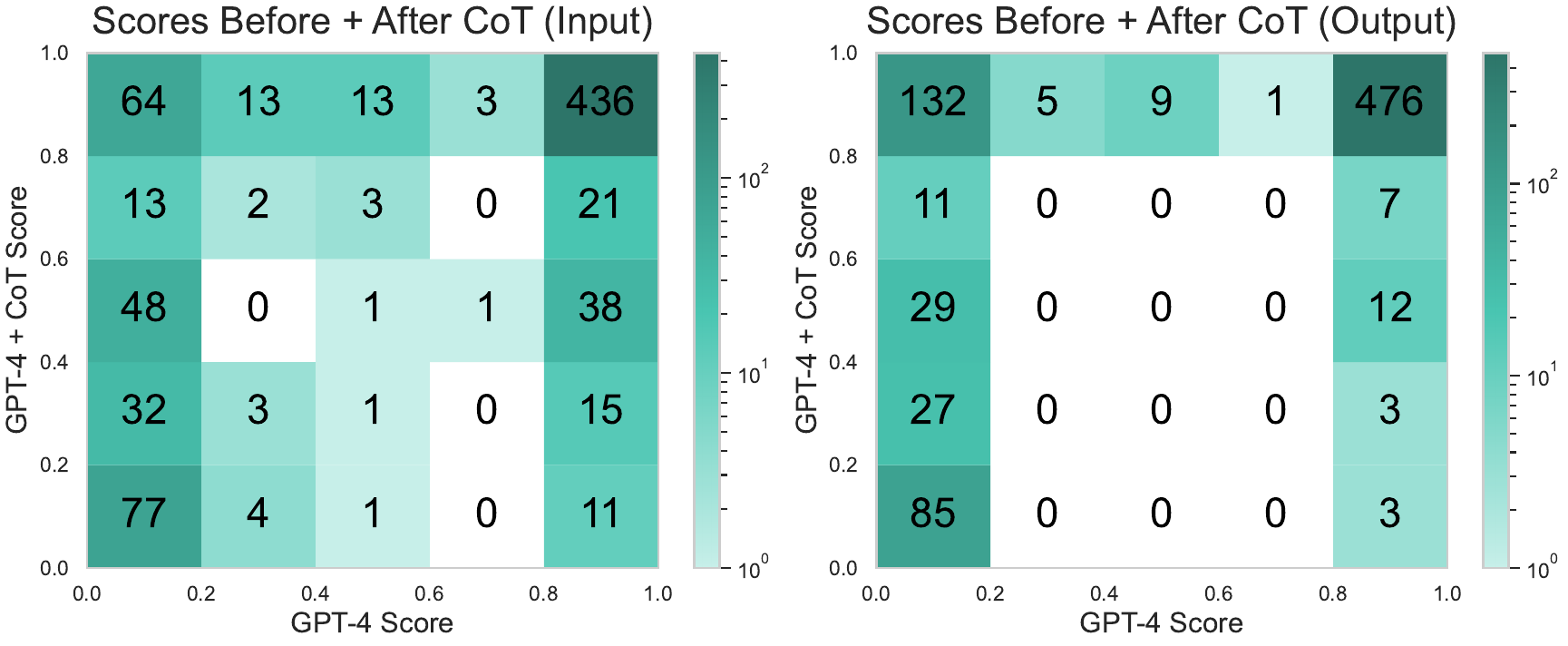}
         \caption{GPT-4}
     \end{subfigure}
     \caption{Confusion Matrix of Direct Prediction vs. CoT Prediction $(T=0.2)$, Granular Version}
     \label{fig:confusion-cot-all-granular}
\end{figure}

In Fig. \ref{fig:confusion-cot-all-granular}, we show a more granular perspective of Fig. \ref{fig:confusion-cot-all}, which again highlights that CoT often decreases the pass@1 score of many samples. Again, we observe a stark difference between the impact of CoT on GPT-4 and other models.

\textbf{Qualitative example of CoT harming performance}: Finally, we show one example of input prediction and one example of output prediction where GPT-4 succeeds without CoT and fails with CoT.
\begin{lstlisting}
# Output Prediction
def f(phone_number):
    while phone_number.find('77777') != -1:
        phone_number = phone_number.replace('77777', 'seg', 1)
    return phone_number
assert f('7747777722') == '774seg22'

# GPT-4 CoT says that '77777' is not in '7747777722', returning '7747777722'
\end{lstlisting}

\begin{lstlisting}
# Input Prediction
def f(mylist):
    revl = mylist[:]
    revl.reverse()
    mylist.sort(reverse=True)
    return mylist == revl
assert f([5, 8]) == True

# GPT-4 CoT correctly says that "we need to provide a list that remains the same when sorted in descending order and when reversed," but then says the list should already be sorted in descending order, returning f([5, 4, 3, 2, 1]).
\end{lstlisting}

\begin{figure}[H]
     \centering
     \begin{subfigure}[b]{0.49\textwidth}
         \centering
         \includegraphics[width=1\textwidth]{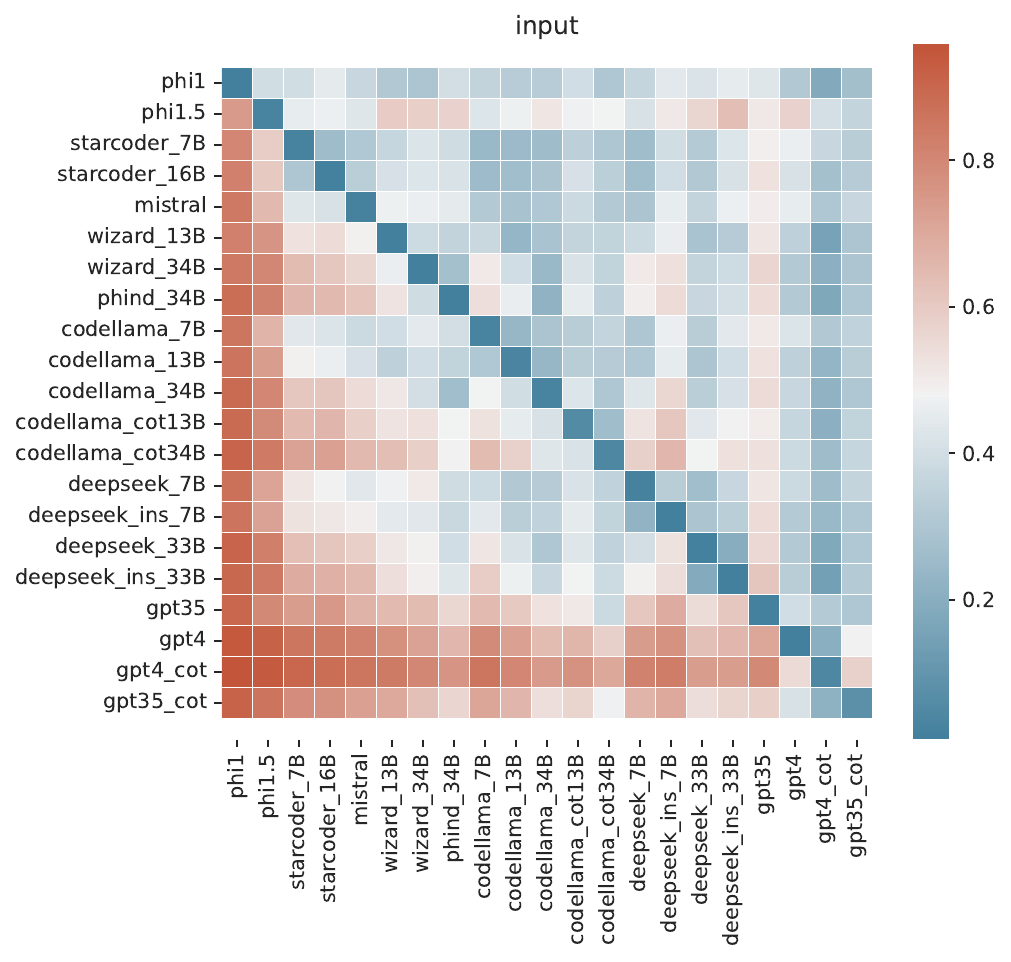}
     \end{subfigure}
     \hfill
     \begin{subfigure}[b]{0.49\textwidth}
         \centering
         \includegraphics[width=1\textwidth]{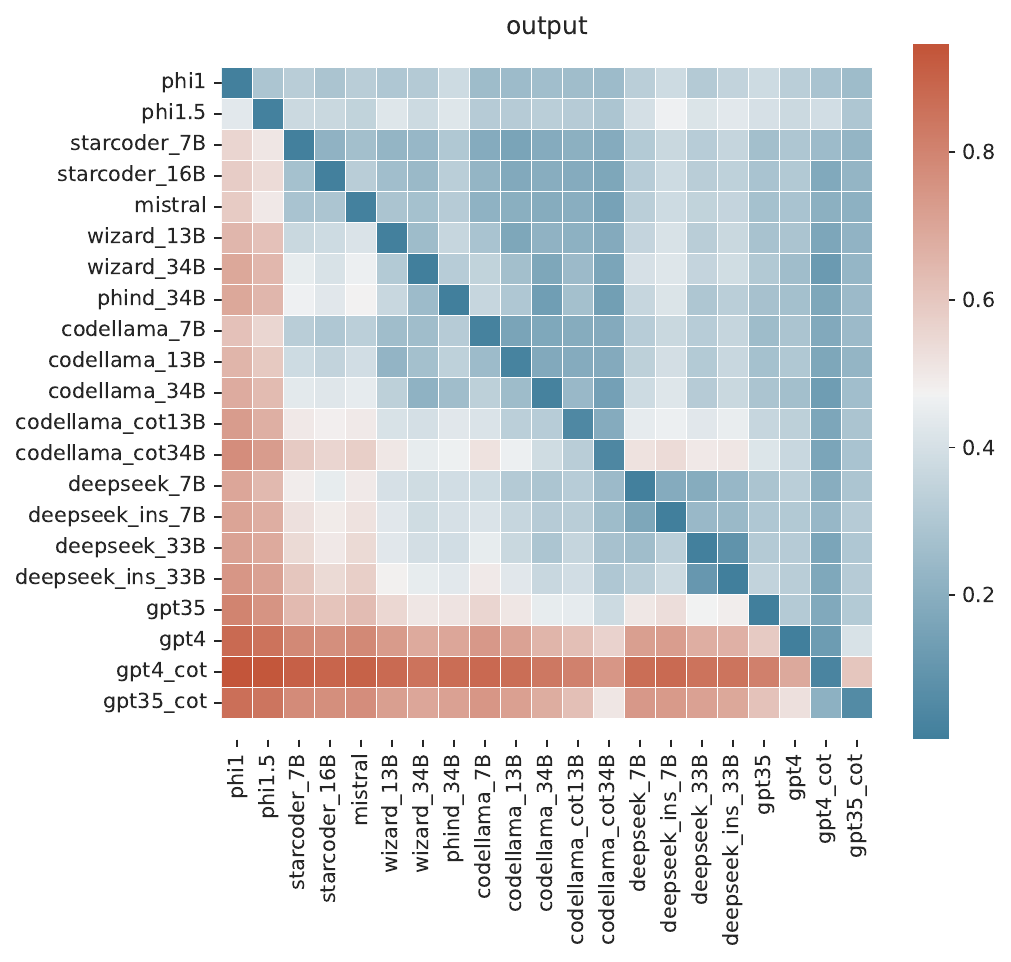}
     \end{subfigure}
     \caption{$P(Y \mid X=0) / P(Y)$ where each $Y$ is the accuracy of models in each row ($X$ for column). }
     \label{fig:error_prob-input-output}
\end{figure}

\textbf{Correlations between failures of different models}: Fig.~\ref{fig:error_prob-input-output} shows $P(Y \mid X=0) / P(Y)$, the accuracy of model $Y$ given that model $X$ fails completely relative to the original accuracy of model $Y$. Although what is hard for a better model tend to be hard for worse models on average, worse models succeeded on some examples where the better models fail completely, showing idiosyncrasies in failures even for the best \textsc{GPT-4 CoT} model.

\subsection{Results on Diversity of Generations} \label{sec:appendix-diversity}

\begin{figure}[H]
     \centering
     \begin{subfigure}[b]{0.49\textwidth}
         \centering
         \includegraphics[width=\textwidth]{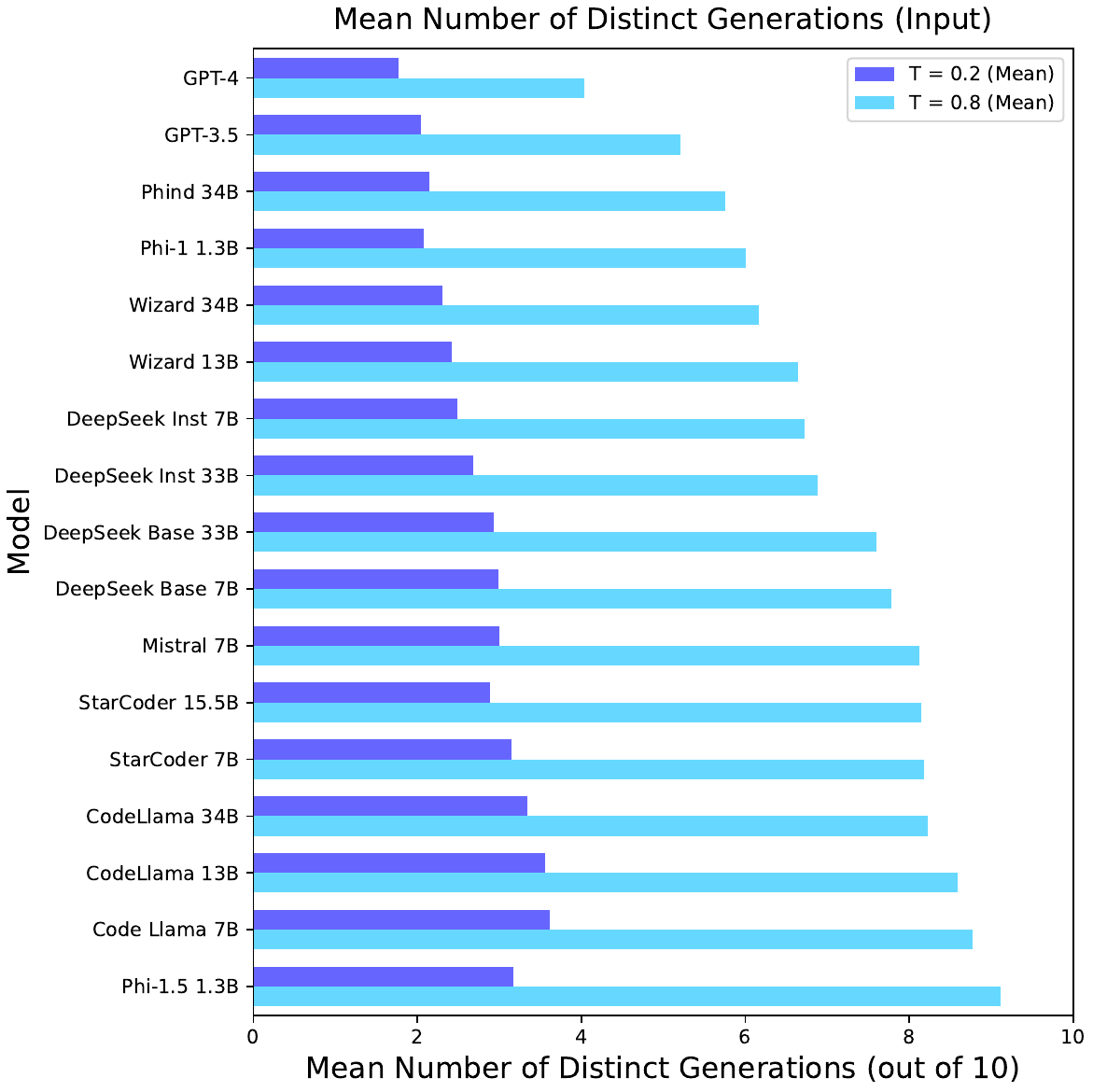}
         \caption{Input prediction}
         \label{fig:sample-diversity-direct-input}
     \end{subfigure}
     \begin{subfigure}[b]{0.49\textwidth}
         \centering
         \includegraphics[width=\textwidth]{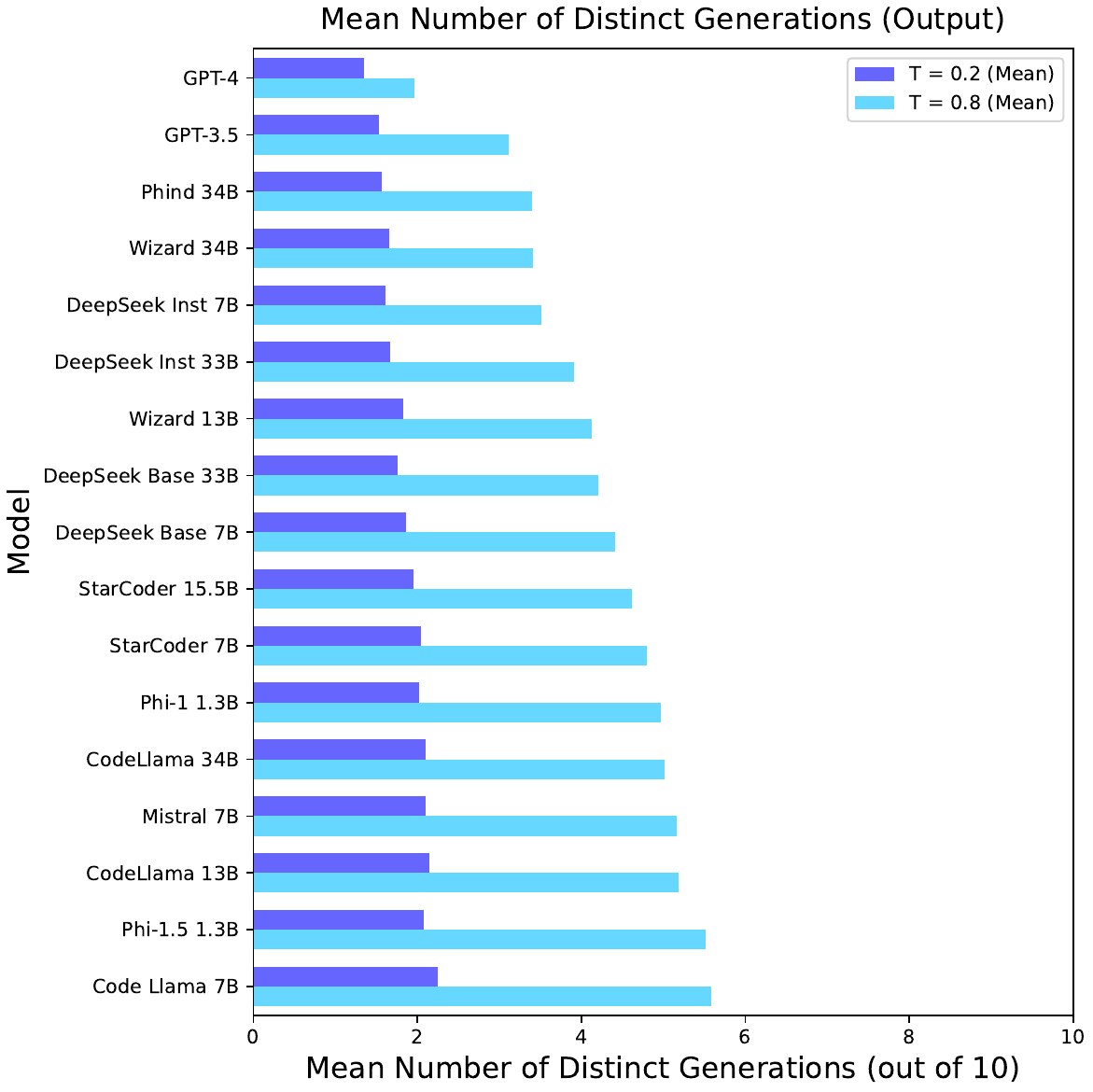}
         \caption{Output prediction}
         \label{fig:sample-diversity-direct-output}
     \end{subfigure}
     \caption{Number of distinct generations of various models (out of 10) at $T=0.2$ and $T=0.8$}
     \label{fig:sample-diversity-direct}
\end{figure}

\textbf{Diversity of generations across models}: Next, we analyze the diversity of generated inputs and outputs across various models (without regard to correctness). In Fig. \ref{fig:sample-diversity-direct}, we plot the mean and median number of unique answers generated across samples of \benchmark for a selection of evaluated models. First, we observe the drastic increase in diversity between using $T=0.2$ and $T=0.8$. Second, by comparing Fig. \ref{fig:sample-diversity-direct-input} with Fig. \ref{fig:sample-diversity-direct-output}, we note that input prediction generally has a larger diversity of generations than output prediction. This may be due to the fact that output prediction only has one correct answer, but input prediction may have multiple correct answers. Third, we observe that at the same temperatures, Code Llama models have the highest diversity, while distilled models like Phind and WizardCoder have a lower diversity.

\begin{figure}[H]
     \centering
     \begin{subfigure}[b]{0.49\textwidth}
         \centering
         \includegraphics[width=\textwidth]{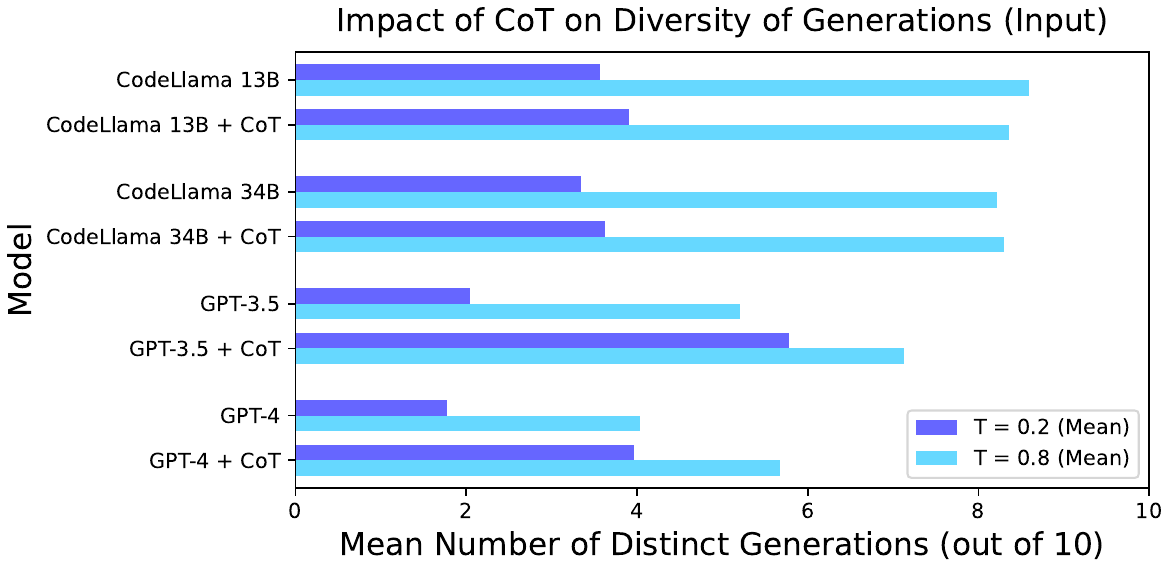}
         \caption{Input prediction}
         \label{fig:sample-diversity-cot-input}
     \end{subfigure}
     \hfill
     \begin{subfigure}[b]{0.49\textwidth}
         \centering
         \includegraphics[width=\textwidth]{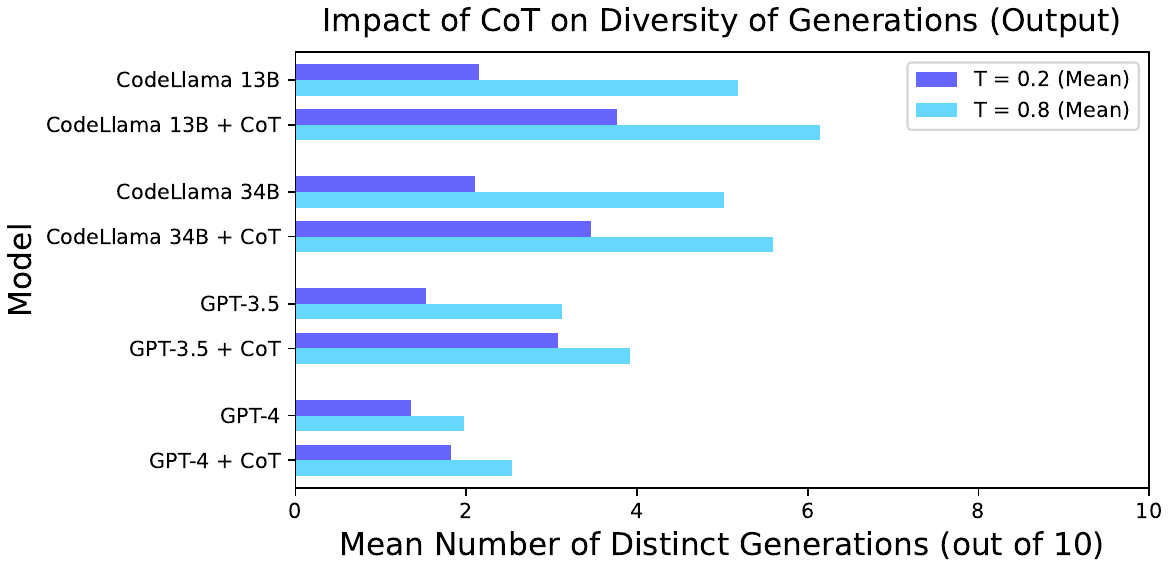}
         \caption{Output prediction}
         \label{fig:sample-diversity-cot-output}
     \end{subfigure}
     \caption{Number of distinct generations of various models (normalized to be out of 10) at $T=0.2$ and $T=0.8$ with and without CoT. We observe that CoT increases the diversity of generations.}
     \label{fig:sample-diversity-cot}
\end{figure}

\textbf{CoT increase the diversity of generations}: In Fig. \ref{fig:sample-diversity-cot}, for each model, we plot the average number of distinct generations across all samples, where different chains of thought with the same input or output prediction are considered identical.  We see again the trend of input prediction generations being more diverse than output prediction generations. Interestingly, we observe that using CoT increases the diversity at both temperatures. 

\begin{figure}[H]
     \centering
     \includegraphics[width=\textwidth]{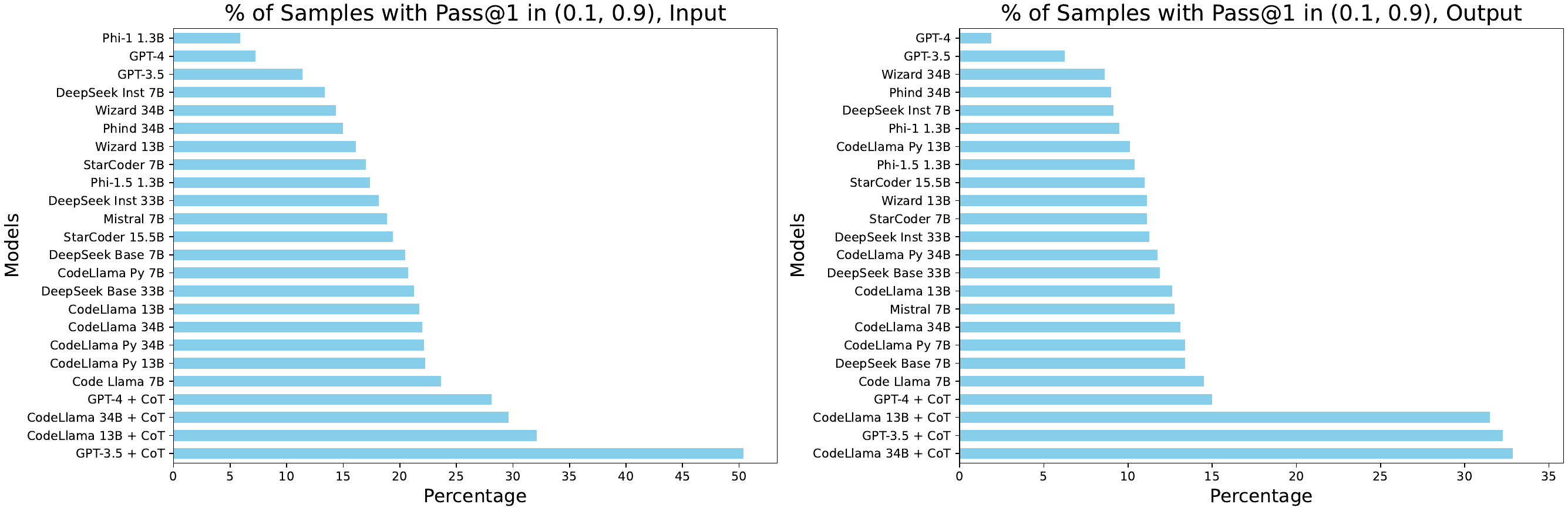}
     \caption{Percentage of samples where pass@1 score is in $(0.1, 0.9)$, exclusive.}
     \label{fig:sample-frequency-all}
\end{figure}

\textbf{Functional diversity via distribution of pass@1 scores}: In Fig.~\ref{fig:sample-frequency-all}, for each model, we plot the percentage of samples where the pass@1 score $(T=0.2)$ is between $0.1$ and $0.9$, exclusive, indicating that the sample is neither "too easy" nor "too hard" for the model. This is a measure of functional diversity because models with more diversity are likely to generate both correct and incorrect samples, leading to more intermediate pass@1 scores. We make a few observations relatively consistent with our prior observations. First, the percentages are relatively low across the board, indicating that at a temperature of $T=0.2$, models are generally producing a majority of correct or a majority of incorrect outputs. Second, distilled models have a much lower functional diversity than base models, for example comparing Phind 34B to CodeLlama 34B or DeepSeek Instruct 33B to DeepSeek Base 33B. Third, CoT greatly increases the functional diversity of models, which is very evident when looking at GPT-3.5 and GPT-4. 

\subsection{Difficulty of Benchmark Samples}
\begin{figure}[h]
    \centering
    \includegraphics[width=\textwidth]{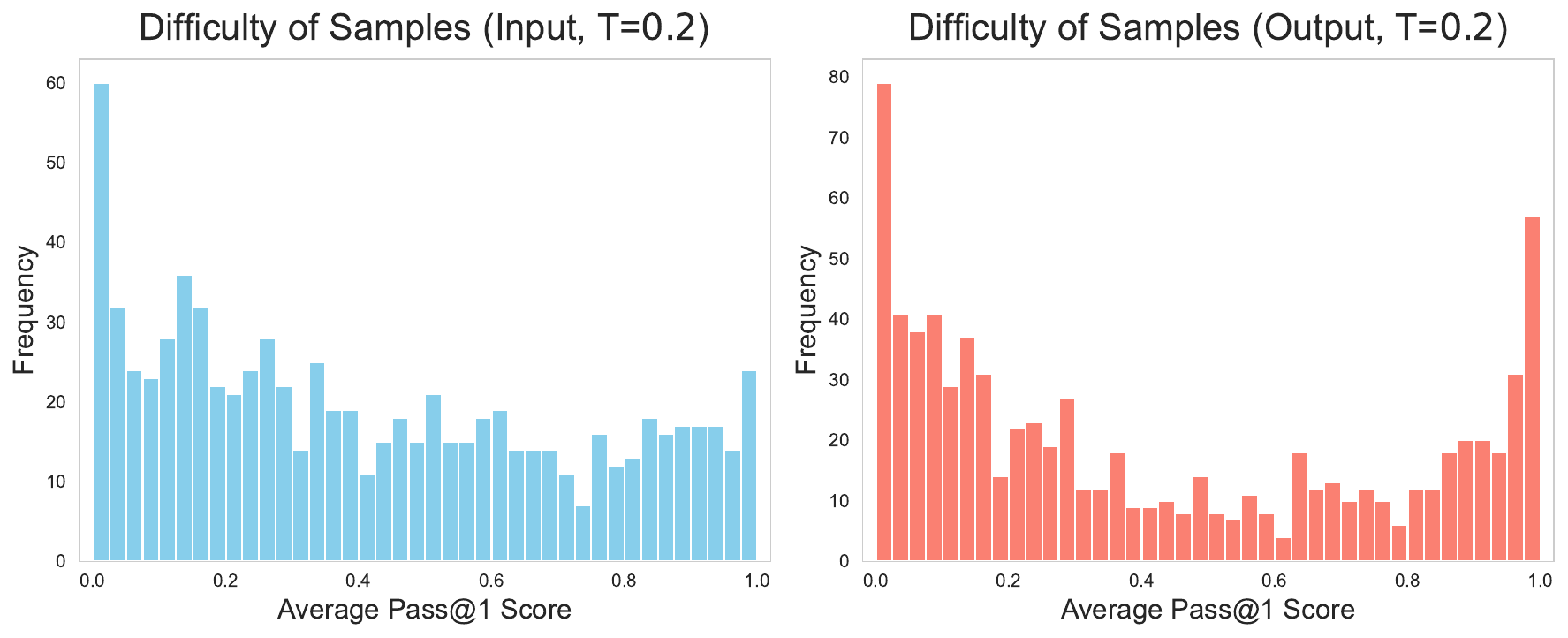}
    \caption{Difficulty of all samples of our benchmark, averaged across all models $(T=0.2)$}
    \label{fig:benchmark-sample-difficulty}
\end{figure}

\textbf{Distribution of sample difficulties}: In Fig. \ref{fig:benchmark-sample-difficulty}, we show the average pass@1 score across all models for $T=0.2$ in order to get a sense of the difficulty distribution of \benchmark.

\begin{figure}[h]
    \centering
     \begin{subfigure}[b]{0.49\textwidth}
         \centering
         \includegraphics[width=\textwidth]{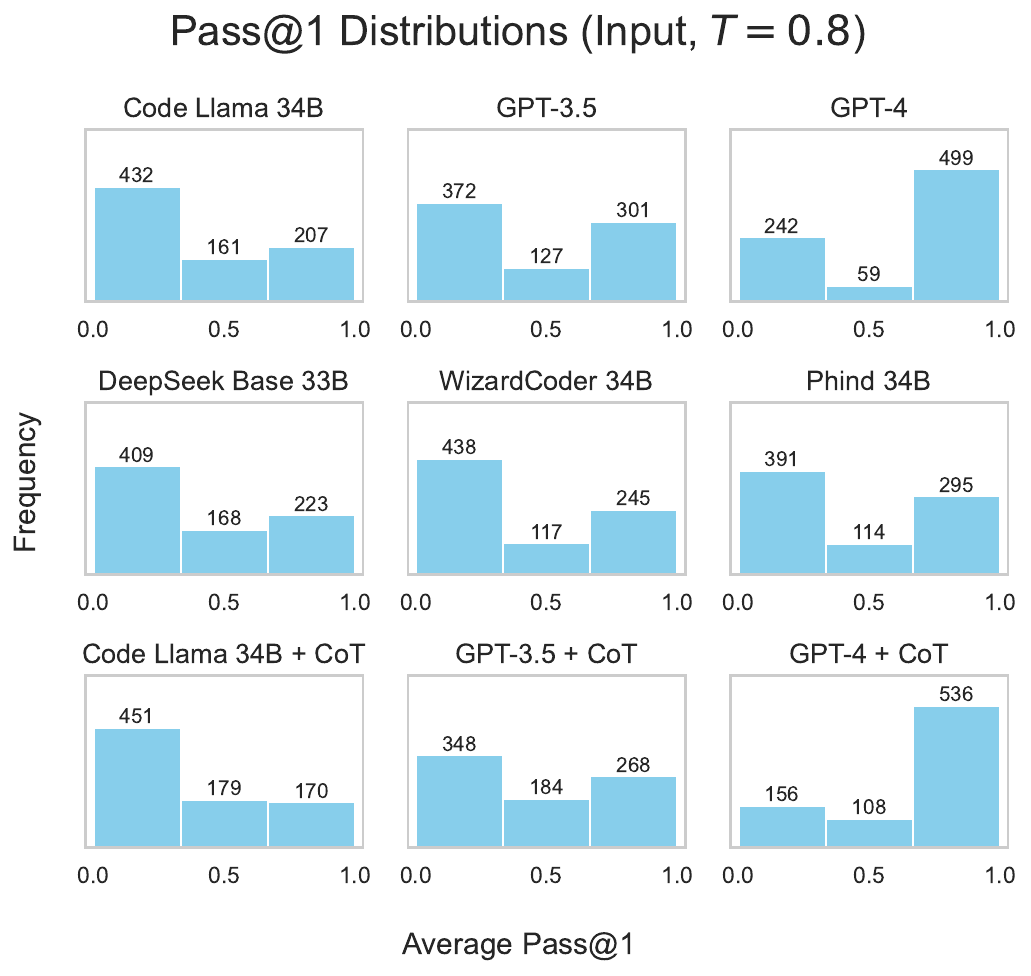}
     \end{subfigure}
     \hfill
     \begin{subfigure}[b]{0.49\textwidth}
         \centering
         \includegraphics[width=\textwidth]{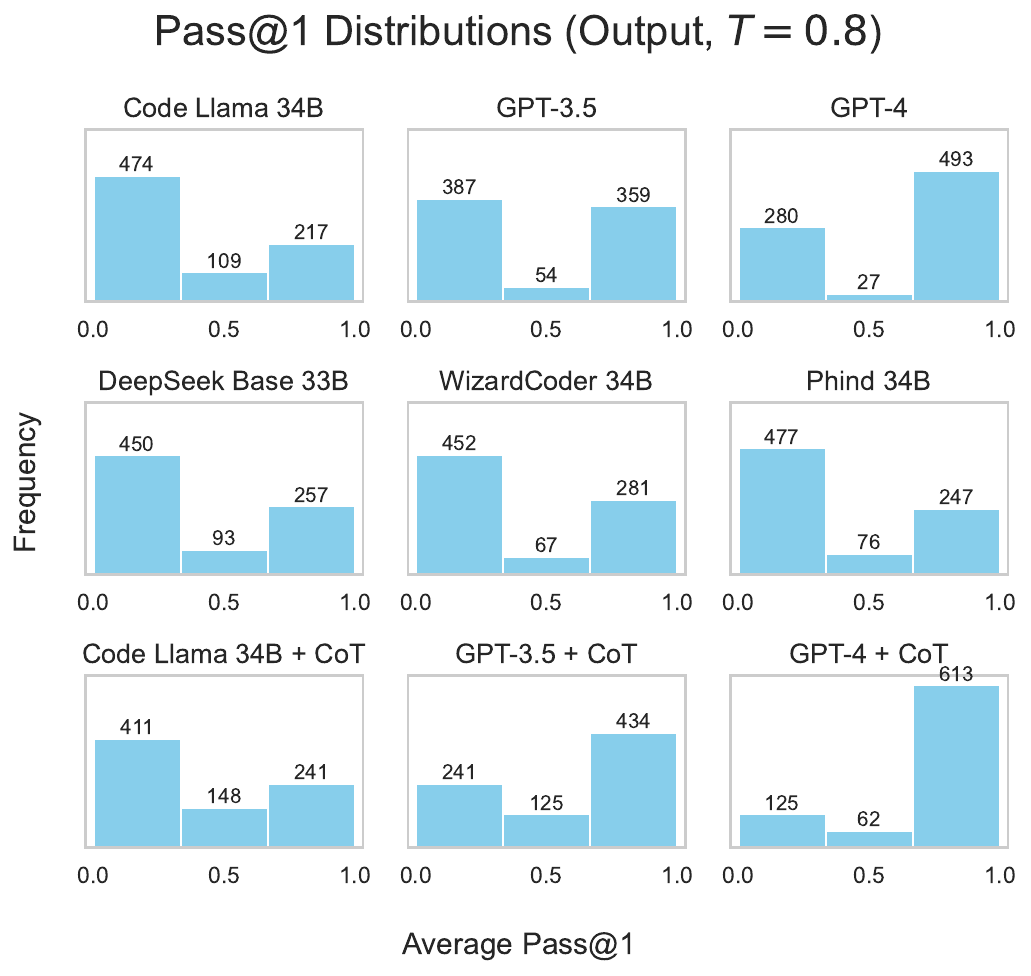}
     \end{subfigure}
     \caption{Pass@1 Distributions of Selected Models}
     \label{fig:pass1-distributions-selected}
\end{figure}

In Fig. \ref{fig:pass1-distributions-selected}, we show the pass@1 distributions of a few of the best-performing models at $T=0.8$. Compared to the overall distribution, the distribution appears to be more bimodal. The output prediction distribution is more bimodal than the input prediction distribution, perhaps reflecting the differences in the tasks themselves. We also see the familiar trend of CoT increasing the number of samples with intermediate scores (in this case between $0.25$ and $0.75$).

\begin{figure}[H]
     \centering
     \includegraphics[width=\textwidth]{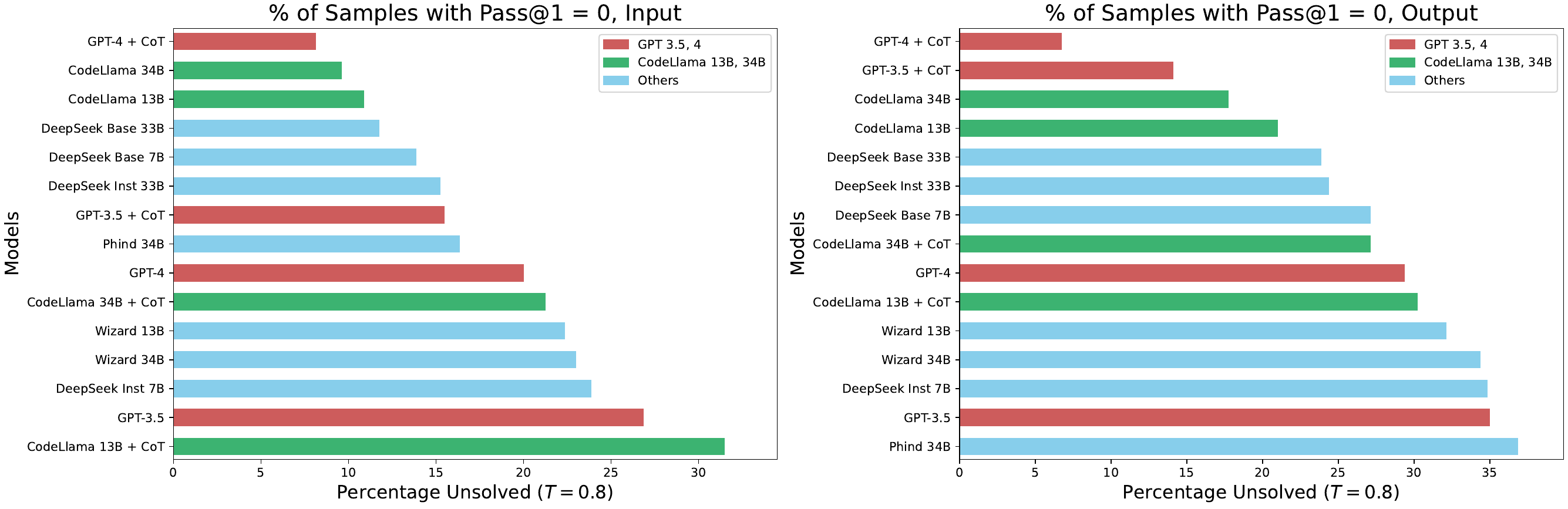}
     \caption{Percentage of samples unsolved, where pass@1 is $0$ ($T=0.8$)}
     \label{fig:sample-frequency-unsolved}
\end{figure}

\begin{figure}[H]
     \centering
     \includegraphics[width=\textwidth]{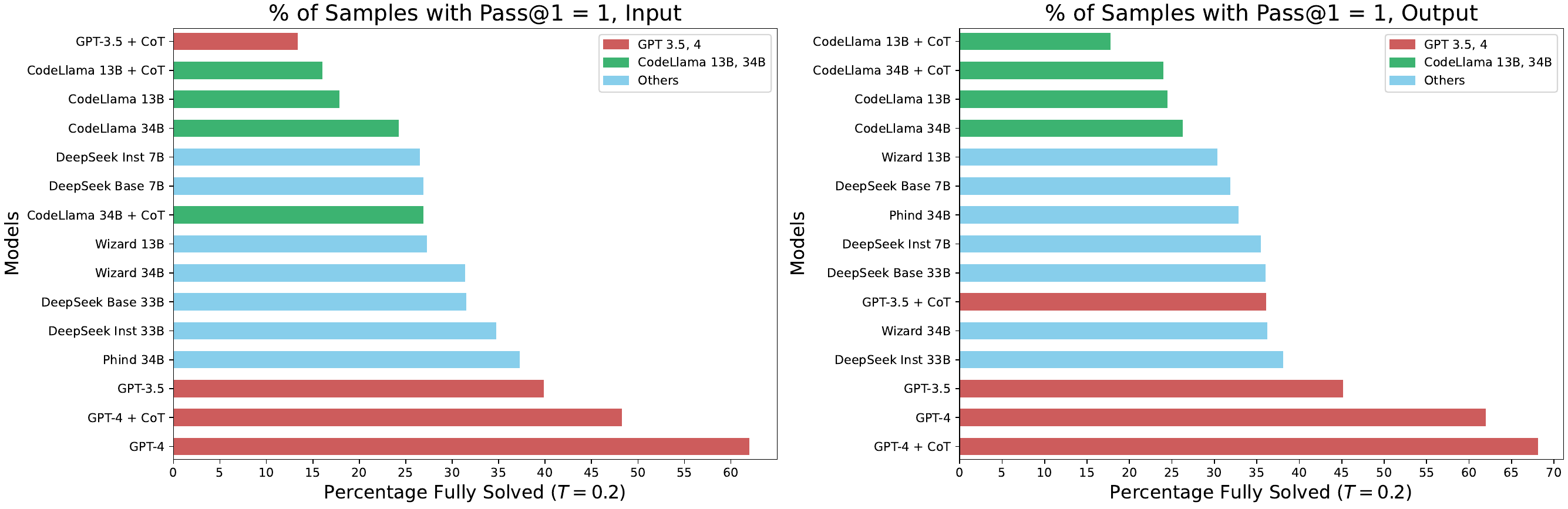}
     \caption{Percentage of samples fully solved, where pass@1 score is $1$ ($T=0.2$)}
     
     \label{fig:sample-frequency-fully-solved}
\end{figure}

\textbf{Fully solved and unsolved samples}: In Figs. \ref{fig:sample-frequency-unsolved} and \ref{fig:sample-frequency-fully-solved}, we examine a different metric, the percentage of examples with pass@1 score equal to $0$ and $1$, respectively, at $T=0.8$. In a sense, this metric captures the ability of models to solve problems. It is also related to diversity, as with a higher diversity, the likelihood of solving the problem may increase. A few observations arise from looking at this metric.

Fig. \ref{fig:sample-frequency-unsolved}, shows the percentage of samples that are completely unsolved by each model, i.e. with $0$ pass@1. We analyze this metric for $T=0.8$, because it leads to more diversity, which would improve this metric. First, when considering non-CoT modes, while GPT-3.5 and GPT-4 (red) are the two best-performing models at pass@1, they perform considerably worse at this metric than models such as Code Llama 34B and DeepSeek Base 33B. Second, instruction-tuned and distilled models (DeepSeek Instruct, Phind, WizardCoder) perform worse than their base counterparts, suggesting that their diversity may have been stifled from adherence to their instruction tuning datasets. Third, we observe that for the two Code Llama models, CoT actually makes this metric worse, but for GPT models, CoT makes it better. For GPT models, we hypothesize that this may be due to the increased diversity of CoT. 

In contrast, Fig. \ref{fig:sample-frequency-fully-solved} shows the percentage of samples that models get fully correct, i.e. with a perfect pass@1. We analyze this metric for $T=0.2$, as it would lead to more consistency, improving this metric. First, we see that GPT-4 excels, achieving over 60\% for both input and output prediction. Second, when comparing base models with instruction tuned models, we see a trend matching the one before: since instruction tuned models are more consistent, they score better on this metric. Third, for output prediction, even though GPT-4 + CoT generally increases diversity, we see that consistency is not sacrificed!

\subsection{Impact of Anonymizing Functions} \label{sec:appendix-anonymization}
As a small ablation to understand the effect of variable names on execution ability, we also test CodeLlama 7B, 13B, and 34B on an anonymized version of a subset of the benchmark, where variable names are replaced with \texttt{x1, x2, ...} identifiers. An example of an anonymized function is shown in Listing \ref{lst:benchmark-anonymized-sample}. We use the same few-shot prompt without anonymization and report both pass@1 ($T=0.2$) and pass@5 ($T=0.8$) results on the anonymized benchmark with $N=10$ samples. The results are shown in Table \ref{tab:benchmark-results-anonymous}. This strengthens the case against memorization affects.

\begin{lstlisting}[caption={Sample of benchmark and anonymized version},label={lst:benchmark-anonymized-sample}, captionpos=t, breaklines=true]
Original:
def f(s):
    nums = ''.join(filter(lambda c:c.isdecimal(), s))
    if nums == '': return 'none'
    m = max([int(num) for num in nums.split(',')])
    return str(m)
assert f('01,001') == '1001'

Anonymized:
def f(x0):
    x1 = ''.join(filter(lambda x2: x2.isdecimal(), x0))
    if x1 == '':
        return 'none'
    x3 = max([int(x4) for x4 in x1.split(',')])
    return str(x3)
assert f('01,001') == '1001'
\end{lstlisting}

\begin{table}[htbp]
    \centering
    \caption{Impact of Anonymization on \benchmark}
    \begin{tabular}{cccccc}
        \toprule
        \multirow{2}{*}{\textbf{Model}} & \multirow{2}{*}{\textbf{Anonymized}} & \multicolumn{2}{c}{\textbf{Input Prediction}} & \multicolumn{2}{c}{\textbf{Output Prediction}} \\
        \cmidrule(lr){3-4} \cmidrule(lr){5-6}
        & & \textbf{Pass@1} & \textbf{Pass@5} & \textbf{Pass@1} & \textbf{Pass@5} \\
        \midrule
        \multirow{3}{*}{CodeLlama 7B} 
	& \xmark & 36.6\% & 48.0\% & 36.4\% & 43.5\% \\
        & \cmark & 37.5\% & 53.3\% & 34.0\% & 46.9\% \\
        & $\Delta$ & \green{+0.9\%} & \green{+5.3\%} & \red{-2.4\%} & \green{+3.4\%} \\
        \midrule
        \multirow{3}{*}{CodeLlama 13B} 
        & \xmark & 39.0\% & 50.2\% & 38.3\% & 44.7\% \\
        & \cmark & 40.0\% & 55.8\% & 36.1\% & 50.6\% \\
        & $\Delta$ & \green{+1.0\%} & \green{+5.6\%} & \red{-2.2\%} & \green{+5.9\%} \\
        \midrule
        \multirow{3}{*}{CodeLlama 34B} 
        & \xmark & 46.5\% & 57.4\% & 41.1\% & 47.5\% \\
        & \cmark & 48.0\% & 63.8\% & 39.1\% & 54.0\% \\
        & $\Delta$ & \green{+1.5\%} & \green{+6.4\%} & \red{-2.0\%} & \green{+6.5\%} \\
        \bottomrule
    \end{tabular}
    \label{tab:benchmark-results-anonymous}
\end{table}

\subsection{Impact of Data-Generating Model}
In the early phases of this work, we were concerned that using Code Llama 34B to generate the benchmark would give the model an unfair advantage. Therefore, we checked the performance of a few models when generating data with Code Llama 13B, GPT-3.5, and GPT-4. The results are shown in Table \ref{tab:benchmark-results-data-generating}. 

These samples were generated using a different prompt and a much more relaxed filter, so the raw scores differ from those in the main text. Across all datasets, we see that the relative ordering of Code Llama 13B, Code Llama 34B, and GPT-3.5 are preserved. We also observed that generating data with GPT-3.5 led to a significantly easier benchmark. After looking at a few samples manually, we believe this is because the resulting inputs are much more predictable and guessable, such as \texttt{f("abcde")} rather than \texttt{f("mai2!")}. Including few-shot examples with random inputs did not improve this issue, and we believe this is an artifact of instruction tuning. We believe that together with the anonymization results in Appendix \ref{sec:appendix-anonymization}, these results provide some evidence that evaluating a model on its own generated data does not seem to provide it a significant advantage.

\begin{table}[H]
    \centering
    \caption{Impact of Data Generating Model}
    \begin{tabular}{cccc}
        \toprule
        \textbf{Data Model} & \textbf{Evaluation Model} & \textbf{Input Pass@1} & \textbf{Output Pass@1} \\
        \midrule
        CL 13B & CL 13B & 28.1\% & 28.4\% \\
        CL 13B & CL 34B & 33.8\% & 29.2\% \\
        \midrule
        CL 34B & CL 13B & 25.1\% & 24.3\% \\
        CL 34B & CL 34B & 29.9\% & 25.4\% \\
        CL 34B & GPT-3.5 & 40.5\% & 36.6\% \\
        \midrule
        GPT-3.5 & CL 13B & 42.3\% & 49.7\% \\
        GPT-3.5 & CL 34B & 52.1\% & 50.7\% \\
        GPT-3.5 & GPT-3.5 & 67.1\% & 67.2\% \\
        \midrule
        GPT-4 & CL 13B & 28.1\% & 42.4\% \\
        GPT-4 & CL 34B & 37.0\% & 44.6\% \\
        \bottomrule
    \end{tabular}
    \label{tab:benchmark-results-data-generating}
\end{table}

\subsection{Fine-tuning} \label{subsec:appendix-finetuning}
We discover three interesting insights from fine-tuning. In the main text, we only discuss insight 3. As a refresher, we fine-tuned \codellamalarge on 138889 samples of Python functions distilled with the procedure outlined in Sec. \ref{sec:benchmark-construction}, without filtering. For the output prediction task, the model was fine-tuned on assertions of the form \texttt{assert f(input) == output}, and for the input prediction task, the model was fine-tuned on assertions of the form \texttt{assert output == f(input)}. During evaluation time, the fine-tuned model was asked to complete assertions of the same format as given in fine-tuning. 

\begin{figure}[H]
     \centering
     \begin{subfigure}[b]{0.48\textwidth}
         \centering
         \includegraphics[scale=0.4]{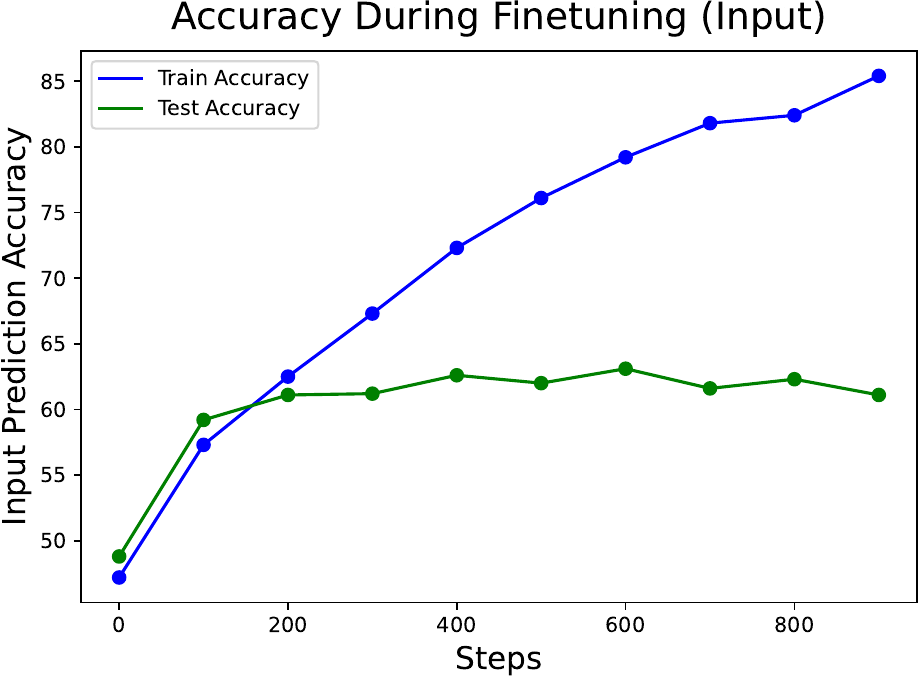}
         \caption{Input prediction}
         \label{fig:finetuning-accuracy-plot-input}
     \end{subfigure}
     \hfill
     \begin{subfigure}[b]{0.48\textwidth}
         \centering
         \includegraphics[scale=0.4]{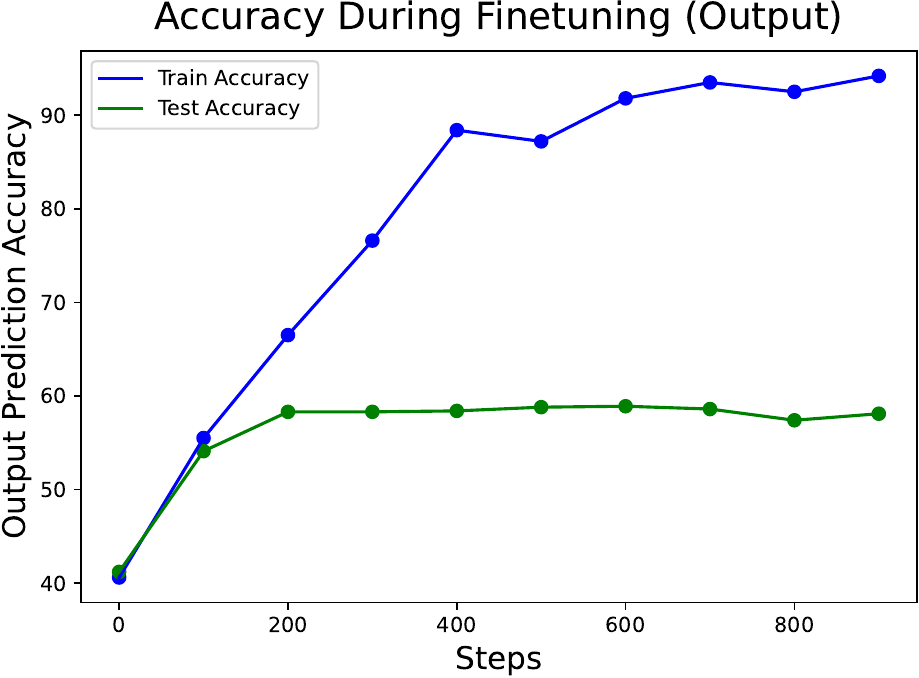}
         \caption{Output prediction}
         \label{fig:finetuning-accuracy-plot-output}
     \end{subfigure}
     \caption{Train accuracy (500 random samples) and test accuracy (superset of \benchmark) while finetuning. For both tasks, there is improvement, and the model steadily fits the training set while plateauing on the testing set.}
     \label{fig:finetuning-accuracy-plot}
\end{figure}

\textbf{1. Direct fine-tuning leads to modest performance improvements}: In the first setup, we analyze a stronger decontamination setup than that in the main text. Specifically, we remove samples that match functions used in the benchmark, even if the input-output pairs are different. In Fig. \ref{fig:finetuning-accuracy-plot}, we show the train and test accuracy of the model during the finetuning process. For ease of evaluation, the train accuracy is reported on a random subset of 500 samples from the finetuning set. The reported test accuracy is on a superset of \benchmark. 

First, we observe that fine-tuning is able to significantly increase performance on both input and output prediction tasks. Second, we observe that while the training accuracy is steadily increasing and the model is able to overfit the training set, the testing accuracy plateaus relatively quickly. This suggesting that simple fine-tuning may not be enough to achieve near-perfect scores on \benchmark. Third, we observe that it is easier to overfit the training set on the output prediction benchmark than on the input prediction benchmark. We hypothesize this may be partially due to the fact that \texttt{assert output == f(input)} is a less natural format for assertions and partially due to the fact that input prediction requires a more sophisticated level of reasoning compared to output prediction.

\begin{figure}[H]
     \centering
     \begin{subfigure}[b]{0.48\textwidth}
         \centering
         \includegraphics[scale=0.4]{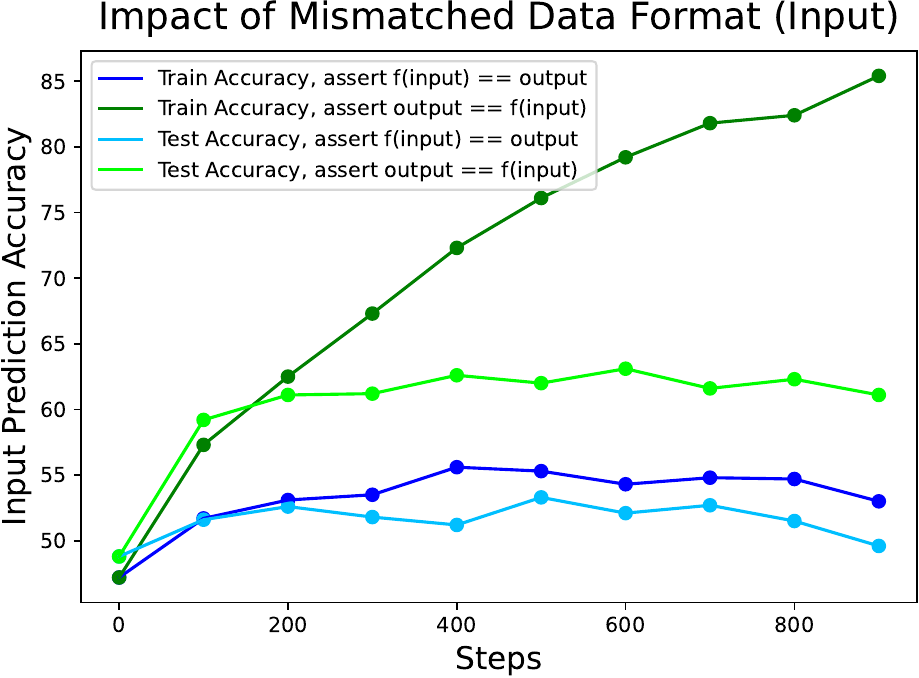}
         \caption{Input prediction}
         \label{fig:finetuning-accuracy-format-plot-input}
     \end{subfigure}
     \hfill
     \begin{subfigure}[b]{0.48\textwidth}
         \centering
         \includegraphics[scale=0.4]{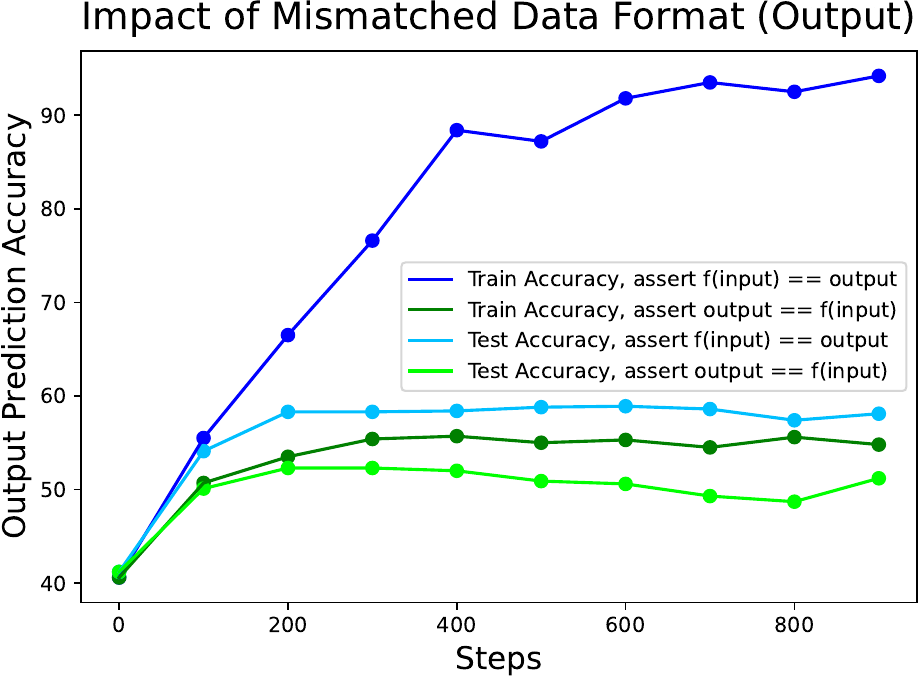}
         \caption{Output prediction}
         \label{fig:finetuning-accuracy-format-plot-output}
     \end{subfigure}
     \caption{Aligning the fine-tuning data format with the evaluation data format is very important for benchmark performance.}
     \label{fig:finetuning-accuracy-format-plot}
\end{figure}

\textbf{2. The format of fine-tuning data greatly impacts its effectiveness}: We also discovered that it is important that the finetuning assertions be formatted in the same way as when evaluating the model at test time. As evidence of this, we fine-tune \codellamalarge with two different sets of assertions, one on \texttt{assert output == f(input)} assertions and the other on \texttt{assert f(input) == output} assertions. We compare the accuracy of the two finetuned models on both input and output prediction in Fig. \ref{fig:finetuning-accuracy-format-plot}. We observe that when the format of the fine-tuning data and the testing data are different, the model even has difficulty overfitting the training set, showing that it may not have fully learned the equivalence of the two formats and the meaning of the \texttt{==} operator. This is perhaps another example of the ``reversal curse'' of LLMs \citep{berglund2023reversal}. The corresponding testing accuracy also plateaued at a lower accuracy when the format was misaligned. For example, in Fig. \ref{fig:finetuning-accuracy-format-plot-input}, comparing the light green line with the light blue line shows almost a 10\% difference in testing accuracy for input prediction when trained on a misaligned format. That being said, fine-tuning still improved performance relative to the base model, even with a mismatched format, showing that the fine-tuning with a mismatched format did still instill some information into the model.

\begin{figure}[H]
     \centering
     \begin{subfigure}[b]{0.48\textwidth}
         \centering
         \includegraphics[scale=0.4]{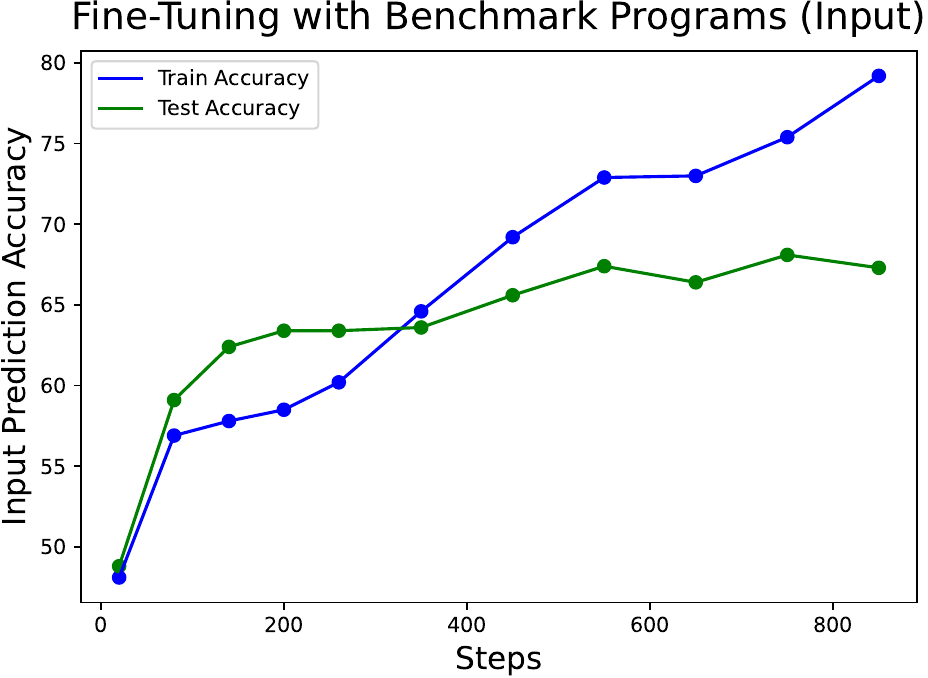}
         \caption{Input prediction}
         \label{fig:finetuning-accuracy-samples-plot-input}
     \end{subfigure}
     \hfill
     \begin{subfigure}[b]{0.48\textwidth}
         \centering
         \includegraphics[scale=0.4]{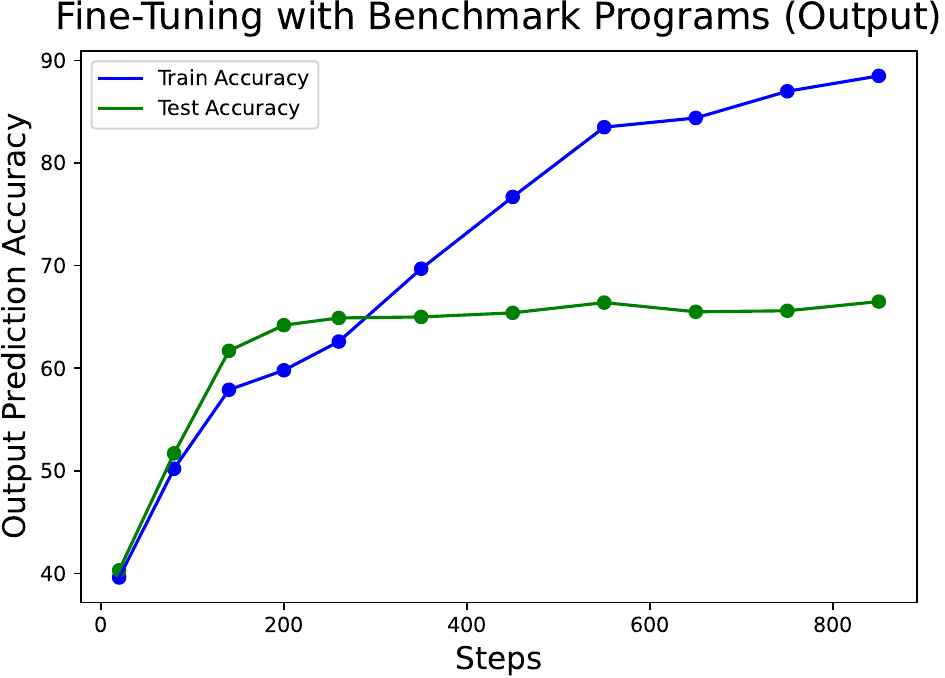}
         \caption{Output prediction}
         \label{fig:finetuning-accuracy-samples-plot-output}
     \end{subfigure}
     \caption{Finetuning 50\% on the original finetuning set and 50\% on "cheating" data}
     \label{fig:finetuning-accuracy-samples-plot}
\end{figure}

\textbf{3. Including benchmark programs still cannot improve test accuracy beyond 70\%}: Finally, we explore the upper limits of fine-tuning on functions and assertions via a "cheating" setup. We curate a small set of \textit{7259 samples consisting only of programs in the benchmark} but with different input-output pairs. We finetune on a mixture of 50\% of the original finetuning set and 50\% of this new set, showing the training and testing accuracy over time in Fig. \ref{fig:finetuning-accuracy-samples-plot}. Despite finetuning on programs very similar to the benchmark, we still observe a plateauing effect in the test accuracy, suggesting that our execution tasks may be too difficult to learn from this simple fine-tuning scheme. Therefore, we suggest a few more fine-tuning ideas for improving our benchmark in Sec. \ref{sec:limitations-future-work}.

% \begin{figure}
%      \centering
%      \begin{subfigure}[b]{0.48\textwidth}
%          \centering
%          \includegraphics[scale=0.4]{figs/finetuning/finetuning_input_samples.pdf}
%          \caption{Input prediction}
%          \label{fig:plots:a}
%      \end{subfigure}
%      \hfill
%      \begin{subfigure}[b]{0.48\textwidth}
%          \centering
%          \includegraphics[scale=0.4]{figs/finetuning/finetuning_output_samples.pdf}
%          \caption{Output prediction}
%          \label{fig:plots:b}
%      \end{subfigure}
%      \caption{Including programs from the benchmark dataset leads to a small boost.}
%      \label{fig:finetuning-accuracy-samples-plot}
% \end{figure}

\section{Prompts}
In this section, we list all the prompts we use throughout the paper. Other than ensuring that generations could be parsed properly, all prompts were not optimized towards any particular models.

\subsection{Benchmark Generation Few-Shot Prompts}
\begin{lstlisting}[caption={All few-shot examples used for benchmark generation},label={lst:benchmark-generation-fewshot}, captionpos=t, breaklines=true]
string_1 = """[TASK]
str.split
[/TASK]
[PYTHON]
def f(text):
    words = text.split()
    result = []
    for i in range(len(words)):
        if i % 2 == 0:
            result.append(words[i][::-1])
        else:
            result.append(words[i].upper())
    return ' '.join(result)
[/PYTHON]
[TEST]
assert f("am7 fiDfd n") == ??
assert f("bnasadl") == ??
assert f("a j c n x X k") == ??
assert f("98 bask2 asoijdf9") = ??
assert f("") == ??
[/TEST]"""

string_2 = """[TASK]
str.capitalize
[/TASK]
[PYTHON]
def f(text):
    a = []
    words = text.split(' ')
    for i in range(len(words)):
        if words[i][0].isdigit():
            return 'no'
        if i%2 == 0:
            a.append(words[i].capitalize())
        else:
            a.append(words[i])
    return ' '.join(a)
[/PYTHON]
[TEST]
assert f("20xk flkawhf") == ??
assert f("lkw hj sfaibw fi 9") == ??
assert f("abbot 2929 mbpu") == ??
assert f("rotor zisxrs fh29nx") == ??
assert f("pxk 5 bxD 9") == ??
[/TEST]"""

string_3 = """[TASK]
str.rindex
[/TASK]
[PYTHON]
def f(text, char):
    index = text.rindex(char)
    result = list(text)
    while index > 0:
        result[index] = result[index-1]
        result[index-1] = char
        index -= 2
    return ''.join(result)
[/PYTHON]
[TEST]
assert f('mnjs krupa', 'u') == ??
assert f('kqwomn0xj', 'm') == ??
assert f('qpfi jzm', 'j') == ??
assert f('102x0zoq', '0') == ??
assert f('nzu  ei,', 'e') == ??
[/TEST]"""

string_4 = """[TASK]
str.rpartition
[/TASK]
[PYTHON]
def f(text, char):
    if char in text:
        pref, char, suff = text.rpartition(char)
        suff = suff[:-len(char)] + char + suff[len(char):]
        return suff + pref
    return text
[/PYTHON]
[TEST]
assert f('smswfwe-r', '-') == ??
assert f(',wpzpppdl/', 'p') == ??
assert f('9284701', '2') == ??
assert f('nvizoh2ja', 'c') == ??
assert f('aaa0a1', 'a') == ??
[/TEST]"""

string_5 = """[TASK]
str.center
[/TASK]
[PYTHON]
def f(text):
    ls = list(text)
    for i in range(1, len(ls) - 1):
        ls.insert(i, '+')
    return ''.join(ls).center((len(ls) - 1) * 2)
[/PYTHON]
[TEST]
assert f('lynel') == ??
assert f('nzoh') == ??
assert f('u') == ??
assert f('anfsoixz') == ??
assert f('xzd') == ??
[/TEST]"""

list_1 = """[TASK]
list.pop
[/TASK]
[PYTHON]
def f(names, num):
    queue = names
    while len(queue) > 1:
        for _ in range(num):
            queue.append(queue.pop(0))
        queue.pop(0)
    return queue.pop()
[/PYTHON]
[TEST]
assert f(['aiwn', 'xke', 'mpwiy'], 2) == ??
assert f(['y', 'z', 'cc', '2', '5', '.', 'zksdfjn'], 7) == ??
assert f(['98bfaj', 'cn11', 'fakldj', 'tjasl', 'a'], 10) == ??
assert f(['aghbvm'], 1) == ??
assert f(['mnv', 'fjw', 'fnk'], 0) == ??
[/TEST]"""

list_2 = """[TASK]
list.insert
[/TASK]
[PYTHON]
def f(text, position, value):
    length = len(text)
    index = position % (length + 1)
    if position < 0 or index < 0:
        index = length // 2
    new_text = list(text)
    new_text.insert(index, value)
    return ''.join(new_text)
[/PYTHON]
[TEST]
assert f('h grateful k', 3, 'h') == ??
assert f('umjwi', -5, 'm') == ??
assert f('coscifysu', 0, 'd') == ??
assert f('fnmart', 4, 'o') == ??
assert f('rzti', -1, 'a') == ??
[/TEST]"""

list_3 = """[TASK]
list.remove
[/TASK]
[PYTHON]
def f(array, elem):
    array.reverse()
    try:
        while elem in array:
            array.remove(elem)
    finally:
        array.reverse()
    return array
[/PYTHON]
[TEST]
assert f([-1, 2, 1, -8, 2], 2) == ??
assert f([], 2) == ??
assert f([1], 1) == ??
assert f([3, 6, 4, -2, 5], 4) == ??
assert f([3, 2, 1, 2, 7, 1], 1) == ??
[/TEST]"""

list_4 = """[TASK]
list.append
[/TASK]
[PYTHON]
def f(nums):
    count = len(nums)
    for i in range(-count+1, 0):
        nums.append(nums[i])
    return nums
[/PYTHON]
[TEST]
assert f([2, 6, 1, 3, 1]) == ??
assert f([7, 1, 2, 6, 0, 2]) == ??
assert f([4, 3, 2, 1, 2, -1, 4, 2]) == ??
assert f([0, 6, 2, -1, -2]) == ??
assert f([-6, -2, 1, -3, 0, 1]) == ??
[/TEST]"""

list_5 = """[TASK]
list.index
[/TASK]
[PYTHON]
def f(nums, swap1, swap2):
    i1 = nums.index(swap1)
    i2 = nums.index(swap2)
    nums[i1], nums[i2], nums[i1 + 1], nums[i2 + 1] = nums[i2], nums[i1], nums[i2 + 1], nums[i1 + 1]
    return nums
[/PYTHON]
[TEST]
assert f([6, 2, 1, 3, 4, 5], 3, 4) == ??
assert f([1, 1, 5, 3, 1, 2], 1, 2) == ??
assert f([1, 2, 1, 4, 1], 4, 2) == ??
assert f([6, 2, 3, 1, 7, 5, 7], 3, 7) == ??
assert f([2, 8, 8, 3, 8, 3, 9], 3, 2) == ??
[/TEST]"""
\end{lstlisting}

\subsection{Direct Prediction Prompts} \label{sec:appendix-direct-prompts}
In Listings \ref{lst:benchmark-direct-input-prompt-llama}, \ref{lst:benchmark-direct-input-prompt-gpt}, \ref{lst:benchmark-direct-output-prompt-llama}, \ref{lst:benchmark-direct-output-prompt-gpt}, and \ref{lst:benchmark-direct-output-prompt-phind}, we include the prompts we use for our evaluation. We use a few-shot prompt for all models other than GPT models. For many models, we observed that using the zero-shot prompt leads to a generation that are not in a easily parsable format, and including the few-shot examples led to predictable formatting. For fairness, we also measured the performance of several few-shot prompts on the GPT models for a randomly sampled subset of the benchmark (instead of the full benchmark for cost reasons). However, we observed a decrease in performance compared to the zero-shot prompts for both input prediction and output prediction. Therefore, we decided to use the zero-shot prompt for GPT models and report numbers using that prompt. In addition, we use a separate output prediction prompt for Phind because the prompt in Listing  \ref{lst:benchmark-direct-output-prompt-llama} often led to explanation text before completing the assert. 

\begin{lstlisting}[caption={Input Prediction (non-GPT)},label={lst:benchmark-direct-input-prompt-llama}, captionpos=t, breaklines=true]
You will be given a function f and an output in the form f(??) == output. Find any input such that executing f on the input leads to the given output. There may be multiple answers, but you should only output one. Think step by step before arriving at an answer. Finally, surround the answer, with no additional words, with [ANSWER] and [/ANSWER] tags. Express your answer as a function call that when executed will give the output.

[PYTHON]
def f(my_list):
    count = 0
    for i in my_list:
        if len(i) % 2 == 0:
            count += 1
    return count
assert f(??) == 3
[/PYTHON]
[ANSWER]
f(["mq", "px", "zy"])
[/ANSWER]

[PYTHON]
def f(s1, s2):
    return s1 + s2
assert f(??) == "banana"
[/PYTHON]
[ANSWER]
f("ba", "nana")
[/ANSWER]

[PYTHON]
{function}
assert f(??) == {output}
[/PYTHON]
[ANSWER]
\end{lstlisting}

\begin{lstlisting}[caption={Input Prediction (GPT)},label={lst:benchmark-direct-input-prompt-gpt}, captionpos=t, breaklines=true]
You will be given a function f and an output in the form output == f(??). Output the completion of the last line so that the code will run without errors by finding any input such that executing f on the input leads to the given output. There may be multiple answers, and you can output any one. Do NOT output any additional information.

{function}
assert {output} == f(
\end{lstlisting}

\begin{lstlisting}[caption={Output Prediction (non-GPT, non-Phind)},label={lst:benchmark-direct-output-prompt-llama}, captionpos=t, breaklines=true]
Based on the given Python code, which may contain errors, complete the assert statement with the output when executing the code on the given test case. Do NOT output any extra information, even if the function is incorrect or incomplete. Do NOT output a description for the assert.

def f(n):
    return n
assert f(17) == 17

{function}
assert f({input}) ==
\end{lstlisting}

\begin{lstlisting}[caption={Output Prediction (GPT)},label={lst:benchmark-direct-output-prompt-gpt}, captionpos=t, breaklines=true]
Based on the given Python code, which may contain errors, complete the assert statement with the output when executing the code on the given test case. Do not output any extra information, even if the function is incorrect or incomplete.

{function}
assert f({input}) == 
\end{lstlisting}

\begin{lstlisting}[caption={Output Prediction (Phind)},label={lst:benchmark-direct-output-prompt-phind}, captionpos=t, breaklines=true]
Based on the given Python code, which may contain errors, complete the assert statement with the output when executing the code on the given test case. Do NOT output any extra information, even if the function is incorrect or incomplete. Output "# done" after the assertion.

def f(n):
    return n
assert f(17) == 17 # done

{function}
assert f({input}) ==
\end{lstlisting}

\subsection{Chain of Thought Prompts}\label{sec:appendix-cot-prompts}
Below, we include the prompts we use for the chain of thought experiments. For the same reasons mentioned in Appendix \ref{sec:appendix-direct-prompts}, use a one-shot prompt for Code Llama models and a zero-shot prompt for GPT models. 

\begin{lstlisting}[caption={CoT input prediction prompt (Code Llama)},label={lst:benchmark-cot-input-prompt-llama}, captionpos=t, breaklines=true]
You will be given a function f and an output in the form f(??) == output. Your task is to find any input such that executing f on the input leads to the given output. There may be multiple answers, but only output one. First, think step by step. Then, surround the answer with [ANSWER] and [/ANSWER] tags. Express your answer as a function call that when executed will give the output.

def f(x):
    return x + 1
assert f(??) == 17

To find an input such that executing f on the input leads to the given output, we can work backwards from the given assertion. We know that f(??) == 17. 

Since the function f(x) returns x + 1, for f(??) to be equal to 17, the value of ?? should be 16. 

Therefore, the function call that will give the output as 17 is:
[ANSWER]f(16)[/ANSWER]    

{function}
assert f(??) == {output}
\end{lstlisting}

\begin{lstlisting}[caption={CoT input prediction prompt (GPT)},label={lst:benchmark-cot-input-prompt-gpt}, captionpos=t, breaklines=true]
You will be given a function f and an output in the form f(??) == output. Your task is to find any input such that executing f on the input leads to the given output. There may be multiple answers, but only output one. First, think step by step. Then, surround the answer with [ANSWER] and [/ANSWER] tags. Express your answer as a function call that when executed will give the output.

{function}
assert f(??) == {output}
\end{lstlisting}

\begin{lstlisting}[caption={CoT output prediction prompt (Code Llama)},label={lst:benchmark-cot-output-prompt-llama}, captionpos=t, breaklines=true]
You are given a function and an input. Complete the assertion with the output of executing the function on the input. First, reason step by step before arriving at an answer. Then, surround the answer as an assertion with [ANSWER] and [/ANSWER] tags.

def f(s):
    return s + "a"
assert f("hi") == ??

The function f takes a string s as input and returns the concatenation of s with the string "a".

To determine the output of executing the function f on the input "hi", we need to concatenate "hi" with "a". 

Therefore, the output of executing the function f on the input "hi" is "hia".

[ANSWER]assert f("hi") == "hia"[/ANSWER]

{function}
assert f(input) == ??
\end{lstlisting}

\begin{lstlisting}[caption={CoT output prediction prompt (GPT)},label={lst:benchmark-cot-output-prompt-gpt}, captionpos=t, breaklines=true]
What should the output of this code be so that the assertion is correct? Reason step by step before arriving at an answer. Finally, surround the answer, with no additional words, with [ANSWER] and [/ANSWER] tags.

{function}
\end{lstlisting}
\section{Qualitative Analysis} \label{sec:appendix-qualitative}
In this section, we see some examples of interesting successes and failures of the best performing model, GPT-4, with and without CoT. GPT-4 is relatively sensitive to its prompt, and slight tweaks in the prompt may lead correct examples to fail or incorrect examples to succeed. However, we believe that these examples are nevertheless interesting and reveal insights into the operating modes of GPT-4. Note that some of these examples may not be in the benchmark and were taken from a larger set of generated examples.

\subsection{Output Prediction without CoT}

\subsubsection{GPT-4 Successes without CoT, Output Prediction}
Even without CoT, we found that GPT-4 achieves impressively high pass@1 scores on output prediction. We highlight a few GPT-4 successes below that we found impressive, suggesting that GPT-4 has the capability to perform somewhat complex reasoning and code execution. 

\begin{lstlisting}[language=python]
def f(text):
    if ',' in text:
        before, _, after = text.partition(',')
        return after + ' ' + before
    return ',' + text.partition(' ')[-1] + ' 0'
assert f('244, 105, -90') == ' 105, -90 244'

# GPT-3.5 output: '-90 244'
# CodeLlama 34B output: '244, 105, -90 0'
\end{lstlisting}

\begin{lstlisting}[language=python]
def f(text):
    text = text.lower()
    count = 0
    for char in text:
        if char.isalpha():
            count += 1
    return count
assert f("The computer factory") == 18

# GPT-3.5 output: 3
# CodeLlama 34B output: 16
\end{lstlisting}

\begin{lstlisting}[language=python]
def f(text):
    d = {}
    updated = []
    for ch in text:
        if ch in d:
            d[ch] += 1
        else:
            d[ch] = 1
    while len(d) != 0:
        el = d.popitem()
        for i in range(el[1]):
            updated.append(el[0])
    return ''.join(updated)
assert f('pdrq') == 'qrdp'

# GPT-3.5 output: 'pdrq'
# CodeLlama 34B output: 'qprd'
\end{lstlisting}

\begin{lstlisting}[language=python]
def f(a, b):
    b.reverse()
    c = b.copy()
    b.extend(a.copy())
    b.extend(c)
    return b
assert f([5, 2, 3], [4, 9, 3, 1]) == [1, 3, 9, 4, 5, 2, 3, 1, 3, 9, 4]

# GPT-3.5 output: [1, 3, 9, 4, 5, 2, 3]
# CodeLlama 34B output: [4, 9, 3, 1, 5, 2, 3, 4, 9, 3, 1]
\end{lstlisting}

\begin{lstlisting}[language=python]
def f(s):
    ret = ';'.join(sorted([c for c in s if c.isalnum()]))
    return ret
assert f('%*^8938a(6^' * 3) == '3;3;3;6;6;6;8;8;8;8;8;8;9;9;9;a;a;a'

# GPT-3.5 and CodeLlama 34B both do not terminate after 500 tokens
\end{lstlisting}

\begin{lstlisting}[language=python]
def f(nums, a, b):
    new_nums = []
    for n in nums:
         if n < a or n > b:
             new_nums.append(n)
    new_nums.sort()
    new_nums.extend(nums)
    return new_nums
assert f([25, 44, 24, 22, 38, 5, 35, 15], 20, 44) == [5, 15, 25, 44, 24, 22, 38, 5, 35, 15]

# GPT-3.5 output: [5, 15, 22, 24, 25, 35, 38, 44, 25, 44, 24, 22, 38, 5, 35, 15]
# CodeLlama 34B output: [5, 15, 22, 24, 25, 35, 38, 44, 25, 44, 24, 22, 38, 5, 35, 15]
\end{lstlisting}

\subsubsection{GPT-4 Failures without CoT, Output Prediction}
We still find a set of relatively simple failures on output prediction, which we expect would be relatively simple without CoT.

\begin{lstlisting}[language=python]
def f(nums):
    nums.reverse()
    return "".join(map(str, nums))
assert f([-1, 9, 3, 1, -2]) == '-2139-1'

# GPT-4 output: "-2, 1, 3, 9, -1"
\end{lstlisting}

\begin{lstlisting}[language=python]
def f(nums, num):
    for i in nums:
        if nums[i]==num:
            return num
    return 'Not Found'
assert f({'elad': 186, 'colton': 162, '12': 5}, 5) == '5'

# GPT-4 output: 'Not found'
\end{lstlisting}

\begin{lstlisting}[language=python]
def f(text):
    dups = list(text)
    dups.append(dups[0])
    return ''.join(dups)
assert f('u') == 'uu'

# GPT-4 output: 'u'
\end{lstlisting}

\begin{lstlisting}[language=python]
def f(match, fill, n):
    return fill[:n] + match
assert f('9', '8', 2) == '89'

# GPT-4 output: '889'
\end{lstlisting}

\begin{lstlisting}[language=python]
def f(string, prefix):
    if string.startswith(prefix):
        return string.removeprefix(prefix)
    return string
assert f("Vipra", "via") == 'Vipra'

# GPT-4 output: ""
\end{lstlisting}

\subsection{Input Prediction without CoT}
Similarly, we highlight examples from input prediction.
\subsubsection{GPT-4 Successes without CoT, Input Prediction}
\begin{lstlisting}[language=python]
def f(l, elems):
    l.reverse()
    l.extend(elems)
    l.extend(l)
    l.reverse()
    l.reverse()
    del l[(len(l)-1):]
    return l
assert f([], [-1, 2, 7, 2, 8]) == [-1, 2, 7, 2, 8, -1, 2, 7, 2]

# GPT-3.5 output: f([2, 7, 2, 8], [-1])
# CodeLlama 34B output: f([-1, 2, 7, 2, 8], [-1, 2, 7, 2])
\end{lstlisting}

\begin{lstlisting}[language=python]
def f(text, position):
    length = len(text)
    index = position % length
    if position < 0 or index < 0:
        index = length // 2
    new_text = list(text)
    new_text.pop(index)
    return ''.join(new_text)
assert f('voxnzcuo', 7) == 'voxnzcu'

# GPT-3.5 output: f('voxnzcu', 42)
# CodeLlama 34B output: f('voxnzcu', -4)
\end{lstlisting}

\begin{lstlisting}[language=python]
def f(data, num):
    new_dict = {}
    temp = list(data.items())
    for i in range(len(temp) - 1, num - 1, -1):
        new_dict[temp[i]] = None
    return temp[num:] + list(new_dict.items())
assert f({2: 10, 3: 1}, 0) == [(2, 10), (3, 1), ((3, 1), None), ((2, 10), None)]

# GPT-3.5 output: f({(2, 10): None, (3, 1): None}, 0)
# CodeLlama 34B output: f({2: 10, 3: 1}, 2)
\end{lstlisting}

\begin{lstlisting}[language=python]
def f(nums):
    nums = [y for y in nums if y > 0]
    if len(nums) <= 3:
        return nums
    nums.reverse()
    half = len(nums)//2
    return nums[:half] + [0]*5 + nums[half:]
assert f([10, -1, 3, 2, -5, -6, 2, 6]) == [6, 2, 0, 0, 0, 0, 0, 2, 3, 10]

# GPT-3.5 output: f([6, 2, 0, 0, 0, 0, 0, 2, 3, 10])
# CodeLlama 34B output: f([6, 2, 3, 10])
\end{lstlisting}

\begin{lstlisting}[language=python]
def f(text):
    upper = 0
    lower = 0
    for i in range(len(text)):
        if text[i].islower():
            lower+=1
        else:
            upper+=1
    if lower > upper:
        return text.swapcase()
    return text
assert f('iAv,sual6eaParntEnthuhsisth,{kasae}') == 'IaV,SUAL6EApARNTeNTHUHSISTH,{KASAE}'

# GPT-3.5 output: f("\'iAv,sual6eaparnteNthuhsisth,{kasae}\'")
# CodeLlama 34B output: f("IaV,sual6eApArnTeNthuhsiStH,{KASAE}")
\end{lstlisting}

\subsubsection{GPT-4 Failures without CoT, Input Prediction}

\begin{lstlisting}[language=python]
def f(name):
    return [name[0], name[1][::-1][0]]
assert f(??) == ['m', 'a']

# GPT-4 output: f(['m', 'apple'])
\end{lstlisting}

\begin{lstlisting}[language=python]
def f(nums):
    if 0 in nums:
        nums.pop(nums.index(0))
        nums.insert(0, 0)
    if 1 in nums:
        nums.pop(nums.index(1))
        nums.append(1)
    return nums
assert f(??) == [0, 0, 1, 0, 1]

# GPT-4 output: f([0, 0, 1, 0, 1])
\end{lstlisting}

\begin{lstlisting}[language=python]
def f(text, speaker):
    while text.startswith(speaker):
        text = text[len(speaker):]
    return text
assert f(??) == 'Do you know who the other was? [NEGMENDS]'

# GPT-4 output: f('Do you know who the other was? [NEGMENDS]', '')
\end{lstlisting}

\begin{lstlisting}[language=python]
def f(text):
    if text.isdecimal():
        return text
    else:
        s = ''.join(sorted(text))
        if s.isdecimal():
            return 'true'
        else:
            return 'false'
assert f(??) == '133476815'

# GPT-4 output: 134678315
\end{lstlisting}

\begin{lstlisting}[language=python]
def f(length, text):
    if len(text) == length:
        return text[::-1]
    return False
assert f(??) == False

# GPT-4 output: f(5, "Hello")
\end{lstlisting}

\subsection{Output Prediction with CoT}
\label{sub:output_cot}
\subsubsection{GPT-4 Successes with CoT, Output Prediction}
We see that when equipped with CoT, GPT-4 has some abilities to simulate an execution process, similar to what was seen in \citep{nye2021show}.

In this example, we see both the ability to reason about the flow of the code and the ability to count uppercase letters.
\begin{lstlisting}
def f(string):
    upper = 0
    for c in string:
        if c.isupper():
            upper += 1
    return upper * (2,1)[upper % 2]
assert f('PoIOarTvpoead') == 8

GPT-4: The function f(string) counts the number of uppercase letters in the input string. If the number of uppercase letters is even, it multiplies this count by 2. If the number of uppercase letters is odd, it multiplies this count by 1.

In the string 'PoIOarTvpoead', there are 4 uppercase letters: 'P', 'I', 'O', and 'T'. Since 4 is an even number, the function will return 4 * 2 = 8.

So, the correct output of the function for the input 'PoIOarTvpoead' is 8.

[ANSWER] 8 [/ANSWER]
\end{lstlisting}

In the next two examples, we see the ability of GPT-4 with CoT to do relatively complex string manipulation tasks.
\begin{lstlisting}
def f(text):
    replaced = list(text)
    while ' ' in replaced: replaced.remove(' ')
    while '@' in replaced: replaced.remove('@')
    while '&' in replaced: replaced.remove('&')
    return ''.join(replaced)
assert f('&C C#BiOt &[(15-5%#-u&1@t') == 'CC#BiOt[(15-5%#-u1t'

GPT-4: The function `f(text)` takes a string as input and returns a new string where all spaces (' '), '@' symbols, and '&' symbols have been removed. 

Let's apply this function to the string '&C C#BiOt &[(15-5%#-u&1@t':

1. Remove all spaces: '&CC#BiOt&[(15-5%#-u&1@t'
2. Remove all '@' symbols: '&CC#BiOt&[(15-5%#-u&1t'
3. Remove all '&' symbols: 'CC#BiOt[(15-5%#-u1t'

So, the output of the function for the given string should be 'CC#BiOt[(15-5%#-u1t'.

[ANSWER] 'CC#BiOt[(15-5%#-u1t' [/ANSWER]
\end{lstlisting}

\begin{lstlisting}
def f(widths):
    reversed_names = []
    for skill in sorted(widths, key=widths.get, reverse=True):
        reversed_names.append("{0}={1}".format(widths[skill], skill))
    return '&'.join(reversed_names)
assert f({'craftsmanship':2, 'upholstery':1, 'tailoring':3}) == '3=tailoring&2=craftsmanship&1=upholstery'

GPT-4: The function f takes a dictionary as an argument. The dictionary keys are skill names and the values are their corresponding widths. The function sorts the dictionary based on the values in descending order. Then it creates a list of strings where each string is a combination of the value and the key separated by an equal sign. Finally, it joins all the strings in the list with an ampersand (&) and returns the resulting string.

Let's break down the steps:

1. The dictionary {'craftsmanship':2, 'upholstery':1, 'tailoring':3} is sorted based on the values in descending order, resulting in the list ['tailoring', 'craftsmanship', 'upholstery'].

2. For each skill in the list, a string is created with the format "{value}={key}". The resulting list is ['3=tailoring', '2=craftsmanship', '1=upholstery'].

3. The strings in the list are joined with an ampersand (&), resulting in the string '3=tailoring&2=craftsmanship&1=upholstery'.

So, the output of the function f({'craftsmanship':2, 'upholstery':1, 'tailoring':3}) should be '3=tailoring&2=craftsmanship&1=upholstery'.

[ANSWER]'3=tailoring&2=craftsmanship&1=upholstery'[/ANSWER]
\end{lstlisting}

In the next two examples, we see the ability to simulate and reason about for/while loops.
\begin{lstlisting}
def f(nums):
    count = 0
    for i in range(len(nums)):
        if len(nums) == 0:
            break
        if count % 2 == 0:
            nums.pop()
        else:
            nums.pop(0)
        count += 1
    return nums
assert f([3, 2, 0, 0, 2, 3]) == []

GPT-4: The function `f(nums)` takes a list of numbers as input. It initializes a counter `count` to 0. Then it enters a loop that runs for the length of the list. 

In each iteration, it checks if the length of the list is 0. If it is, it breaks the loop. This condition will never be true because the loop runs for the initial length of the list and not the current length of the list.

Then it checks if the counter is even. If it is, it removes the last element from the list. If it's not, it removes the first element from the list. 

After each iteration, it increments the counter by 1.

So, for the input list [3, 2, 0, 0, 2, 3], the function will perform the following operations:

- Remove the last element (3) because the counter is 0 (even). The list becomes [3, 2, 0, 0, 2].
- Remove the first element (3) because the counter is 1 (odd). The list becomes [2, 0, 0, 2].
- Remove the last element (2) because the counter is 2 (even). The list becomes [2, 0, 0].
- Remove the first element (2) because the counter is 3 (odd). The list becomes [0, 0].
- Remove the last element (0) because the counter is 4 (even). The list becomes [0].
- Remove the first element (0) because the counter is 5 (odd). The list becomes [].

So, the function should return an empty list for the input [3, 2, 0, 0, 2, 3].

[ANSWER]
[]
[/ANSWER]
\end{lstlisting}

\begin{lstlisting}
def f(number):
    multiplier = 2
    power = 1
    while number > 0:
        remainder = number % multiplier
        number //= multiplier
        if remainder == 0:
            power += 1         
        if number == 1:
            return f'{2 ** power}-free'
assert f(8) == '16-free'

# GPT-4: The function f takes a number as input and checks if it is divisible by 2. If it is, it increments the power by 1 and continues to divide the number by 2 until the number is no longer divisible by 2. The function then returns a string that says the number is "2 to the power of (power + 1)-free".

In the case of the number 8, the function will divide 8 by 2 three times before it is no longer divisible by 2. This means the power will be incremented three times, from 1 to 4. Therefore, the function will return the string "2 to the power of 4-free", or "16-free".

So, the output of the code should be "16-free" for the assertion to be correct.

[ANSWER]"16-free"[/ANSWER]
\end{lstlisting}

\subsubsection{GPT-4 Failures with CoT, Output Prediction} \label{sec:appendix-gpt4-cot-output}
Now, we highlight some failure cases of GPT-4 output prediction with CoT. One interesting failure mode is that the CoT makes a mistake on information that GPT-4 seems to know when prompted independently. For example, in the first example below, the CoT insists that 6173 is less tan 1000. However, when prompted independently, GPT-4 knows this is not true. We provide a few examples of these paired mistakes with proof of knowledge below. The outputs are abbreviated for brevity and to highlight the failures.

The first example shows a failure in number comparison.
\begin{lstlisting}
Prompt: What is the output of the code?

def f(num):
    if 0 < num < 1000 and num != 6174:
        return 'Half Life'
    return 'Not found'
assert f(6173) == ??

GPT-4: ... In the case of `f(6173)`, the number 6173 is !*greater than 0, less than 1000*!, and not equal to 6174, so the function will return 'Half Life'...
\end{lstlisting}

\begin{lstlisting}
Prompt: If num = 6173, is it true that 0 < num < 1000 and num != 6174?

GPT-4: No, the statement is not true. The variable num is assigned the value 6173, so the condition 0 < num < 1000 is false because 6173 is not less than 1000. The second condition, num != 6174, is true because 6173 is indeed not equal to 6174. However, since the first condition is false, the overall statement is false.
\end{lstlisting}

This example is showing a semantic misunderstanding of \texttt{str.islower}.
\begin{lstlisting}
Prompt: What is the output of the code?

def f(text):
    if not text.islower():
        temp_text = ''
        for i in range(len(text)):
            if text[i].islower():
                temp_text += text[i]
        return temp_text
    return text
assert f("816a") == ??

GPT-4: ... !*The function `f` takes a string `text` as input and checks if the string is not all lowercase using the `islower()` method *!...

...!*In the given assertion, the input string is `"816a"`. This string is not all lowercase because it contains digits, which are neither lowercase nor uppercase. Therefore, the function will go into the loop and check each character*!
...
\end{lstlisting}

\begin{lstlisting}
Prompt: What is "816a".islower()?

GPT-4: ... The string "816a" contains digits and a lowercase alphabetic character 'a'. Since the only alphabetic character in the string is lowercase, the `islower()` method would return `True` when called on this string...
\end{lstlisting}

This example shows a failure with sorting \texttt{"h e l l o"}, which seems like it could be a tokenization artifact.

\begin{lstlisting}
Prompt: What is the output of this code?

def f(str):
    return str and ''.join(sorted(str))
assert f("h e l l o") == ??

GPT-4: ...
1. The input string `"h e l l o"` is truthy, so the condition `str` is `True`.
!*2. The `sorted` function is called on the string, which results in the list `[' ', ' ', ' ', 'e', 'h', 'l', 'l', 'o']`.*!
...
\end{lstlisting}

\begin{lstlisting}
Prompt: In Python, what is sorted("h e l l o")?

GPT-4: ...Here's what happens:

```python
result = sorted("h e l l o")
```

The `result` will be a list that contains the characters of the string, including spaces, sorted in ascending order based on their Unicode code points. The space character has a lower code point than any of the letters, so it will appear first in the sorted list. The result will look like this:

```python
[' ', ' ', ' ', ' ', 'e', 'h', 'l', 'l', 'o']
```
\end{lstlisting}

This example shows a failure with the \texttt{str.strip} method.

\begin{lstlisting}
Prompt: What is the output of this code?

def f(text, strip_chars):
    return text[::-1].strip(strip_chars)[::-1]
assert f('tcmfsmj', 'cfj') == ??

GPT-4: The function `f` takes two arguments: `text` and `strip_chars`. It reverses the `text` string, strips characters from both ends that are found in `strip_chars`, and then reverses the string back to its original order.

Let's break down what happens step by step with the input `f('tcmfsmj', 'cfj')`:

1. The string `'tcmfsmj'` is reversed, resulting in `'jmsfmct'`.
2. The `strip` method is called with `'cfj'` as the argument, which removes any leading or trailing characters found in `'cfj'` from the reversed string `'jmsfmct'`.
!*3. The leading `'j'` and the trailing `'c'` and `'t'` are stripped from the string, resulting in `'msfm'`.*!
4. The stripped string `'msfm'` is then reversed again, resulting in `'mfsm'`.

Therefore, the assertion should be:

```python
assert f('tcmfsmj', 'cfj') == 'mfsm'
```

The output of the code, if the assertion passes, is nothing since assertions do not produce output if the condition is true.
\end{lstlisting}

\begin{lstlisting}
Prompt: In Python, what is "jmsfmct".strip("cfj")?

GPT-4: ... Here's what happens step by step:

- The first character 'j' is in the set of characters to remove ('cfj'), so it is removed.
- The next character 'm' is not in the set of characters to remove, so the stripping stops at the beginning of the string.
- Looking at the end of the string, the last character 't' is not in the set of characters to remove, so the stripping stops at the end of the string without removing anything.

The resulting string after the `strip` operation will be `"msfmct"`.
\end{lstlisting}

In this example, we observe a list indexing error.
\begin{lstlisting}
Prompt: What is the output of this code?

def f(nums):
    count = len(nums)
    for i in range(-count+1, 0):
        nums.extend([nums[i], nums[i]])
    return nums
assert f([0, 6, 2, -1, -2]) == ??

GPT-4: ...
- Initial list: `[0, 6, 2, -1, -2]`
- i = -4: Extend with two copies of `nums[-4]` which is `6`. New list: `[0, 6, 2, -1, -2, 6, 6]`
- i = -3: Extend with two copies of !*`nums[-3]` which is `2`.*! New list: `[0, 6, 2, -1, -2, 6, 6, 2, 2]`
...
\end{lstlisting}

\begin{lstlisting}
Prompt: In Python, what is the value at index -3 of [0, 6, 2, -1, -2, 6, 6]?

GPT-4: In Python, negative indices count from the end of the list, with -1 being the last element. To find the value at index -3, we count three places from the end of the list.

Given the list:

```
[0, 6, 2, -1, -2, 6, 6]
```

The indices of the elements are as follows:

```
 0   1   2   3   4   5   6  (positive indices)
-7  -6  -5  -4  -3  -2  -1  (negative indices)
```

So, the value at index -3 is `-2`.
\end{lstlisting}

In this example, we see a string indexing error.

\begin{lstlisting}
Prompt: In Python, what is the output of this code?

def f(text):
    for i in range(-len(text), 0):
        if text[i].islower():
            return text[:i] + text[i].swapcase() + text[-i:]
    return text
assert f('Boo5') == ??

GPT-4: ...
- When `i = -3`, `text[i]` is `'o'`, which is lowercase. So the function will return a new string according to the rule above...

!*- `text[-i:]` is `'o5'`*! (the substring from the character after the current character to the end of the string).

Therefore, the function will return `'BOo5'`...
\end{lstlisting}

\begin{lstlisting}
Prompt: In Python, if text = "Boo5" and i = -3, what is text[-i:]?

GPT-4: ...Given `text = "Boo5"` and `i = -3`, let's calculate `-i`:

`-i = -(-3) = 3`

So, `text[-i:]` is equivalent to `text[3:]`, which means we start from the 3rd index (keeping in mind that indexing starts at 0) and take all characters from that position to the end of the string.

The character at index 3 in the string "Boo5" is '5'. Therefore, `text[-i:]` will return '5'.
\end{lstlisting}

\subsection{Input Prediction with CoT}
\subsubsection{GPT-4 Successes with CoT, Output Prediction}

The following examples highlights code reasoning abilities.
\begin{lstlisting}
def f(text, size):
    counter = len(text)
    for i in range(size-int(size%2)):
        text = ' '+text+' '
        counter += 2
        if counter >= size:
            return text
assert f(??) == '     7     '

GPT-4: The function f takes two arguments: a text and a size. It adds spaces to the beginning and end of the text until the length of the text is greater than or equal to the size. The function stops adding spaces when the length of the text is greater than or equal to the size.

In this case, the output is '     7     ', which is a string of length 11. The text in the middle is '7', which is of length 1. The function adds 5 spaces to the beginning and end of the text to reach a total length of 11.

So, the input to the function should be '7' and 11.

[ANSWER]
f('7', 11)
[/ANSWER]
\end{lstlisting}

\begin{lstlisting}
def f(text):
    uppercase_index = text.find('A')
    if uppercase_index >= 0:
        return text[:uppercase_index] + text[text.find('a') + 1 :]
    else:
        return ''.join(sorted(text))
assert f(??) == '   DEGHIVjkptx'

GPT-4: The function f takes a string as input. It first checks if there is an uppercase 'A' in the string. If there is, it removes the first occurrence of lowercase 'a' from the string and returns the string. If there is no uppercase 'A', it sorts the string in alphabetical order and returns it.

In this case, the output is '   DEGHIVjkptx', which is a sorted string. This means that the input string did not contain an uppercase 'A', and the function sorted the string.

To get this output, we can input a string that contains the same characters but in a different order. For example, the string 'xtpkjVIGHED   ' contains the same characters and when sorted by the function, it will give the output '   DEGHIVjkptx'.

So, the answer is 'xtpkjVIGHED   '.

[ANSWER]
'xtpkjVIGHED   '
[/ANSWER]    
\end{lstlisting}

The next example shows some ability in hexadecimal conversion. This may be partially due to memorization and data contamination effects, but this exact hex string was not found after a cursory Google search.

\begin{lstlisting}
def f(text):
    hex_string = text.encode('utf-16')
    result = "0x"
    for c in hex_string:
        result += "%.2X" % c
    return result
assert f(??) == '0xFFFE66006E007700200052006F006D0065002100'

GPT-4: The function f takes a string as input, encodes it in utf-16 format, and then converts each character into a hexadecimal representation. The output '0xFFFE66006E007700200052006F006D0065002100' is the hexadecimal representation of a string in utf-16 format.

To find the input, we need to convert the hexadecimal representation back to a string. We can do this by first removing the '0x' prefix and then converting each pair of hexadecimal digits back to a character.

The hexadecimal string 'FFFE66006E007700200052006F006D0065002100' corresponds to the utf-16 encoded string 'nw Rome!'. 

So, the input to the function f that will give the output '0xFFFE66006E007700200052006F006D0065002100' is 'nw Rome!'.

[ANSWER] f('nw Rome!') [/ANSWER]
\end{lstlisting}

Just like we see in output prediction, GPT-4 is able to reason about non-trivial string manipulation code.
\begin{lstlisting}
def f(text):
    text = text.split(',')
    text.pop(0)
    text.insert(0, text.pop(text.index('T')))
    return 'T' + ',' + ','.join(text)
assert f(??) == 'T,T,Sspp,G ,.tB,Vxk,Cct'

GPT-4: The function f takes a string as input, splits it into a list of substrings divided by commas, removes the first element of the list, finds the index of the substring 'T', removes it from its current position and inserts it at the beginning of the list. Finally, it joins the list into a string separated by commas and adds 'T' at the beginning of the string.

To get the output 'T,T,Sspp,G ,.tB,Vxk,Cct', we need to have 'T' as the second element in the list after splitting the input string by commas. The rest of the elements can be in any order. So, one possible input could be 'any,T,Sspp,G ,.tB,Vxk,Cct'.

Let's check this:

f('any,T,Sspp,G ,.tB,Vxk,Cct') will split the input into ['any', 'T', 'Sspp', 'G ', '.tB', 'Vxk', 'Cct'], remove the first element to get ['T', 'Sspp', 'G ', '.tB', 'Vxk', 'Cct'], find the index of 'T' (which is 0), remove it from the list and insert it at the beginning to get the same list ['T', 'Sspp', 'G ', '.tB', 'Vxk', 'Cct'], join the list into a string 'T,Sspp,G ,.tB,Vxk,Cct' and add 'T' at the beginning to get 'T,T,Sspp,G ,.tB,Vxk,Cct', which is the desired output.

So, the answer is 'any,T,Sspp,G ,.tB,Vxk,Cct'.

[ANSWER] 'any,T,Sspp,G ,.tB,Vxk,Cct' [/ANSWER]
\end{lstlisting}

In the following example, GPT-4 shows simple code analysis abilities, understanding that the first case will produce an error.
\begin{lstlisting}
def f(nums):
    count = len(nums)
    if count == 0:
        nums = [0] * int(nums.pop())
    elif count % 2 == 0:
        nums.clear()
    else:
        del nums[:count//2:]
    return nums
assert f(??) == []

GPT-4: The function f takes a list of numbers as input. If the length of the list is 0, it tries to pop an element from the list, which will cause an error because the list is empty. If the length of the list is even, it clears the list. If the length of the list is odd, it deletes the first half of the list. 

To get an empty list as output, we can input an even-length list. 

For example, we can input a list with two elements: [1, 2]. 

So, the answer is [1, 2].

[ANSWER]
[1, 2]
[/ANSWER]
\end{lstlisting}
\subsubsection{GPT-4 Failures with CoT, Input Prediction}
As with output prediction, also notice some relatively simple failures of CoT. One observation is that GPT-4 often tries to check the incorrect input against the assertion and says that it passes, even when it does not. We believe this is not necessarily just due to the lack of execution ability, but a prior that the solution it found is more likely to be correct.

In these first three examples, the model goes directly to a wrong answer without much step-by-step reasoning.

\begin{lstlisting}
def f(text):
    odd = ''
    even = ''
    for i, c in enumerate(text):
        if i % 2 == 0:
            even += c
        else:
            odd += c
    return even + odd.lower()
assert f(??) == 'Mmohamt'

GPT-4: The function f takes a string as input and separates the characters at even and odd indices. It then concatenates the characters at even indices with the characters at odd indices converted to lowercase. 

!*To get the output 'Mmohamt', we can see that the even indexed characters are 'M', 'o', 'a', 't' and the odd indexed characters are 'm', 'h', 'm'.*!

So, the input string can be 'Mmohamt'. 

[ANSWER]
f('Mmohamt')
[/ANSWER]
\end{lstlisting}

\begin{lstlisting}
def f(text):
    string = ''
    for char in text:
        string += char + char.lower()
    return string
assert f(??) == 'llaallaakk'

GPT-4: The function f takes a string as input and for each character in the string, it appends the character itself and its lowercase version to a new string. The function then returns this new string.

Given the output 'llaallaakk', we can see that each character is repeated twice, once in its original form and once in lowercase. !*This means that the input string must have been 'LAK'.*!

So, the input that leads to the given output when passed to function f is 'LAK'.

[ANSWER]
'LAK'
[/ANSWER]
\end{lstlisting}

\begin{lstlisting}
def f(nums, i):
    nums.append(nums.pop(i % len(nums)))
    nums[i % len(nums)] = '?'
    split = nums[::2]
    split.remove('?')
    return split + nums[1::2]
assert f(??) == [2, 4, 4]

GPT-4: The function f takes a list of numbers and an index as input. It pops the element at the index (modulo the length of the list) and appends it to the end of the list. Then it replaces the element at the same index with a '?'. It then splits the list into two lists: one with elements at even indices and one with elements at odd indices. It removes the '?' from the first list and returns the concatenation of the two lists.

To get the output [2, 4, 4], we need to find a list and an index such that after popping the element at the index and appending it to the end, and replacing the element at the index with '?', the list split into even and odd indexed elements (with '?' removed) gives [2, 4, 4].

!*One possible input could be the list [2, 4, 4, 4] and the index 1.*! After popping the element at index 1 and appending it to the end, the list becomes [2, 4, 4, 4]. Replacing the element at index 1 with '?' gives [2, '?', 4, 4]. Splitting into even and odd indexed elements gives [2, 4] and ['?', 4]. Removing '?' from the first list gives [2, 4], and concatenating the two lists gives [2, 4, 4].

So, the input f([2, 4, 4, 4], 1) will give the output [2, 4, 4].
\end{lstlisting}

The following example is a bit tricky, because the naming of the variables does not match the semantics of str.removeprefix. 
\begin{lstlisting}
def f(prefix, s):
    return str.removeprefix(prefix, s)
assert f(??) =='hymi'

GPT-4: First, we need to understand what the function does. !*The function f takes two arguments: a prefix and a string s. It then removes the prefix from the string s and returns the result.*!

Given that the output is 'hymi', we can infer that the string s must have ended with 'hymi' after the prefix was removed. Therefore, we can choose any prefix and append 'hymi' to it to form the string s.

For example, if we choose the prefix to be 'hello', then the string s would be 'hellohymi'. When the prefix 'hello' is removed from 'hellohymi', the result is 'hymi', which is the desired output.

So, the input to the function f that would give the output 'hymi' is ('hello', 'hellohymi').

[ANSWER]
('hello', 'hellohymi')
[/ANSWER]
\end{lstlisting}

The following example shows a case where GPT-4 "checks" the forward execution incorrectly.
\begin{lstlisting}
def f(text):
    count = len(text)
    for i in range(-count+1, 0):
        text = text + text[i]
    return text
assert f(??) =='wlace Alc l  '

GPT-4: The function f takes a string as input and appends the characters from the end of the string to the string itself. The number of characters appended is equal to the length of the original string minus one. 

To get the output 'wlace Alc l  ', we need to find a string that when appended with its last len(string)-1 characters, gives 'wlace Alc l  '. 

By observing the output, we can see that the last 3 characters are spaces. This means that the last 3 characters of the original string must also be spaces. 

The first character of the output is 'w', which is the first character of the original string. The second character of the output is 'l', which is the last character of the original string. 

From this, we can infer that the original string is 'w l  '. 

Let's check this:

!*f('w l  ') = 'w l  ' + ' l  ' = 'wlace Alc l  '}*!

So, the input that gives the output 'wlace Alc l  ' when passed to the function f is 'w l  '. 

So, the answer is:

[ANSWER] 'w l  ' [/ANSWER]
\end{lstlisting}

\subsubsection{GPT-4 Failures with CoT, Examples Requiring Insights}
Unlike output prediction, where stepping through the code line by line suffices, some input prediction examples require insights and observations that can be difficult for models like GPT-4 to make. These examples are a bit more challenging, and passing them can require making these insights. Here, we highlight some of these examples, corresponding relevant insights, and GPT-4's failure modes.

In the following example, \texttt{new\_nums} consists of a sorted portion and an unsorted portion. One insight is that the sorted portion must be a subarray of the unsorted portion. GPT-4 always takes \texttt{[5, 15, 25, 44]} to be the sorted portion and \texttt{[24, 22, 38, 5, 35, 15]} to be the unsorted portion, which cannot be the case because 44 (from the sorted portion) is not in the unsorted portion.

\begin{lstlisting}[language=python]
def f(nums, a, b):
    new_nums = []
    for n in nums:
         if n < a or n > b:
             new_nums.append(n)
    new_nums.sort()
    new_nums.extend(nums)
    return new_nums
assert f(??) == [5, 15, 25, 44, 24, 22, 38, 5, 35, 15]
\end{lstlisting}

In the following example, the simplest solution is to bypass the while loop completely, but the model does not find it. However, the model chooses an output like \texttt{"baec"} because it back-translates the "a" to "i" using the translation table. Unfortunately, it does not take into account that other characters also get translated in the translation table.
\begin{lstlisting}[language=python]
def f(input_string):
    table = str.maketrans('aioe', 'ioua')
    while 'a' in input_string or 'A' in input_string:
        input_string = input_string.translate(table)
    return input_string
assert f(??) == 'biec'
\end{lstlisting}

In the following example, one must notice that \texttt{x} cannot end with a number and \texttt{y} cannot start with a number. Reasoning about \texttt{x+y = 'nisou79-85233'}, GPT-4 incorrectly deduces that the \texttt{y} portion must be \texttt{85233}, while in reality, it must be \texttt{79-85233} because \texttt{x} cannot end with a digit. 

\begin{lstlisting}[language=python]
def f(urn):
    pieces = urn.split('--')
    x = pieces[0].rstrip('0123456789')
    y = pieces[1].lstrip('0123456789')
    return x+y
assert f(??) == 'nisou79-85233'
\end{lstlisting}

In the following example, it is very easy to solve the task by choosing \texttt{source} and \texttt{target} to make \texttt{target in source} false, going directly to the last line. However, GPT-4 tries to go down the more complicated path.
\begin{lstlisting}[language=python]
def f(source, target):
    if target in source:
        prefix, suff = source.split(target)
        prefix = prefix[:len(prefix) // 2] + target + prefix[len(prefix) // 2:]
        return prefix + suff
    return source
assert f('abcjava', 'python') =='abcjava'
\end{lstlisting}

In the following example, an important insight is that the last line of the function is a \texttt{"....".join()}, but that \texttt{"...."} does not occur in the output. This confuses the model.

\begin{lstlisting}[language=python]
def f(letters):
    letters_only = letters.strip("., !?*")
    return "....".join(letters_only.split(" "))
assert f(??) =='h,e,l,l,o,wo,r,ld'
\end{lstlisting}

\subsection{Simple and Isolated GPT-4 Failures} \label{appendix:simple-isolated-failures}
Some of the failures from both input prediction and output prediction can be isolated to capabilities on smaller subtasks. Here, we highlight some interesting subtasks, as well as GPT-4's response to these subtasks. 

\begin{lstlisting}
Prompt: What is a string containing 'a' three times, 'b' three times, 'c' twice, 'd' three times, and 'z' twice?

GPT-4: A string containing the specified number of each letter can be constructed in many different ways, as the order of the characters is not specified. Here is one example of such a string:

!*`aaabbbccdzz`*!
\end{lstlisting}

\begin{lstlisting}
Prompt: In Python, what is " BaB ".rfind(" B ")?

GPT-4: In Python, the `rfind()` method of a string object returns the highest index of the substring if found in the given string. If the substring is not found, it returns -1.

The string `" BaB "` has a space, followed by a capital 'B', followed by another space as the substring we are looking for with `rfind(" B ")`.

Let's break down the string `" BaB "`:

- Index 0: space (' ')
- Index 1: 'B'
- Index 2: 'a'
- Index 3: 'B'
- Index 4: space (' ')

We are looking for the substring `" B "` (space, 'B', space). !*This substring appears only once in the string, starting at index 3 and ending at index 5 (if we consider the space after the last 'B').*!

Therefore, `" BaB ".rfind(" B ")` will return 3, because that is the highest index at which the substring `" B "` starts.
\end{lstlisting}

\begin{lstlisting}
Prompt: In Python, if I have a string s = 'iabnm~~~~~~~~~~', what is s[1::2]?

GPT-4: In Python, the slicing operation `s[start:stop:step]` allows you to extract a substring from a string `s` by specifying a starting index `start`, a stopping index `stop`, and a step `step`. If `start` is omitted, it defaults to the beginning of the string (index 0). If `stop` is omitted, it defaults to the end of the string. If `step` is omitted, it defaults to 1.

In your example, `s[1::2]` means that you want to start at index 1 of the string `s`, go until the end of the string (since no stop index is provided), and select every second character (because the step is 2).

Given the string `s = 'iabnm~~~~~~~~~~'`, let's break down what `s[1::2]` will return:

- Start at index 1: 'a'
!*- Every second character from index 1: 'abnm~~~~~~~~~~' -> 'an~'*!

So, `s[1::2]` will return the string `'an~'`.
\end{lstlisting}

\begin{lstlisting}
Prompt: In Python, what is "+".join(['*', '+', 'n', 'z', 'o', 'h'])?

GPT-4: In Python, the `join` method is used to concatenate a sequence of strings with a specified separator. In the example you provided, `"+".join(['*', '+', 'n', 'z', 'o', 'h'])`, the `join` method is called on the string `"+"`, which is the separator, and it concatenates the elements of the list `['*', '+', 'n', 'z', 'o', 'h']` with a plus sign between each element.

The result of this operation would be a single string where each element from the list is separated by a plus sign:

!*```
*+*+n+z+o+h
```*!

Each element in the list is a string, and they are joined together in the order they appear in the list, with the `+` character inserted between each pair of elements.
\end{lstlisting}

\begin{lstlisting}
Prompt: In Python, if text = "!123Leap and the net will appear" and res = 123, what is text[len(str(res)):]?

GPT-4: In the given Python code snippet, you have a string `text` and an integer `res`.

```python
text = "!123Leap and the net will appear"
res = 123
```

You want to know what `text[len(str(res)):]` evaluates to. Let's break it down step by step:

1. `str(res)` converts the integer `res` to a string, which is `"123"`.
2. `len(str(res))` calculates the length of the string `"123"`, which is `3`.
3. `text[len(str(res)):]` slices the string `text` starting from the index equal to the length of the string representation of `res` (which is `3`) to the end of the string.

So, `text[len(str(res)):]` will return the substring of `text` starting from the 4th character (since indexing in Python starts at 0) to the end of the string.

!*The 4th character in the string `text` is the character `'L'` (the first three characters are `"!12"`), so the result will be:

```python
"Leap and the net will appear"
```*!

This is the substring of `text` starting from the 4th character to the end.
\end{lstlisting}

\begin{lstlisting}
Prompt: In Python, what is "pomodoro".replace("or", "pomodoro")?

GPT-4: In Python, the `replace` method is used to replace occurrences of a substring within a string with another substring. When you call `"pomodoro".replace("or", "pomodoro")`, you are asking Python to replace every occurrence of the substring `"or"` in the string `"pomodoro"` with the substring `"pomodoro"`.

Here's what happens step by step:

1. Python finds the first occurrence of `"or"` in `"pomodoro"`, which is at the end of the string.
2. !*Python replaces this occurrence with `"pomodoro"`, resulting in the string `"pomodopomodoro"`.*!

There is only one occurrence of `"or"` in the original string, so the final result of the `replace` method in this case is `"pomodopomodoro"`.
\end{lstlisting}

\subsection{Failing Gems}
Next, we identify a collection of examples that we find GPT-4 often fails on, even with CoT. Some of these examples overlap with examples shown above, but others are new. Overall, we believe this collection of examples provides an interesting lens to understand the behaviour of GPT-4 on input and output prediction.

\subsubsection{Failing Gems, Output Prediction}
\begin{lstlisting}[breaklines=true, language=python]
def f(nums):
    for i in range(-len(nums), 0):
        nums.insert(-i, nums[i])
    return nums
assert f([-6, -2, 1, -3, 0, 1]) == [-6, -6, -2, 1, 1, 1, -3, 0, 0, 1, 1, -6]

def f(text):
    if not text.islower():
        temp_text = ''
        for i in range(len(text)):
            if text[i].islower():
                temp_text += text[i]
        return temp_text
    return text
assert f("816a") == '816a'

def f(list, separator):
    text = separator.join(list)
    return separator.join(reversed(text))
assert f(['is', 'it', 'top'], '@') == 'p@o@t@@@t@i@@@s@i'

def f(text, res):
    for c in '*\n"':
        text = text.replace(c, '!' + str(res))
    if text.startswith('!'):
        text = text[len(str(res)):]
    return text
assert f('"Leap and the net will appear', 123) == '3Leap and the net will appear'

def f(num):
    if 0 < num < 1000 and num != 6174:
        return 'Half Life'
    return 'Not found'
assert f(6173) == 'Not found'

def f(date):
    return  date[6:] + date[4:6] + date[0:4]
assert f("08-10-2009") == '20090-08-1'

def f(text, suffix):
    if suffix and suffix[-1] in text:
        return f(text.rstrip(suffix[-1]), suffix[:-1])
    else:
        return text
assert f('rpyttc', 'cyt') == 'rpytt'

def f(s, x):
    count = 0
    for i, c in enumerate(s):
        if x in s[i:] and x not in s[:i]:
            count += 1
    return count
assert f('fvyijrtwrjrsasgt', 'g') == 15

def f(text):
    segments = text.split()
    for i in range(len(segments)):
        segments[i] = segments[i][0].upper() + segments[i][1:-1] + segments[i][-1].upper()
    return ' '.join(segments)
assert f("hey !") == 'HeY !!'

def f(pattern, items):
    result = []
    for text in items:
        pos = text.rfind(pattern)
        if pos >= 0:
            result.append(pos)

    return result
assert f(" B ", [" bBb ", " BaB ", " bB", " bBbB ", " bbb"]) == []

def f(str):
    return str and ''.join(sorted(str))
assert f("h e l l o") == '    ehllo'

def f(t):
    return t.replace('or', t.center(len(t), 'o'))
assert f("pomodoro") == 'pomodpomodoroo'
\end{lstlisting}

\subsubsection{Failing Gems, Input Prediction}
\begin{lstlisting}[breaklines=true, language=python]
def f(dimension):
    dinline = str(dimension)[1:].zfill(2)
    return dinline[0] * int(dinline[1])
assert f(??) == 'kkkkk'

def f(text):
    for elem in text:
        if elem.isupper():
            try:
                text.remove(elem)
            except ValueError:
                pass
    return text
assert f(??) == ''

def f(text):
    ls = list(text)
    for i in range(0, len(ls)):
        if ls[i]!='+':
            ls.insert(i, '+')
            ls.insert(i, '*')
            break
    return '+'.join(ls)
assert f('nzoh') == '*+++n+z+o+h'

def f(text):
    new_text = list(text)
    dict = {}
    for char in new_text:
        dict[char] = new_text.count(char)
    return dict
assert f('aaabbbccdddzz') == {'a': 3, 'b': 3, 'c': 2, 'd': 3, 'z': 2}

def f(text):
    odd = ''
    even = ''
    for i, c in enumerate(text):
        if i % 2 == 0:
            even += c
        else:
            odd += c
    return even + odd.lower()
assert f('Mammoth') == 'Mmohamt'

def f(nums, i):
    nums.append(nums.pop(i % len(nums)))
    nums[i % len(nums)] = '?'
    split = nums[::2]
    split.remove('?')
    return split + nums[1::2]
assert f([4, 2, 4, 2], 0) == [2, 4, 4]

def f(prefix, s):
    return str.removeprefix(prefix, s)
assert f('hymi', 'hymifulhxhzpnyihyf') == 'hymi'

def f(text):
    if ',' in text:
        before, _, after = text.partition(',')
        return after + ' ' + before
    return ',' + text.partition(' ')[-1] + ' 0'
assert f('244, 105, -90') == ' 105, -90 244'

def f(s):
    return '{}{}{}'.format(s[3:], s[2], s[5:8])
assert f('jbucwc') == 'cwcuc'

def f(nums):
    for i in range(len(nums)):
        nums.insert(i, nums[i]**2)
    return nums
assert f([1, 2, 4]) == [1, 1, 1, 1, 2, 4]

def f(c, text):
    t = c
    for c in reversed(text):
        t = c + t*2
        t = c + t
    return t + text
assert f(';?', 'i') == 'ii;?;?i'

def f(nums, location, item):
    if len(nums) >= location and 0 <= location:
        return nums.insert(location, item)
    return nums
assert f([1, 2, 3, 4, 5, 6], -5, -5) == [1, 2, 3, 4, 5, 6]

def f(text):
    return max(text.find(ch) for ch in 'aeiou')
assert f("qsqgijwmmhbchoj") == 13
\end{lstlisting}

\end{appendices}

\end{document}